\documentclass[final,authoryear,12pt]{elsarticle}
\usepackage{natbib}
\usepackage{amssymb, amsmath}
\usepackage{amsfonts}
\usepackage{amsthm}
\usepackage{graphics}
\usepackage{bm}
\usepackage{float}
\usepackage{breqn}
\usepackage{appendix}
\usepackage{comment}
\usepackage[multiple]{footmisc}
\usepackage{verbatim}
\usepackage{breqn}
\usepackage{rotating}
\usepackage{color}
\usepackage{subcaption}
\usepackage{booktabs}
\usepackage{multirow}
\usepackage[top=2.5cm, bottom=2.5cm, left=3.0cm, right=2.5cm]{geometry}
\input{ee.sty}
\makeatletter
\renewcommand\section{\@startsection{section}{1}{\z@}%
  {-18\p@ \@plus -6\p@ \@minus -3\p@}%
  {9\p@ \@plus 6\p@ \@minus 3\p@}%
  {\normalsize\bfseries\boldmath}}
\renewcommand\subsection{\@startsection{subsection}{2}{\z@}%
  {-12\p@ \@plus -6\p@ \@minus -3\p@}%
  {3\p@ \@plus 6\p@ \@minus 3\p@}%
  {\normalfont\normalsize\itshape}}
\renewcommand\subsubsection{\@startsection{subsubsection}{3}{\z@}%
  {-12\p@ \@plus -6\p@ \@minus -3\p@}%
  {\p@}%
  {\normalfont\normalsize\itshape}}
\makeatother

\newtheorem{lemma}{Lemma}
\newtheorem{proposition}{Proposition}
\newtheorem{corollary}{Corollary}
\newtheorem{assumption}{Assumption}[section]
\journal{Preliminary and incomplete, please do not cite.}
\theoremstyle{definition}

\floatstyle{ruled}
\newfloat{algorithm}{htbp}{loa}
\floatname{algorithm}{Algorithm}

\begin{document}
\makeatletter
    \def\ps@pprintTitle{
      \let\@oddhead\@empty
      \let\@evenhead\@empty
      \def\@oddfoot{\it\hfil\today}
      \let\@evenfoot\@empty
    }
\bibpunct{(}{)}{,}{a}{}{,}
\bibliographystyle{ecta}

\begin{frontmatter}
\title{Almost Uniform Bias Correction for\\ Persistent Panel Local Projections\tnoteref{label1}}
\tnotetext[label1]{This paper replaces an earlier paper titled \emph{Dynamic Panel Data, Half Panel Jackknife and (near) Uniform Inference}. We would like to thank Nick Brown, Maurice Bun, Harold Chiang, James Duffy, Gerdie Everaert, Ulrich Hounyo, Elena Manresa, Bent Nielsen, Martin Weidner, Kenneth West, Jeffrey Wooldridge, as well as participants at the 2025 EC2 conference in Lugano, the Econometrics Workshop in Oxford, the New York Camp Econometrics XX, and seminars at Oxford, UW Madison, U Albany, Emory, Dalian, Florida State U, and Bank of Lithuania, for helpful comments on various versions of the paper. Financial support from the Dutch Research Council (NWO) under research grants $451-17-002$ and VI.Vidi.231E.030 is gratefully acknowledged by the second author.}
\author[a]{Geert Dhaene}
\author[b]{Art\={u}ras Juodis\corref{cor1}}
\cortext[cor1]{Corresponding author. Amsterdam School of Economics, University of Amsterdam, Roetersstraat 11, 1018 WB Amsterdam, the Netherlands. Email: a.juodis@uva.nl.}

\address[a]{Faculty of Economics and Business, KU Leuven}
\address[b]{Amsterdam School of Economics, University of Amsterdam and Tinbergen Institute}

\begin{abstract}
Panel local projections are widely used to estimate impulse responses in policy analysis. This paper examines the properties of the fixed-effect estimator and various bias-correction methods in panel local projections, focusing on settings with autoregressive control variables. We analyze the bias and variance of these estimators while allowing for persistence in both the target and control variables. Our theoretical findings indicate that the half-panel jackknife has desirable bias-reducing properties beyond the stationary setup. For empirical researchers, we recommend using \emph{bias-correction-adjusted} standard errors to account for the sampling uncertainty associated with the half-panel jackknife estimator, regardless of the time-series dimension of the data.
\end{abstract}

\end{frontmatter}

\section{Introduction}
\label{section:intro}
The impulse response function (IRF) characterizes the dynamic effects of external shocks on economic variables of interest. Historically, IRFs have been estimated using vector autoregressive (VAR) models. In contrast, the local projection (LP) method introduced by \citet{Jorda2005} estimates IRFs directly by regressing the future dependent variable at each horizon on the current values of the regressors using ordinary least squares (OLS). \citet{JordaTaylor2025} review the state of the art in the local projection methodology and discuss best practices, possible extensions of LP methods, and their limitations.

The use of local projections is also gaining ground in the panel-data context. One particularly active field in which the panel LP methodology has been used extensively is macro-finance. In this literature, panel LPs have played an important methodological role in establishing the link between financial crises and economic contractions; see, for example, \citet{mei2023nickellbiaspanellocal} for a review. In many of these applications, both the time-series ($T$) and cross-sectional ($N$) dimensions are fairly small or moderate (say, at most 50 to 100), while the empirical specifications usually include a set of predetermined (or weakly exogenous) control variables. Typically, this set includes multiple lags of the target variable. Irrespective of the dimensions and models considered, the standard fixed-effect (FE) estimator is commonly used as the default estimator. As a result, many of the empirical specifications can be viewed as special cases of general linear panel-data models with weakly exogenous regressors, with dynamic (autoregressive) models constituting a particular case. Hence, the FE estimator for this class of problems is not immune to the well-known \citet{Nickell} (or weak-exogeneity) bias.

Despite the empirical popularity of the panel LP method, the connection between the panel LP literature and the dynamic panel-data literature has long been largely overlooked. This can partly be explained by the fact that typical datasets analyzed using panel LP methods have a relatively large $T$ dimension and a small $N$ dimension. This contrasts with prototypical microeconometric applications, in which $T$ is small and $N$ is generally large. These types of datasets largely shaped the early dynamic panel-data literature; see, for example, \citet{Holts1988}, \citet{AB1991}, and \citet{BB1998}, among many others.

At the same time, as first noted by \citet{HerbstJohannsen2024}, the time-series dimension $T$ is generally not large enough to justify ignoring the bias of the FE estimator, or the OLS estimator more generally, in the presence of weakly exogenous regressors. Consequently, the bias described by \citet{Nickell} for panel autoregressive models is directly relevant for typical panel LP applications. \citet{HerbstJohannsen2024} therefore suggest using a large-$T$ \emph{analytical} bias correction to alleviate the finite-sample bias of the FE estimator.

Analytical bias corrections of this type have been used and proposed in the large-$N$, large-$T$ panel-data literature by, for example, \citet{hahnkuerst2011}, \citet{GalvaoKato2014}, and \citet{MoonWeidnerET}, and can be traced back to the seminal work of \citet{kiviet1995}. As an alternative to the analytical bias correction of the FE estimator in the LP setting, \citet{mei2023nickellbiaspanellocal} suggest using the half-panel jackknife (HPJ) procedure of \citet{SPJ} as a general-purpose bias-correction tool for panel LPs. By construction, this procedure requires no tuning parameter and is completely agnostic about the properties of the underlying data-generating process (DGP).

\paragraph{This paper} We follow the work of \citet{HerbstJohannsen2024} and study the bias and variance properties of the FE estimator and the corresponding bias-correction procedures in a stylized setting with a single persistent autoregressive control variable.

The motivation for this setting has both theoretical and empirical origins. From a theoretical perspective, the simulation results in \citet{HerbstJohannsen2024} indicate that the performance of analytical bias-correction procedures deteriorates as either the autoregressive coefficient $\alpha$ increases or $T$, the time-series dimension, decreases.\footnote{The same observation was made by \citet{mei2023nickellbiaspanellocal} for the HPJ procedure they consider.}
From an empirical perspective, persistence in the main variable of interest, in the form of a large autoregressive parameter $\alpha$, translates into an IRF with a relatively slow rate of decay. This feature is documented by \citet{mei2023nickellbiaspanellocal} for several datasets and is revisited in this paper. We study the asymptotic properties under the simplifying assumption that the corresponding autoregressive parameter approaches one as the time-series dimension grows. This induces a drifting sequence of DGPs and allows us to characterize the relevant subset, or the full range, of the parameter space; see, for example, \citet{ECTA:ECTA1257}.

The contribution of this paper is \emph{twofold}.

\textit{i)} We demonstrate that neither of the bias-correction methods considered delivers uniformly valid inference over the entire parameter space of the autoregressive coefficient, $\alpha \in (-1,1]$. The parameter space over which the HPJ correction is valid is generally larger than that for the analytical bias correction of \citet{HerbstJohannsen2024}.

\textit{ii)} We show how asymptotically valid confidence intervals that fully account for the finite-sample variability associated with bias correction can be constructed when the HPJ bias correction is effective. Our Monte Carlo results indicate practical benefits from using such adjusted standard errors (SEs) for datasets with dimensions typical of empirical applications.

Our results indicate limitations of the currently available bias-correction methods. Hence, our primary aim is to raise awareness among empirical researchers that standard practices for bias correction, including those implemented in standard software, even if generally applicable from a theoretical point of view, may require modification to account for the additional variability associated with bias correction in the panel LP setting. We do not aim to introduce a new econometric method to resolve the Nickell bias or to address the nonuniformity of HPJ-based bias-correction methods. Nor do we aim to compare or compete with ``uniform'' inference procedures developed for specific panel-data models, which are generally difficult to generalize or implement in the presence of general (unrestricted) control variables; see, for example, \citet{doi:10.1108/S0731-905320140000033008}, \citet{ChaoPhilips2017}, and \citet{BunKleibergen2010}. 

\paragraph{Organization} The paper is organized as follows. In Section \ref{section:main}, we present the model and the half-panel jackknife estimator. Asymptotic properties of the FE estimator and its bias-corrected variants in the panel LP setting are investigated in Section \ref{section:asymptotic}, while in Section \ref{section:applicability}, we provide some guidance for empirical researchers based on our theoretical results. In Section \ref{section:empirical}, we use real data to illustrate our findings, while in Section \ref{section:montecarlo}, the finite-sample properties of the methods are explored using simulations. Finally, Section \ref{section:conclusions} concludes. \ref{appendix:inference} discusses limitations of the conventional standard errors in our context. Proofs of our theoretical results are given in \ref{appendix::proofs}.
\section{The panel LP setup}
\label{section:main}
\subsection{The setting}
In what follows, we present the panel LP setting of \citet{HerbstJohannsen2024}. The panel LP model consists of a set of regression models for the target variable $y_{i,t}$, indexed by the impulse response horizon $h$:
\begin{equation}
\label{eq::model}
y_{i,t+h}
=
\eta_{i}^{(h)}
+
\vx_{i,t}'\vbeta^{(h)}
+
u_{i,t}^{(h)},
\end{equation}
for $i=1,\ldots,N$, $t=1,\ldots,T-h$, and $h=0,\ldots,H$. Here, $\eta_{i}^{(h)}$ is a unit-specific unobserved intercept, usually referred to as a ``fixed effect''; $\vx_{i,t}:=\left(x_{i,t},\vc_{i,t-1}'\right)'$ is a vector of observed variables containing the observed or identified scalar structural shock $x_{i,t}$ and a set of control variables $\vc_{i,t-1}$. We partition the coefficient vector as
\begin{equation*}
\vbeta^{(h)}
=
\left(\theta^{(h)},\vdelta^{(h)\prime}\right)'.
\end{equation*}
The first element of the sequence $\{\vbeta^{(h)}\}_{h=0}^{H}$ traces out the impulse response function of interest, which we denote by $\vtheta^{(H)}=(\theta^{(0)},\ldots,\theta^{(H)})'$. We denote the sequence of coefficient vectors associated with the control variables by $\{\vdelta^{(h)}\}_{h=0}^{H}$. Although these coefficients are not of primary interest in LP analysis, they play an important role in what follows. Throughout this paper, we assume that the maximum projection horizon $H$ is fixed and does not increase with the sample sizes $N$ and $T$.\footnote{
In recent independent work, \citet{everaert2026LP} shows that large-$N$, large-$T$ approximations, including those considered in this paper, can be misleading when $H/T$ is sizeable in finite samples. To address this issue, an alternative fixed-$T$ approximation and a corresponding bias-corrected estimator are proposed. Unlike the present paper, however, \citet{everaert2026LP} explicitly assumes time-series stationarity of the data, an assumption that we do not impose. Consequently, neither paper considers the more general setting. Rather, the two papers are complementary: \citet{everaert2026LP} provides bias correction under stationarity, whereas we do not impose this restriction.}

As is common in the empirical literature, we estimate each $\vbeta^{(h)}$ by OLS, treating $\eta_{i}^{(h)}$ as parameters. That is, we use the fixed-effect, or within-group, estimator:
\begin{equation*}
\widehat{\vbeta}_{FE}^{(h)}
=
\left(
\frac{1}{N}\sum_{i=1}^{N}
\frac{1}{T-h}\sum_{t=1}^{T-h}
\widetilde{\vx}_{i,t}\widetilde{\vx}_{i,t}'
\right)^{-1}
\left(
\frac{1}{N}\sum_{i=1}^{N}
\frac{1}{T-h}\sum_{t=1}^{T-h}
\widetilde{\vx}_{i,t}\widetilde{y}_{i,t+h}
\right),
\end{equation*}
where
\begin{equation*}
\widetilde{\vx}_{i,t}
:=
\vx_{i,t}
-
\frac{1}{T-h}\sum_{s=1}^{T-h}\vx_{i,s},
\end{equation*}
and $\widetilde{y}_{i,t+h}$ is defined analogously.

In the remainder of the paper, we follow \citet{https://doi.org/10.1002/jae.2333} and \citet{HerbstJohannsen2024} and assume that the iid\ shocks $x_{i,t}$ are directly observed by the empirical researcher for all $i$ and $t$. We discuss how our results can be extended beyond this setting in Section \ref{ssection::alternativeDGP}.

\subsection{The weak-exogeneity bias}

We now study the fixed-$T$ inconsistency of the FE estimator in the LP context and explain how this result motivates the bias-correction methods proposed in the literature.

The FE estimator $\widehat{\vbeta}_{FE}^{(h)}$ admits the following expansion:
\begin{equation}
\label{eq::FE_expansion}
\widehat{\vbeta}_{FE}^{(h)}
-
\vbeta_{0}^{(h)}
-
\E[\mB_{i,T}^{(h)}]^{-1}\E[\va_{i,T}^{(h)}]
=
\left(
\frac{1}{N}\sum_{i=1}^{N}\mB_{i,T}^{(h)}
\right)^{-1}
\left(
\frac{1}{N}\sum_{i=1}^{N}\vs_{i,T}^{(h)}
\right),
\end{equation}
where $\vbeta_{0}^{(h)}$ denotes the true coefficient vector and
\begin{equation*}
\mB_{i,T}^{(h)}
:=
\frac{1}{T-h}\sum_{t=1}^{T-h}
\widetilde{\vx}_{i,t}\widetilde{\vx}_{i,t}',
\quad
\va_{i,T}^{(h)}
:=
\frac{1}{T-h}\sum_{t=1}^{T-h}
\widetilde{\vx}_{i,t}\widetilde{u}_{i,t}^{(h)},
\end{equation*}
with
\begin{equation}
\label{eq::influence}
\vs_{i,T}^{(h)}
:=
\va_{i,T}^{(h)}
-
\E[\va_{i,T}^{(h)}]
-
\left(
\mB_{i,T}^{(h)}
-
\E[\mB_{i,T}^{(h)}]
\right)
\E[\mB_{i,T}^{(h)}]^{-1}
\E[\va_{i,T}^{(h)}].
\end{equation}
Here, we explicitly use the assumption that the cross-sectional units are identically distributed.\footnote{Throughout this paper, we maintain the assumption that the cross-sectional units $i$ are independent and identically distributed.}

By construction, the component $\E[\va_{i,T}^{(h)}]$ is given by
\begin{align*}
\E[\va_{i,T}^{(h)}]
&=
\E\left[
\frac{1}{T-h}\sum_{t=1}^{T-h}
\vx_{i,t}u_{i,t}^{(h)}
\right]
-
\frac{1}{(T-h)^2}
\sum_{t=1}^{T-h}\sum_{s=1}^{T-h}
\E\left[
\vx_{i,t}u_{i,s}^{(h)}
\right].
\label{eq::bias_score}
\end{align*}
Although the first term is zero under correct model specification,\footnote{The same applies if the corresponding coefficients are interpreted as pseudo-true values; see, e.g., \citet{GalvaoKato2014} and \citet{JuodisMiss2020}.} the second term can be nonzero when the regressors $\vx_{i,t}$ are weakly exogenous with respect to the LP error term $u_{i,t}^{(h)}$, that is,\footnote{Here, we follow \citet{https://doi.org/10.3982/QE2622} and \citet{mei2023nickellbiaspanellocal} and formulate the weak-exogeneity conditions in terms of \emph{conditional} moments, as opposed to \citet{HerbstJohannsen2024}, who use unconditional moments implied by the Wold decomposition theorem.}
\begin{equation*}
\E\left[
u_{i,t}^{(h)}
\mid
\{\vx_{i,s}\}_{s=1}^{q}
\right]
=
0,
\quad
q=1,\ldots,t.
\end{equation*}

This generally renders the FE estimator inconsistent for any fixed value of $T$. With this in mind, define
\begin{equation*}
\vb_{T}^{(h)}
:=
\E[\mB_{i,T}^{(h)}]^{-1}
\E[\va_{i,T}^{(h)}].
\end{equation*}
It is common in the panel-data literature to refer to this term as the \emph{bias} of the FE estimator. In the presence of weakly exogenous regressors, the fixed-$T$ inconsistency of the FE estimator is an instance of the incidental-parameter problem of \citet{neyman1948}. It has been studied extensively in the special case of panel autoregressive models; see \citet{Nickell}, \citet{kiviet1995}, \citet{Hahn2002}, \citet{Bun2005}, \citet{Moreira2009}, and \citet{Dhaene}, among others.

By construction, $\E[\vs_{i,T}^{(h)}]=\vzeros$, while $\vs_{i,T}^{(h)}$ is independently and identically distributed across cross-sectional units. Hence, in the asymptotic regime in which $N$ increases while $T$ remains fixed, equation \eqref{eq::FE_expansion} can be used directly to derive the asymptotic distribution of $\widehat{\vbeta}_{FE}^{(h)}$ around the pseudo-true value $\vbeta_{*}^{(h)}:=\vbeta_{0}^{(h)}+\vb_{T}^{(h)}$ (subject to the usual regularity conditions).

Moreover, under appropriate regularity conditions (see, for example, \citet{GalvaoKato2014} and \citet{JuodisMiss2020}), the bias term $\vb_{T}^{(h)}$ admits the expansion
\begin{equation}
\label{eq::bias_expansion}
\vb_{T}^{(h)}
=
\frac{1}{T}\vb^{(h,1)}
+
\largeO(R_{T}),
\end{equation}
where $\vb^{(h,1)}$ is the leading, first-order bias term, which does not depend on $T$, and $\largeO(R_{T})$ is a remainder term of smaller order than $T^{-1}$. Combining this expansion with the asymptotic normality implied by the fixed-$T$ expansion in \eqref{eq::FE_expansion}, the results in \citet{GalvaoKato2014}, under appropriate covariance-stationarity restrictions, yield
\begin{equation}
\label{eq::galvao_kato}
\sqrt{NT}
\left(
\widehat{\vbeta}_{FE}^{(h)}
-
\vbeta_{0}^{(h)}
\right)
\dto
\calN\left(
\sqrt{\rho}\vb^{(h,1)},
\lim_{T\to\infty}
\E[\mB_{i,T}^{(h)}]^{-1}
\E\left[
T\vs_{i,T}^{(h)}(\vs_{i,T}^{(h)})'
\right]
\E[\mB_{i,T}^{(h)}]^{-1}
\right),
\end{equation}
as $N,T\to\infty$ jointly, with $N/T\to\rho\in[0,\infty)$.

Thus, unlike in the fixed-$T$ setting, where asymptotically normal inference is generally conducted around the pseudo-true value $\vbeta_{*}^{(h)}:=\vbeta_{0}^{(h)}+\vb_{T}^{(h)}$, the large-$T$ asymptotic distribution is centered around the true value $\vbeta_{0}^{(h)}$ but is contaminated by a noncentrality parameter.

\subsection{Bias correction}
The usual asymptotically normal inference based on equation \eqref{eq::galvao_kato}, which ignores the noncentrality parameter induced by the weak-exogeneity bias, is not reliable unless $N/T\to 0$. This motivates a class of bias-corrected FE estimators, denoted by $\widetilde{\vbeta}_{BC}^{(h)}$, that satisfy
\begin{equation*}
\sqrt{NT}
\left(
\widetilde{\vbeta}_{BC}^{(h)}
-
\vbeta_{0}^{(h)}
\right)
\dto
\calN\left(
\vzeros,
\lim_{T\to\infty}
\E[\mB_{i,T}^{(h)}]^{-1}
\E\left[
T\vs_{i,T}^{(h)}(\vs_{i,T}^{(h)})'
\right]
\E[\mB_{i,T}^{(h)}]^{-1}
\right).
\end{equation*}
This class includes the bias-correction method of \citet{HerbstJohannsen2024} and the half-panel jackknife estimator of \citet{SPJ}, as advocated by \citet{mei2023nickellbiaspanellocal}. These are the two methods considered in the remainder of this paper.

Restricting attention to the covariance-stationary setting, \citet{HerbstJohannsen2024} show that the leading term in the bias of the FE estimator $\widehat{\theta}_{FE}^{(h)}$ is given by
\begin{equation*}
b_{\theta}^{(h,1)}
=
-
\sum_{q=0}^{h-1}
\theta_{0}^{(q)},
\quad
h>0,
\end{equation*}
whereas $b_{\theta}^{(0,1)}=0$. This suggests the following easy-to-use bias-corrected estimator of $\theta^{(h)}$:
\begin{equation*}
\widehat{\theta}_{HJ}^{(h)}
:=
\widehat{\theta}_{FE}^{(h)}
+
\frac{1}{T-h}
\sum_{q=0}^{h-1}
\widehat{\theta}_{FE}^{(q)},
\quad
h>0,
\end{equation*}
with $\widehat{\theta}_{HJ}^{(0)}:=\widehat{\theta}_{FE}^{(0)}$.\footnote{Following \citet{HerbstJohannsen2024}, we include the finite-sample adjustment $T-h$. An alternative formulation is
\begin{equation*}
\widehat{\theta}_{HJ}^{(h)}
:=
\widehat{\theta}_{FE}^{(h)}
+
\frac{1}{T}
\sum_{q=0}^{h-1}
\widehat{\theta}_{FE}^{(q)},
\quad
h>0.
\end{equation*}
\citet{HerbstJohannsen2024} also propose a recursively constructed bias-corrected estimator, labeled BCC. Although this estimator may exhibit superior finite-sample properties at larger values of $h$ in the setting considered by \citet{HerbstJohannsen2024}, it does not address problems discussed in this paper. For simplicity, we therefore omit it from further discussion.}

The HPJ approach of \citet{SPJ} can be motivated by observing that equation \eqref{eq::bias_expansion} is valid for any time-series length. It therefore also applies when estimation is based on only a fraction of the time-series observations. For simplicity, assume that $T$ is even, and define $T_{1}:=T/2$ and $T_{2}:=T-T_{1}$. Then,
\begin{equation*}
\vb_{\calT}^{(h)}
=
\frac{1}{\calT}
\vb^{(h,1)}
+
\largeO(R_{\calT}),
\end{equation*}
for $\calT\in\{T,T_{1},T_{2}\}$. Let $\widehat{\vbeta}_{1,FE}^{(h)}$ and $\widehat{\vbeta}_{2,FE}^{(h)}$ denote the FE estimators of $\vbeta^{(h)}$ based on the first $T_{1}$ observations and the remaining $T_{2}$ observations, respectively.

It then follows that the HPJ estimator,
\begin{equation}
\label{eq::SPJ}
\widehat{\vbeta}_{HPJ}^{(h)}
:=
2\widehat{\vbeta}_{FE}^{(h)}
-
\frac{1}{2}
\left(
\widehat{\vbeta}_{1,FE}^{(h)}
+
\widehat{\vbeta}_{2,FE}^{(h)}
\right),
\end{equation}
eliminates the leading bias term. In particular,
\begin{equation*}
2\vb_{T}^{(h)}
-
\frac{1}{2}
\left(
\vb_{T_{1}}^{(h)}
+
\vb_{T_{2}}^{(h)}
\right)
=
\largeO(R_{T}).
\end{equation*}
Finally, as shown by \citet{SPJ} and \citet{GalvaoKato2014}, the HPJ estimator has the same first-order asymptotic variance as the original FE estimator. Thus, the HPJ removes the leading asymptotic bias at no first-order asymptotic cost.

\subsection{Numerical preview of results}
\label{ssection:numerical_illustration}
The results of \citet{HerbstJohannsen2024} indicate that their bias-correction procedure effectively removes bias in stationary and mildly persistent settings. Similarly, the simulation results in \citet{mei2023nickellbiaspanellocal} indicate that the HPJ approach can successfully remove the leading bias term of the FE estimator in covariance-stationary settings. We now investigate numerically which of these findings carry over to settings with moderately and highly persistent control variables.

\begin{table}[htbp!]
  \renewcommand{\arraystretch}{.9}
  \addtolength{\tabcolsep}{-2pt}
  \centering
  \footnotesize
  \caption{Scaled finite-sample bias in estimating $\theta^{(h)}$.}
\begin{tabular}{ccc|ccc|ccc}
\hline
\hline
&&&\multicolumn{3}{c|}{$h=0$}&\multicolumn{3}{c}{$h=4$}\\
\hline
$T$ & $\alpha_{0}$ && \textnormal{FE} & \textnormal{HJ} & \textnormal{HPJ} & \textnormal{FE} & \textnormal{HJ} & \textnormal{HPJ}\\
\hline
30 & 0.85 && -0.30 & -0.30 &  0.35 & -6.69 & -2.39 &  0.77 \\
30 & 0.95 && -0.67 & -0.67 &  0.22 & -10.89 & -6.06 & -1.17 \\
30 & 1.00 && -0.99 & -0.99 & -0.04 & -14.18 & -9.13 & -3.68 \\
\hline
50 & 0.85 && -0.12 & -0.12 &  0.23 & -4.39 & -1.09 &  1.06 \\
50 & 0.95 && -0.36 & -0.36 &  0.27 & -7.25 & -3.48 &  1.04 \\
50 & 1.00 && -0.75 & -0.75 & -0.00 & -11.02 & -7.07 & -1.65 \\
\hline
\end{tabular}
\caption*{Notes. Results are based on $M=4000$ Monte Carlo replications for the DGP in equation \eqref{eq::numerical_DGP}, with $\theta_{0}=1$ and $N=50$. Bias is scaled by $\sqrt{NT}$. ``FE'' denotes the fixed-effect estimator, ``HJ'' denotes the bias-correction procedure of \citet{HerbstJohannsen2024}, and ``HPJ'' denotes the half-panel jackknife of \citet{SPJ}.}
\label{table::res_illustration_bias_new}
\end{table}

\begin{table}[htbp!]
  \renewcommand{\arraystretch}{.9}
  \addtolength{\tabcolsep}{-2pt}
  \centering
  \footnotesize
  \caption{Coverage rates for $\theta^{(h)}$ using the HJ and HPJ estimators. $N=50$.}
\begin{tabular}{ccc|cccc|cccc}
\hline
\hline
&&&\multicolumn{4}{c|}{$h=0$}&\multicolumn{4}{c}{$h=4$}\\
\hline
$T$ & $\alpha_{0}$ && \textnormal{HJ} & $\mathrm{HPJ}_{\mathrm{GK}}$ & $\mathrm{HPJ}_{\mathrm{MSS}}$ & $\mathrm{HPJ}_{\mathrm{BCA}}$ & \textnormal{HJ} & $\mathrm{HPJ}_{\mathrm{GK}}$ & $\mathrm{HPJ}_{\mathrm{MSS}}$ & $\mathrm{HPJ}_{\mathrm{BCA}}$\\
\hline
30 & 0.85 && 0.82 & 0.92 & 0.93 & 0.93 & 0.94 & 0.84 & 0.91 & 0.94 \\
30 & 0.95 && 0.89 & 0.92 & 0.93 & 0.93 & 0.85 & 0.81 & 0.89 & 0.92 \\
30 & 1.00 && 0.93 & 0.93 & 0.94 & 0.94 & 0.51 & 0.69 & 0.78 & 0.82 \\
\hline
50 & 0.85 && 0.84 & 0.93 & 0.94 & 0.94 & 0.93 & 0.87 & 0.91 & 0.94 \\
50 & 0.95 && 0.88 & 0.93 & 0.93 & 0.94 & 0.92 & 0.86 & 0.91 & 0.94 \\
50 & 1.00 && 0.94 & 0.94 & 0.94 & 0.94 & 0.68 & 0.83 & 0.87 & 0.90 \\
\hline
\end{tabular}
\caption*{Notes. Results are based on $M=4000$ Monte Carlo replications for the DGP in equation \eqref{eq::numerical_DGP}. The nominal coverage rate is $95\%$. For the HJ procedure, confidence intervals are constructed using the percentile bootstrap. For the HPJ estimator, we consider the standard errors of \citet{GalvaoKato2014} ($\mathrm{HPJ}_{\mathrm{GK}}$), \citet{mei2023nickellbiaspanellocal} ($\mathrm{HPJ}_{\mathrm{MSS}}$), and the bias-correction-adjusted standard errors proposed in this paper ($\mathrm{HPJ}_{\mathrm{BCA}}$).}
\label{table::res_illustration_coverage_new}
\end{table}

Following \citet{HerbstJohannsen2024}, we assume that the true DGP is given by
\begin{equation}
\label{eq::numerical_DGP}
y_{i,t}
=
\alpha_{0}y_{i,t-1}
+
(1-\alpha_{0})\mu_{i}
+
\theta_{0}x_{i,t}
+
\epsi_{i,t},
\end{equation}
where
$x_{i,t}\overset{\scriptstyle\text{iid}}{\sim}\calN(0,1)$,
$\epsi_{i,t}\overset{\scriptstyle\text{iid}}{\sim}\calN(0,1)$,
$y_{i,0}=\mu_{i}=0$, and $\theta_{0}=1$, as in \citet{HerbstJohannsen2024}. For this DGP, the implied LP coefficients are
$\theta_{0}^{(h)}=\alpha_{0}^{h}\theta_{0}$
and
$\delta_{0}^{(h)}=\alpha_{0}^{h+1}$,
when $c_{i,t-1}=y_{i,t-1}$ is included as a control variable.

Tables \ref{table::res_illustration_bias_new}--\ref{table::res_illustration_coverage_new} report the bias and coverage results for this DGP. Four patterns are noteworthy:
\begin{enumerate}[(i)]
\item The HJ approach removes a substantial part of the bias of the original FE estimator, but its performance deteriorates as $\alpha_{0}$ increases.

\item At horizon $h=4$, the HPJ estimator generally has a smaller absolute bias than the HJ estimator, especially for larger values of $\alpha_{0}$.

\item At $h>0$, coverage rates based on the HPJ estimator can be poor for larger values of $\alpha_{0}$ when standard errors proposed in the existing literature, including those of \citet{GalvaoKato2014} and \citet{mei2023nickellbiaspanellocal}, are used.

\item At longer horizons, coverage can improve substantially when the bias-correction-adjusted procedure proposed in this paper, $HPJ_{BCA}$, is used.
\end{enumerate}

In Section \ref{section:asymptotic}, we develop a theoretical framework to explain the numerical patterns documented in Tables \ref{table::res_illustration_bias_new}--\ref{table::res_illustration_coverage_new}. In Section \ref{section:applicability}, we discuss in detail how the proposed \emph{bias-correction-adjusted} standard errors can be constructed and why they can improve upon the proposals of \citet{GalvaoKato2014} and \citet{mei2023nickellbiaspanellocal}.
\section{Asymptotic results}
\label{section:asymptotic}
\subsection{Data-generating process}
Consider the following dynamic panel-data model, written in unobserved-components form:
\begin{align}
y_{i,t}
&=
\mu_{i}
+
u_{i,t},
\\
u_{i,t}
&=
\alpha_{0}u_{i,t-1}
+
\theta_{0}x_{i,t}
+
\epsi_{i,t},
\label{eq::GDP1}
\end{align}
for $t\geq 1$. Equivalently, the model can be written as
\begin{equation*}
y_{i,t}
=
\alpha_{0}y_{i,t-1}
+
(1-\alpha_{0})\mu_{i}
+
\theta_{0}x_{i,t}
+
\epsi_{i,t},
\quad
t\geq 1.
\end{equation*}
Alternatively, the DGP admits the following LP representation:
\begin{equation*}
y_{i,t+h}
=
\left(1-\alpha_{0}^{h+1}\right)\mu_{i}
+
\theta_{0}\alpha_{0}^{h}x_{i,t}
+
\alpha_{0}^{h+1}y_{i,t-1}
+
\left(
\alpha_{0}^{h}\epsi_{i,t}
+
\sum_{q=0}^{h-1}
\alpha_{0}^{q}
\left(
\epsi_{i,t+h-q}
+
\theta_{0}x_{i,t+h-q}
\right)
\right),
\end{equation*}
for any $h=0,\ldots,H$ and $t\geq 1$, where an empty sum is defined to be zero. This representation matches the general form of the LP regression in equation \eqref{eq::model}, with all relevant quantities defined by comparison of the two expressions. As in Section \ref{ssection:numerical_illustration}, we assume that $\vc_{i,t-1}=y_{i,t-1}$ (scalar).

To simplify the derivations, we assume that the total number of observations is $T+1+H$, where $H$ is the maximum LP horizon considered. We also use exactly $T$ time-series observations to estimate each $\widehat{\vbeta}^{(h)}$, so that
\begin{equation*}
\mB_{i,T}^{(h)}
=
\mB_{i,T},
\end{equation*}
for any $h=0,\ldots,H$. This entails no asymptotic loss of generality as long as $H$ is fixed.

Following \citet{ECT:9967724}, we consider asymptotic approximations based on drifting sequences of DGPs. In particular, we allow the autoregressive parameter to depend on $T$, writing $\alpha_{0}=\alpha_{T}$.
\begin{assumption}[Drifting DGP]
As $T\to\infty$, $\alpha_0=\alpha_T$ evolves as
\label{ass::drifting}
\begin{equation}
\label{eq::drifting}
\alpha_{T}
=
\exp\left(
-cT^{-\gamma}
\right),
\end{equation}
where $c> 0$ and $\gamma\in[0,\infty]$ are fixed constants, with $\alpha_{T}=1$ when $\gamma=\infty$.
\end{assumption}

We divide the parameter space indexed by $\gamma$ into three regions:
\begin{enumerate}[a)]
\setlength\itemsep{0.1em}
  \item $\gamma\in(1,\infty]$: near-unit-root and exact-unit-root regimes;
  \item $\gamma=1$: the local-to-unity regime;
  \item $\gamma\in[0,1)$: stable and weakly integrated regimes.
\end{enumerate}
As is standard for drifting DGPs, we adopt the functional form in equation \eqref{eq::drifting} for analytical convenience, rather than as a literal description of the true DGP. Other asymptotic characterizations of estimator behavior when $\alpha_{0}$ is close to one are possible. Section \ref{section:montecarlo}, by contrast, examines finite-sample properties for fixed values of $(N,T,\alpha_{0})$. The special case $\gamma=0$ corresponds to the stationary setting considered by \citet{HerbstJohannsen2024} and \citet{mei2023nickellbiaspanellocal}.

\begin{assumption}[Random variables]\
\label{ass::rv}
\begin{enumerate}[i)]
    \item $\epsi_{i,t}\overset{\scriptstyle \text{iid}}{\sim} (0,\sigma_{\epsilon}^{2})$, where $\sigma_{\epsilon}^{2}>0$ and $\E[\epsi_{i,t}^{4}]<C<\infty$.
  \item $x_{i,t}\overset{\scriptstyle \text{iid}}{\sim} (0,\sigma_{x}^{2})$, where $\sigma_{x}^{2}>0$ and $\E[x_{i,t}^{4}]<C<\infty$.
  \item The sequences $\{\epsi_{i,t}\}$ and $\{x_{i,t}\}$ are mutually independent.
  \item $u_{i,0}=0$ for all $i$.
\end{enumerate}
\end{assumption}

Because the model for $y_{i,t}$ is specified in unobserved-components form, the FE estimator is invariant to the location parameter $\mu_{i}$. The assumption $u_{i,0}=0$ is imposed to streamline some of the derivations in the local-to-unity setting. It could instead be replaced by an appropriate bounded-moments condition, as in \citet{https://doi.org/10.3982/ECTA18756}.\footnote{A bounded-moments condition rules out distant initialization in the near-unit-root setting; see, for example, \citet{kruiniger2009}, \citet{doi:10.1080/07474938.2013.833829}, and \citet{BunKleibergen2010}.} Finally, define
\begin{equation*}
\nu_{i,t}
:=
\epsi_{i,t}
+
\theta_{0}x_{i,t},
\quad
\sigma_{\nu}^{2}
:=
\E[\nu_{i,t}^{2}].
\end{equation*}
\subsection{Bias approximations}
The expansions in this subsection are for each fixed admissible pair $(c,\gamma)$ and are
uniform over the finite set of horizons $h=0,\ldots,H$. The following proposition characterizes the leading bias term $b_{\theta}^{(h,1)}$ of the FE estimator $\widehat{\theta}_{FE}^{(h)}$.

\begin{proposition}
\label{lemma::expansionFE}
Under Assumptions \ref{ass::drifting}--\ref{ass::rv}, the bias term $b_{T,\theta}^{(h)}$ of $\widehat{\theta}_{FE}^{(h)}$ satisfies:
\begin{enumerate}[a)]
\setlength\itemsep{0.1em}
  \item For $\gamma\in(1,\infty]$: $
  b_{\theta}^{(h,1)}
  =
  -\theta_{0}
  \left(
  h
  +
  \frac{3}{2}
  \left(
  \frac{\sigma_{\epsi}^{2}}{\sigma_{\nu}^{2}}
  +
  h
  \right)
  \right),
  \quad
  R_{T}
  =
  T^{-(1+\min(\gamma-1,1))}.$

  \item For $\gamma=1$: $
  b_{\theta}^{(h,1)}
  =
  -\theta_{0}
  \left(
  h
  +
  \xi_{\theta}(c)
  \left(
  \frac{\sigma_{\epsi}^{2}}{\sigma_{\nu}^{2}}
  +
  h
  \right)
  \right),
  \quad
  R_{T}
  =
  T^{-2}.$

  \item For $\gamma\in[0,1)$: $
  b_{\theta}^{(h,1)}
  =
  -\theta_{0}
  \sum_{q=0}^{h-1}
  \alpha_{T}^{q},
  \quad
  R_{T}
  =
  T^{-2+\gamma}.$
\end{enumerate}
These results hold as $T\to\infty$ for all $h=0,\ldots,H$, and the function $\xi_{\theta}(x)$, for $x\in\overline{\mathbb{R}}_{+}$, is defined as
\begin{align*}
\xi_{\theta}(x)
&:=
\frac{
4\left(\exp(-x)-1+x\right)^2
}{
x^2\left(\exp(-2x)-1+2x\right)
-
2x\left(
4\exp(-x)-\exp(-2x)-3+2x
\right)
}.
\end{align*}
\end{proposition}
We adopt the convention that
$\sum_{q=0}^{h-1}\alpha_{T}^{q}=0$
when $h=0$, and use the fact that
$\lim_{T\to\infty}\sum_{q=0}^{h-1}\alpha_{T}^{q}=h$ for $\gamma> 0$. However, in the weakly integrated
regime $\gamma\in (0,1)$, we retain the finite-$T$ geometric sum in the leading coefficient, rather
than replacing it by $h$, to preserve the stated remainder rate.

Proposition \ref{lemma::expansionFE} shows that the bias of the FE estimator differs substantially between settings in which the autoregressive coefficient is moderate and those in which it is highly persistent. For $\gamma\geq 1$, the leading bias term is the sum of two components with distinct origins. The first component arises from the $MA(h)$ structure of the LP approximation and corresponds to the stable-regime expression after taking the appropriate limit. The second component arises from the weak-exogeneity bias induced by the control variable $y_{i,t-1}$. Consequently, even at horizon $h=0$, the FE estimator has a nonzero leading bias, unlike in the stable, stationary setting of \citet{HerbstJohannsen2024}. Following the terminology of \citet{10.1257/aer.20221116}, we refer to this component as the \emph{contamination} bias.

Another important implication of Proposition \ref{lemma::expansionFE} is that the magnitude of the leading bias increases with $h$. This result is already known for the stationary case, $\gamma=0$, from \citet{HerbstJohannsen2024}. For $\gamma\geq 1$, this dependence is especially transparent, since the leading bias term becomes an affine function of $h$.

Using analogous arguments, we can formulate a corresponding proposition for the coefficient on the control variable $y_{i,t-1}$.

\begin{proposition}
\label{lemma::expansionFE_control}
Under Assumptions \ref{ass::drifting}--\ref{ass::rv}, the bias term $b_{T,\delta}^{(h)}$ of $\widehat{\delta}_{FE}^{(h)}$ satisfies:
\begin{enumerate}[a)]
\setlength\itemsep{0.1em}
  \item For $\gamma\in(1,\infty]$: $
  b_{\delta}^{(h,1)}
  =
  -3
  \left(
  \frac{\sigma_{\epsi}^{2}}{\sigma_{\nu}^{2}}
  +
  h
  \right),
  \quad
  R_{T}
  =
  T^{-(1+\min(\gamma-1,1))}.$

  \item For $\gamma=1$: $ 
  b_{\delta}^{(h,1)}
  =
  -\xi_{\delta}(c)
  \left(
  \frac{\sigma_{\epsi}^{2}}{\sigma_{\nu}^{2}}
  +
  h
  \right),
  \quad
  R_{T}
  =
  T^{-2}.
  $

  \item For $\gamma\in[0,1)$: $
b_{\delta}^{(h,1)}
  =
  -(1+\alpha_{T})
  \left(
  \frac{\sigma_{\epsi}^{2}}{\sigma_{\nu}^{2}}
  \alpha_{T}^{h}
  +
  \sum_{q=0}^{h-1}
  \alpha_{T}^{q}
  \right),
  \quad
  R_{T}
  =
  T^{-2+\gamma}.$
\end{enumerate}
These results hold as $T\to\infty$ for all $h=0,\ldots,H$, and
\begin{align*}
\xi_{\delta}(x)
&=
\xi_{\theta}(x)
\left(
\frac{\exp(-x)-1+x}{x^{2}}
\right)^{-1},
\quad
x\in\overline{\mathbb{R}}_{+}.
\end{align*}
\end{proposition}

Proposition \ref{lemma::expansionFE_control} is, in its own right, a contribution to the dynamic panel-data literature, as it extends Theorems 2 and 4 of \citet{Hahn2002} for the panel AR(1) setting. In particular, setting $\theta_{0}=0$ and $h=0$ yields their results for the cases $\gamma=0$ and $\gamma=\infty$.

The near-unit-root coefficients in regime (a) and the limiting weakly integrated coefficients for $\gamma\in(0,1)$ can be recovered by taking $x\to 0$ and $x\to\infty$, respectively:\footnote{See, for example, Lemma 2 of \citet{10.2307/2336692} for a related distributional analysis in the time-series setting.}
\begin{align*}
\lim_{x\to 0}\xi_{\delta}(x)
&=
3,
&
\lim_{x\to\infty}\xi_{\delta}(x)
&=
2,
\\
\lim_{x\to 0}\xi_{\theta}(x)
&=
\frac{3}{2},
&
\lim_{x\to\infty}\xi_{\theta}(x)
&=
0.
\end{align*}
To see why these are the relevant limits for $\xi_{\delta}(x)$ when $\gamma\in(0,1)$, observe that $\lim_{T\to\infty}(1+\alpha_{T})=2$. The limiting argument does not apply to $\gamma=0$.

\subsection{Non-uniformity of bias-correction procedures}
\label{ssection::nonuniformity}
In this section, we use the results of Proposition \ref{lemma::expansionFE} to show that neither of the two bias-correction procedures proposed in the literature is uniformly valid under Assumptions \ref{ass::drifting}--\ref{ass::rv}.

Our first result characterizes the leading bias term $b_{\theta_{HJ}}^{(h,1)}$ of the bias-corrected estimator proposed by \citet{HerbstJohannsen2024}.

\begin{corollary}
\label{lemma::expansionHJ}
Under Assumptions \ref{ass::drifting}--\ref{ass::rv}, the bias term $b_{T,\theta_{HJ}}^{(h)}$ associated with the estimator $\widehat{\theta}_{HJ}^{(h)}$ has the following leading term:
\begin{enumerate}[a)]
\setlength\itemsep{0.1em}
  \item For $\gamma\in(1,\infty]$: $
  b_{\theta_{HJ}}^{(h,1)}
  =
  -\theta_{0}
  \frac{3}{2}
  \left(
  \frac{\sigma_{\epsi}^{2}}{\sigma_{\nu}^{2}}
  +
  h
  \right).$

  \item For $\gamma=1$: $
  b_{\theta_{HJ}}^{(h,1)}
  =
  -\theta_{0}
  \xi_{\theta}(c)
  \left(
  \frac{\sigma_{\epsi}^{2}}{\sigma_{\nu}^{2}}
  +
  h
  \right).$

  \item For $\gamma\in[0,1)$: $ 
  b_{\theta_{HJ}}^{(h,1)}
  =
  0.$
\end{enumerate}
These results hold as $T\to\infty$ for all $h=0,\ldots,H$. The remainder terms $R_{T}$ have the same orders as in Proposition \ref{lemma::expansionFE}.
\end{corollary}

Corollary \ref{lemma::expansionHJ} shows that the procedure of \citet{HerbstJohannsen2024} successfully removes the bias of the FE estimator arising from the $MA(h)$ structure of the LP regression errors. However, for $\gamma\geq 1$, a bias component associated with the control variable remains. Consequently, the method of \citet{HerbstJohannsen2024} is not uniformly valid over the parameter space specified in Assumption \ref{ass::drifting}.

The following corollary to Proposition \ref{lemma::expansionFE} shows that the HPJ bias correction is not uniformly valid either.

\begin{corollary}
\label{lemma::expansionHPJ}
Under Assumptions \ref{ass::drifting}--\ref{ass::rv}, the bias term $b_{T,\theta_{HPJ}}^{(h)}$ associated with the HPJ estimator $\widehat{\theta}_{HPJ}^{(h)}$ has the following leading term:
\begin{enumerate}[a)]
\setlength\itemsep{0.1em}
  \item For $\gamma=1$: $
  b_{\theta_{HPJ}}^{(h,1)}
  =
  -\theta_{0}
  \left(
  2\xi_{\theta}(c)
  -
  \left(
  \xi_{\theta}\left(\frac{c}{2}\right)
  +
  \xi_{\theta}^{+}\left(\frac{c}{2}\right)
  \right)
  \right)
  \left(
  \frac{\sigma_{\epsi}^{2}}{\sigma_{\nu}^{2}}
  +
  h
  \right).$

  \item For $\gamma\neq 1$: $
  b_{\theta_{HPJ}}^{(h,1)}
  =
  0.$
\end{enumerate}
These results hold as $T\to\infty$ for all $h=0,\ldots,H$. The remainder terms $R_{T}$ have the same orders as in Proposition \ref{lemma::expansionFE} in all cases, and the function $\xi_{\theta}^{+}(\cdot)$ is defined explicitly in Appendix B.3.
\end{corollary}

Corollary \ref{lemma::expansionHPJ} warrants further discussion. Like the procedure of \citet{HerbstJohannsen2024}, the HPJ approach successfully removes the leading linear bias term arising from the $MA(h)$ structure of the LP regression errors for all values of $\gamma$. Moreover, the jackknife removes the contamination-bias component associated with the control variable when $\gamma\neq 1$, including for $\gamma>1$. Consequently, the HPJ is applicable over a substantially larger region of the parameter space than the bias-correction procedure of \citet{HerbstJohannsen2024}.

The HPJ fails to remove the contamination bias only when $\gamma=1$. This irregular behavior can be understood by considering the term
\begin{align*}
(1-\alpha_{T})
\frac{1}{T}
\sum_{t=0}^{T-2}
\sum_{l=0}^{t}
\alpha_{T}^{l}
&=
1
-
\frac{1}{T}
-
\frac{1}{T}
\frac{\alpha_{T}}{1-\alpha_{T}}
+
\frac{1}{T}
\frac{\alpha_{T}}{1-\alpha_{T}}
\alpha_{T}^{T-1}.
\end{align*}
This expression can be recognized as the scaled numerator of the classical \citet{Nickell} bias term in the panel AR(1) model. When $\gamma=1$, it satisfies
\begin{equation*}
(1-\alpha_{T})
\frac{1}{T}
\sum_{t=0}^{T-2}
\sum_{l=0}^{t}
\alpha_{T}^{l}
\to
1
-
\frac{1}{c}
+
\frac{1}{c}
\exp(-c)
\end{equation*}
as $T\to\infty$. The corresponding quantity based only on the first $T_{1}=T/2$ observations satisfies
\begin{equation*}
(1-\alpha_{T})
\frac{1}{T/2}
\sum_{t=0}^{T/2-2}
\sum_{l=0}^{t}
\alpha_{T}^{l}
\to
1
-
\frac{2}{c}
+
\frac{2}{c}
\exp\left(
-\frac{c}{2}
\right).
\end{equation*}
Thus, within each sub-panel of length $T_{1}=T/2$, the \emph{implicit} local-to-unity parameter is halved. Consequently, the homogeneity condition required for the validity of the HPJ procedure is violated.

Another noteworthy feature of the HPJ bias when $\gamma=1$ is that the leading bias term for the second half of the sample, $\xi_{\theta}^{+}(c/2)$, differs from that for the first half, $\xi_{\theta}(c/2)$. This difference primarily reflects the initial conditions. In the first sub-panel, the initial condition is $u_{i,0}=0$. In the second sub-panel, it is $u_{i,T/2}$, which has a nonnegligible asymptotic variance in the local-to-unity setting.\footnote{This result also provides insight into how the results in Propositions \ref{lemma::expansionFE}--\ref{lemma::expansionFE_control} would change if the assumption $u_{i,0}=0$ were replaced by assumptions allowing for unbounded moments of $u_{i,0}$.}

\subsection{Asymptotic distribution}
In this section, we analyze the asymptotic distributions of the FE estimator $\widehat{\vbeta}_{FE}^{(h)}$ and the HPJ estimator $\widehat{\vbeta}_{HPJ}^{(h)}$. For this purpose, we impose the following assumption on the relative expansion rates of $N$ and $T$.

\begin{assumption}[Expansion rates]\
\label{ass::expansion}
For some $\tau\in (0,\infty)$ and $\kappa> 0$, $N^{\kappa}T^{-1}\to\tau$ as $N,T\to\infty$ jointly.
\end{assumption}

The following proposition summarizes the distributional results for the FE estimator of the coefficient on the main variable of interest, $x_{i,t}$.

\begin{proposition}
\label{proposition::FE_distribution_theta}
Under Assumptions \ref{ass::drifting}--\ref{ass::expansion}, as $N,T\to\infty$,
\begin{equation}
\sqrt{NT}
\left(
\widehat{\theta}_{FE}^{(h)}
-
\theta_{0}^{(h)}
-
\frac{1}{T}b_{\theta}^{(h,1)}
\right)
\dto
\calN\left(
0,
\lim_{T\to\infty}
\frac{
\alpha_{T}^{2h}\sigma_{\epsi}^{2}
+
\sum_{q=0}^{h-1}\alpha_{T}^{2q}\sigma_{\nu}^{2}
}{
\sigma_{x}^{2}
}
\right),
\end{equation}
for all $h=0,\ldots,H$, provided that
\begin{equation*}
\kappa
>
\left\{
\begin{array}{lr}
        3^{-1}, & \text{for } \gamma\in\{1\}\cup[2,\infty],\\
        (3-2\gamma)^{-1}, & \text{for } \gamma\in[0,1),\\
        (2\gamma-1)^{-1}, & \text{for } \gamma\in(1,2).
\end{array}
\right.
\end{equation*}
\end{proposition}

Proposition \ref{proposition::FE_distribution_theta} shows that, apart from the first-order asymptotic bias, the FE estimator $\widehat{\theta}_{FE}^{(h)}$ has an asymptotically normal distribution. However, as indicated by the remainder-term rates in Proposition \ref{lemma::expansionFE}, this asymptotic approximation can be imprecise when $N$ and $T$ are of comparable magnitude, particularly because of the discontinuity at $\gamma=1$, that is, in the local-to-unity setting. This discontinuity also provides another explanation for the failure of the HPJ approach to remove the leading bias term when $\gamma=1$.

The next proposition summarizes the asymptotic distribution of the HPJ estimator.

\begin{proposition}
\label{proposition::HPJ_distribution_theta}
Under Assumptions \ref{ass::drifting}--\ref{ass::expansion}, as $N,T\to\infty$,
\begin{equation*}
\sqrt{NT}
\left(
\widehat{\theta}_{HPJ}^{(h)}
-
\theta_{0}^{(h)}
-
\frac{1}{T}
b_{\theta_{HPJ}}^{(h,1)}
\mathit{I}(\gamma=1)
\right)
\dto
\calN\left(
0,
\lim_{T\to\infty}
\frac{
\alpha_{T}^{2h}\sigma_{\epsi}^{2}
+
\sum_{q=0}^{h-1}\alpha_{T}^{2q}\sigma_{\nu}^{2}
}{
\sigma_{x}^{2}
}
\right)
\end{equation*}
for all $h=0,\ldots,H$ and for $\kappa$ satisfying the conditions in Proposition \ref{proposition::FE_distribution_theta}.
\end{proposition}

Consistent with the stationary results of \citet{GalvaoKato2014}, \citet{SPJ}, and \citet{mei2023nickellbiaspanellocal}, Proposition \ref{proposition::HPJ_distribution_theta} shows that HPJ bias correction has no first-order asymptotic effect on the variance of $\widehat{\theta}_{HPJ}^{(h)}$. Thus, under the stated large-$N$, large-$T$ asymptotic regime, standard errors that are valid for the FE estimator of $\theta^{(h)}$ are also asymptotically valid for the HPJ estimator.

The simulation results in Table \ref{table::res_illustration_coverage_new} in Section \ref{ssection:numerical_illustration} suggest that this first-order asymptotic result does not fully capture the finite-sample behavior of HPJ-based inference. In particular, coverage deteriorates as persistence increases, especially at $h=4$ and for smaller values of $T$. The large-$N$, large-$T$ results in Proposition \ref{proposition::HPJ_distribution_theta} therefore do not explain why coverage can remain below its nominal level even for larger values of $T$.

Because the nonuniformity of the HPJ approach originates from the control-variable-induced component of the bias when $\gamma\geq 1$, it is informative to investigate the properties of the FE and HPJ estimators of $\delta^{(h)}$ in this region. For illustrative purposes, we restrict attention to the special case considered in the following proposition.

\begin{proposition}
\label{proposition::HPJ_distribution_delta}
Let
$\epsi_{i,t}\overset{\scriptstyle\text{iid}}{\sim}\calN(0,1)$
and $\gamma=\infty$. Then, under Assumptions \ref{ass::drifting}--\ref{ass::expansion}, as $N,T\to\infty$,
\begin{equation*}
T\sqrt{N}
\left(
\widehat{\delta}_{HPJ}^{(0)}
-
1
\right)
\dto
\calN\left(
0,
\frac{132}{5}
\right),
\end{equation*}
provided that $\theta_{0}=0$ (known) and $\kappa>1/2$.
\end{proposition}
The main implication of this proposition is that HPJ bias correction has a first-order asymptotically nonnegligible effect on the variance of the estimator of the control-variable coefficient when $\gamma=\infty$. Note that here we assume that the value of $\theta_{0}$ is known to the researcher.

In particular, Theorem 4 of \citet{Hahn2002} implies that
\begin{equation*}
T\sqrt{N}
\left(
\widehat{\delta}_{FE}^{(0)}
-
1
+
\frac{3}{T}
\right)
\dto
\calN\left(
0,
\frac{51}{5}
\right).
\end{equation*}
Thus, Proposition \ref{proposition::HPJ_distribution_delta} contrasts with Proposition \ref{proposition::HPJ_distribution_theta}, which documents no first-order variance effect for $\widehat{\theta}_{HPJ}^{(0)}$.

We view Proposition \ref{proposition::HPJ_distribution_delta} as providing a partial explanation for the patterns documented in Table \ref{table::res_illustration_coverage_new}. Although the first-order asymptotic distribution of $\widehat{\theta}_{HPJ}^{(h)}$ does not require a variance adjustment, the variance of the HPJ estimator of $\delta^{(h)}$ differs from that of the FE estimator. In small and moderately sized samples, the two half-panel components of $\widehat{\vbeta}_{HPJ}^{(h)}$ remain correlated. This motivates the construction of the bias-correction-adjusted standard errors formally introduced in Section \ref{ssection:inference}, which account for this correlation.

We set $\theta_{0}=0$ and $h=0$ in Proposition \ref{proposition::HPJ_distribution_delta} to streamline otherwise tedious derivations of the asymptotic variance. As illustrated in Table \ref{table::res_illustration_coverage_new}, the need for adjusted standard errors is not confined to $h=0$ or $\theta_{0}=0$. Moreover, following \citet{Hahn2002}, we restrict attention to normally distributed disturbances $\epsi_{i,t}$ in order to simplify the derivations of the variance contributions associated with $\mB_{i,T}$. We expect that qualitatively similar conclusions will hold for general $\gamma\geq 1$ and for the feasible estimator when the value of $\theta_{0}=0$ is not known to the researcher.

\section{Takeaways for empirical researchers}
\label{section:applicability}

In this section, we summarize the main implications of the theoretical results developed in Section \ref{section:asymptotic} for empirical practice.

\subsection{Jackknife correction is broadly applicable}
\label{ssection:analytical}

The results in Section \ref{ssection::nonuniformity} show that the HPJ procedure is applicable over a broader range of persistence regimes than the analytical large-$T$ bias correction of \citet{HerbstJohannsen2024}. In particular, the HPJ is applicable not only when $\gamma\in[0,1)$, where the method of \citet{HerbstJohannsen2024} is also valid, but also when $\gamma\in(1,\infty]$. In the latter region, the large-$T$ analytical bias correction fails because of an additional bias component induced by the persistent control variable.

For example, when $h=0$, the bias term derived by \citet{HerbstJohannsen2024} is zero, so their procedure does not recommend any bias correction. In contrast, Proposition \ref{lemma::expansionFE} highlights a nonnegligible bias at $h=0$ when $\gamma\in[1,\infty]$. The breakdown of the analytical bias correction in this setting is akin to Theorem 4 in \citet{Hahn2002}, which whows that an analytical first-order bias correction developed for stable autoregressive dynamics, corresponding to $\gamma=0$ in our notation, is not applicable in the unit-root setting, corresponding to $\gamma=\infty$.\footnote{See \citet{Juodis2013RK} for a related result for the panel VAR(1) model.}

The HPJ approach is not valid at $\gamma=1$. Nevertheless, outside the local-to-unity boundary, it removes both the standard LP bias component and the additional contamination bias induced by persistent controls. It is therefore a useful default bias-correction procedure when the degree of persistence is uncertain, provided that inference accounts for the additional variation introduced by the correction.

\subsection{Bias-correction-adjusted inference is recommended}
\label{ssection:inference}

The results in Section \ref{section:asymptotic} have two implications for the asymptotic distributions of the FE and HPJ estimators:
\begin{enumerate}
\item For the coefficient of interest $\theta^{(h)}$, the FE and HPJ estimators have the same first-order asymptotic variance as $N,T\to\infty$.

\item This result need not apply to the coefficients on the control variables, $\delta^{(h)}$. In particular, the asymptotic variance of the HPJ estimator can exceed that of the FE estimator.
\end{enumerate}

Consequently, inference on $\theta_{0}^{(h)}$ can be substantially distorted in finite samples if the additional variability induced by bias correction is not accounted for; see, for example, Table \ref{table::res_illustration_coverage_new}. We therefore propose bias-correction-adjusted standard errors for the HPJ estimator. \ref{appendix:inference} explains why the methods proposed by \citet{GalvaoKato2014} and \citet{mei2023nickellbiaspanellocal} are less generally applicable than the procedure proposed here.

For the discussion that follows, let $Avar_{\sqrt{N}\vbeta_{FE}^{(h)}}$ denote the asymptotic variance of the appropriately scaled FE estimator $\sqrt{N}\widehat{\vbeta}_{FE}^{(h)}$ in the fixed-$T$ limit, and let $Avar_{\sqrt{N}\vbeta_{HPJ}^{(h)}}$ denote the corresponding quantity for $\sqrt{N}\widehat{\vbeta}_{HPJ}^{(h)}$.

The plug-in clustered covariance matrix (CCM) estimator of $Avar_{\vbeta_{FE}^{(h)}}$ is given by
\begin{align*}
\widehat{Avar}_{\sqrt{N}\vbeta_{FE}^{(h)}}
&=
\left(
\frac{1}{N}
\sum_{i=1}^{N}
\mB_{i,T}^{(h)}
\right)^{-1}
\left(
\frac{1}{N}
\sum_{i=1}^{N}
\widehat{\vs}_{i,T}^{(h)}
\left(
\widehat{\vbeta}_{FE}^{(h)}
\right)
\widehat{\vs}_{i,T}^{(h)}
\left(
\widehat{\vbeta}_{FE}^{(h)}
\right)'
\right)
\left(
\frac{1}{N}
\sum_{i=1}^{N}
\mB_{i,T}^{(h)}
\right)^{-1},
\label{eq::avar_fe}
\end{align*}
where, for any vector $\ddot{\vbeta}$,
\begin{equation}
\label{eq::influence_estimate}
\widehat{\vs}_{i,T}^{(h)}
\left(
\ddot{\vbeta}
\right)
=
\frac{1}{T-h}
\sum_{t=1}^{T-h}
\widetilde{\vx}_{i,t}
\left(
\widetilde{y}_{i,t+h}
-
\widetilde{\vx}_{i,t}'\ddot{\vbeta}
\right).
\end{equation}
For any fixed value of $T$, this estimator consistently estimates the asymptotic variance of $\sqrt{N}\widehat{\vbeta}_{FE}^{(h)}$; see, for example, \citet{Hansen2007597}. Moreover, \citet{GalvaoKato2014} show that in the stationary setup the CCM approach (with additional scaling by $T$) consistently estimates the asymptotic variance of the FE estimator as $N,T\to\infty$, including when the FE estimator is consistent only for a pseudo-true value.

Because the CCM approach applies for any fixed $T$, it can also be used to estimate the asymptotic variances of $\sqrt{N}\widehat{\vbeta}_{1,FE}^{(h)}$ and $\sqrt{N}\widehat{\vbeta}_{2,FE}^{(h)}$, as well as their covariance with the full-sample estimator. Our proposed approach exploits the fact that the HPJ estimator is a linear combination of these three estimators; see equation \eqref{eq::SPJ}. In particular, define
\begin{equation}
\label{eq::baa_ccm}
\widehat{Avar}_{\sqrt{N}\vbeta_{HPJ}^{(h)}}
:=
\frac{1}{N}
\sum_{i=1}^{N}
\widehat{\vq}_{i,T}^{(h)}
\left(
\widehat{\vq}_{i,T}^{(h)}
\right)'
\end{equation}
where
\begin{align*}
\widehat{\vq}_{i,T}^{(h)}
&:=
2
\left(
\frac{1}{N}
\sum_{i=1}^{N}
\mB_{i,T}^{(h)}
\right)^{-1}
\widehat{\vs}_{i,T}^{(h)}
\left(
\widehat{\vbeta}_{FE}^{(h)}
\right)
\notag\\
&\quad
-
\frac{1}{2}
\left[
\left(
\frac{1}{N}
\sum_{i=1}^{N}
\mB_{i,T_{1}}^{(h)}
\right)^{-1}
\widehat{\vs}_{i,T_{1}}^{(h)}
\left(
\widehat{\vbeta}_{1,FE}^{(h)}
\right)
+
\left(
\frac{1}{N}
\sum_{i=1}^{N}
\mB_{i,T_{2}}^{(h)}
\right)^{-1}
\widehat{\vs}_{i,T_{2}}^{(h)}
\left(
\widehat{\vbeta}_{2,FE}^{(h)}
\right)
\right].
\end{align*}
Here, $\widehat{\vs}_{i,T_{1}}^{(h)}(\cdot)$ and $\widehat{\vs}_{i,T_{2}}^{(h)}(\cdot)$ are defined analogously to $\widehat{\vs}_{i,T}^{(h)}(\cdot)$, using the respective half-samples.

Like the standard CCM estimator, this construction consistently estimates the 
variance of $\sqrt{N}\widehat{\vbeta}_{HPJ}^{(h)}$ for any fixed value of $T$. Although the procedure is also designed to remain valid when $T\to\infty$, we do not rely on a large-$T$ approximation to simplify the expression for $\widehat{\vq}_{i,T}^{(h)}$.\footnote{This is the fundamental distinction between our method and that of \citet{mei2023nickellbiaspanellocal}; see \ref{appendix:inference}.} Because the construction explicitly accounts for the bias-correction structure of the HPJ estimator, we refer to the resulting standard errors as \emph{bias-correction-adjusted} standard errors. Note that such standard errors only aim at correctly estimating the estimation variability, while true-parameter coverage of confidence intervals requires the remaining bias to be negligible at the applicable rate or to be corrected.

The procedure accounts both for the additional sample variation induced by HPJ bias correction when $T$ is fixed and for the finite-sample and asymptotic variance contributions of the denominator $\mB_{i,T}^{(h)}$. The latter contributions are central to the result in Proposition \ref{proposition::HPJ_distribution_delta}. The proposed variance estimator is therefore designed to remain valid across the persistence regimes indexed by $\gamma$.

Finally, some bootstrap procedures considered in the literature can also deliver confidence intervals with an appropriate width. Examples include the pairs bootstrap considered by \citet{SPJ}, \citet{GalvaoKato2014}, and \citet{goncalves2013}, as well as the score-based wild bootstrap of \citet{JuodisMiss2020}.

\subsection{Robustness to general shock processes}
\label{ssection::alternativeDGP}

Our exposition follows the framework of \citet{https://doi.org/10.1002/jae.2333} and \citet{HerbstJohannsen2024}, in which the iid structural shocks $x_{i,t}$ are directly observed by the empirical researcher. In many applications, however, this assumption may be restrictive. For example, following \citet{https://doi.org/10.3982/ECTA18756} and \citet{mei2023nickellbiaspanellocal}, one may instead consider a setting in which the observed vector $\left(y_{i,t},x_{i,t}\right)'$ is generated by a panel VAR($p$). In this case, the control vector $\vc_{i,t-1}$ contains additional lags of $\left(y_{i,t},x_{i,t}\right)'$. It is therefore natural to consider the implications of the results in Section \ref{section:asymptotic} for this more general setting.\footnote{The importance of accommodating dynamic feedback is also emphasized by \citet{Klosin2024}, who studies the bias induced by omitting lagged outcomes in static FE regressions. In contrast, our focus is on LP specifications in which lagged outcomes are included, but may be highly persistent.}

The mechanisms highlighted in Section \ref{section:asymptotic} are likely to continue to apply in this setting. Propositions \ref{lemma::expansionFE}--\ref{lemma::expansionFE_control} show that the weak-exogeneity bias depends on the persistence of all regressors in $\vx_{i,t}$, rather than solely on the persistence of the shock of interest $x_{i,t}$. Thus, in a richer dynamic specification, the HPJ may fail to remove the leading bias when one or more regressors are local-to-unity, corresponding to $\gamma=1$. Similar considerations apply to both $\widehat{\theta}^{(h)}$ and $\widehat{\delta}^{(h)}$. This includes settings with serially dependent shocks and a richer set of controls, which may be nonstationary or cointegrated.

The same reasoning applies to the variance effects of HPJ bias correction. The special-case result in Proposition \ref{proposition::HPJ_distribution_delta} suggests that bias correction can have first-order variance effects when an estimator converges at the fast, unit-root-type rate $T\sqrt{N}$. By contrast, such effects need not arise for estimators converging at the standard panel rate $\sqrt{NT}$, as in Proposition \ref{proposition::HPJ_distribution_theta}. When the shock process $x_{i,t}$ is stable, a $\sqrt{NT}$ convergence rate can often be obtained by appropriately lag-augmenting the LP specification; see, for example, \citet{TODA1995225} and \citet{https://doi.org/10.3982/ECTA18756}.\footnote{For the stylized DGP considered in this paper, lag augmentation is equivalent to including $y_{i,t-2}$ as an additional, irrelevant control variable.}

At the same time, the results in Table \ref{table::res_illustration_coverage_new} and the discussion in Section \ref{ssection:inference} show that inference should account for the relevant convergence rate and for the additional variability induced by bias correction. Researchers may therefore use the bias-correction-adjusted standard errors proposed in this paper or an appropriate bootstrap-based procedure.

\section{Empirical illustrations}
\label{section:empirical}

In this section, we revisit selected empirical findings of \citet{mei2023nickellbiaspanellocal} on the use of the HPJ approach in panel LP applications. Whereas their analysis compares the FE and HPJ estimators in terms of point estimates and associated confidence intervals, we focus exclusively on the properties of alternative confidence intervals for the HPJ estimator. Our empirical illustrations draw on the studies of \citet{Cerra2008} and \citet{Romer2017}, which span a range of persistence levels in the outcome and control variables. For simplicity, we occasionally modify the original datasets to obtain balanced panels.\footnote{We do not revisit the other empirical illustrations considered by \citet{mei2023nickellbiaspanellocal}, because those applications use standard errors clustered in both the cross-sectional and time-series dimensions. Extending the proposed procedure to this setting is beyond the scope of this paper.}

\subsection{Currency crises: \citet{Cerra2008}}

\citet{Cerra2008} study the effect of currency crises on economic growth using a cross-country dataset. Their shock variable of interest is based on the exchange-market pressure index (EMPI). The EMPI is defined as the sum of the percentage depreciation of the exchange rate and the percentage loss of foreign-exchange reserves, which facilitates cross-country comparability. A currency-crisis dummy is assigned to a country-year when the EMPI lies in the upper quartile of all observations in the panel.

Using this measure, denoted by $x_{i,t}^{CC}$, we consider the following LP specification, as in \citet{mei2023nickellbiaspanellocal}:
\begin{equation*}
\Delta_{h}y_{i,t+h}
=
\eta_{i}^{(h)}
+
\theta^{(h)}x_{i,t}^{CC}
+
\vc_{i,t-1}'\vdelta^{(h)}
+
u_{i,t}^{(h)},
\end{equation*}
for $h=1,\ldots,10$, where the vector of control variables takes the form
\begin{equation*}
\vc_{i,t-1}
:=
\left(
x_{i,t-1}^{CC},
\ldots,
x_{i,t-4}^{CC},
\Delta y_{i,t-1},
\ldots,
\Delta y_{i,t-4}
\right)'.
\end{equation*}
The dependent variable $\Delta_{h}y_{i,t+h}$ denotes the change in the logarithm of real GDP between years $t-1$ and $t+h$.

Because $\vc_{i,t-1}$ contains four lags of both the crisis indicator and GDP growth, 1965 is the first year used in estimation. However, owing to missing observations for some countries, we restrict attention to a balanced subsample of $76$ countries for which estimation is possible from 1967 to 1998 at $h=1$, with appropriate adjustments for $h>1$.\footnote{In this application, the asymptotic approximation may deteriorate at longer horizons $h$, as $T$ is relatively small.}

\begin{figure}[htbp!]
    \centering
    \includegraphics[width=0.9\linewidth]{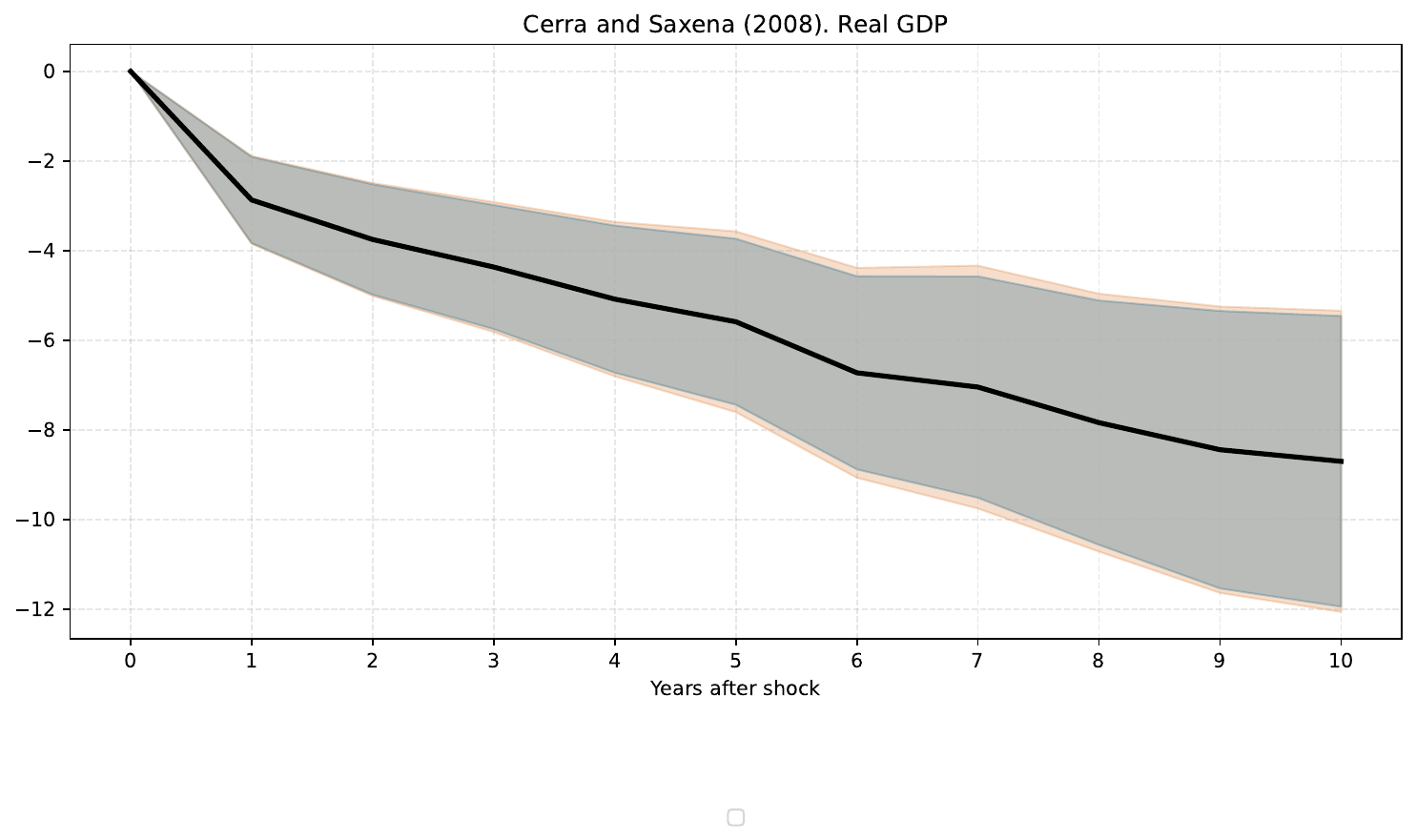}
    \caption{Impulse responses of real GDP to currency crises: \citet{Cerra2008}.}
    \caption*{\footnotesize Notes. This figure presents impulse responses of cumulative changes in the logarithm of real GDP to currency crises. The black solid line represents the HPJ estimate. The blue shaded area represents $95\%$ confidence intervals based on the standard errors of \citet{mei2023nickellbiaspanellocal}. The orange shaded area represents confidence intervals based on the bias-correction-adjusted standard errors. The grey area indicates the overlap between the two confidence intervals.}
    \label{fig::CS_2008}
\end{figure}

As documented by \citet{mei2023nickellbiaspanellocal}, the HPJ estimates imply a more severe and protracted decline in GDP following a currency crisis, suggesting that conventional FE estimates may substantially understate the economic costs of such crises. We re-examine the extent to which these findings depend on the choice of standard errors. The HPJ results are visualized in Figure \ref{fig::CS_2008} and summarized in Table \ref{tab:CS_2008}.\footnote{Comparison with the full-sample results of \citet{mei2023nickellbiaspanellocal} indicates that the main HPJ patterns are preserved in our balanced subsample: the impulse response is downward sloping, and the confidence intervals widen as $h$ increases.}

At all horizons, the confidence intervals based on our procedure are marginally wider than those based on the method of \citet{mei2023nickellbiaspanellocal}. Table \ref{tab:CS_2008} confirms this finding. This likely reflects the limited persistence of the GDP-growth controls and the crisis indicator in this application. At the same time, the HPJ standard errors of \citet{mei2023nickellbiaspanellocal} are generally substantially larger than those obtained using either the FE-based CCM or the HPJ-CCM procedure of \citet{GalvaoKato2014}.

\begin{table}[htbp!]
  \renewcommand{\arraystretch}{.9}
  \addtolength{\tabcolsep}{-2pt}
  \centering
  \footnotesize
  \caption{Ratios of standard errors: \citet{Cerra2008}.}
\begin{tabular}{c|cccccccccc}
\hline
\hline
&\multicolumn{10}{c}{$h$}\\
\hline
&1&2&3&4&5&6&7&8&9&10\\
\hline
FE      & 0.9421 & 0.8902 & 0.8722 & 0.8282 & 0.7914 & 0.7550 & 0.7413 & 0.6883 & 0.6541 & 0.6447 \\
GK2014  & 0.9597 & 0.9188 & 0.9049 & 0.8558 & 0.8055 & 0.7808 & 0.7552 & 0.6964 & 0.6595 & 0.6713 \\
MSS2026 & 1      & 1      & 1      & 1      & 1      & 1      & 1      & 1      & 1      & 1 \\
BCA     & 1.0192 & 1.0241 & 1.0473 & 1.0494 & 1.0890 & 1.0871 & 1.0965 & 1.0557 & 1.0321 & 1.0361 \\
\hline
\hline
\end{tabular}
\caption*{Notes. $h$ denotes the forecast horizon. All entries report ratios of the corresponding standard errors relative to those reported by \citet{mei2023nickellbiaspanellocal}. ``FE'' refers to the clustered covariance matrix (CCM) approach of \citet{doi:10.1111/j.1468-0084.1987.mp49004006.x}, based on FE residuals. ``GK2014'' refers to the HPJ-CCM approach following \citet{GalvaoKato2014}. ``MSS2026'' corresponds to the method of \citet{mei2023nickellbiaspanellocal}, and ``BCA'' denotes the bias-correction-adjusted procedure proposed in Section \ref{ssection:inference}.}
\label{tab:CS_2008}
\end{table}

\subsection{Financial distress: \citet{Romer2017}}

\citet{Romer2017} develop a semiannual index of financial distress for $24$ advanced economies over the period 1967--2012. The index is based on narrative evidence on credit-supply disruptions reported in the \emph{OECD Economic Outlook}. It distinguishes five levels of crisis severity---credit disruption, minor, moderate, major, and extreme---each subdivided into minus, regular, and plus categories. This produces a 16-point scale ranging from $0$, indicating no distress, to $15$, indicating an extreme crisis.

Using this measure, denoted by $x_{i,t}^{FD}$, \citet{mei2023nickellbiaspanellocal} estimate the following panel LP specification:
\begin{equation}
\label{eq::PLP_RR}
y_{i,t+h}
=
\eta_{i}^{(h)}
+
g_{t}^{(h)}
+
\theta^{(h)}x_{i,t}^{FD}
+
\vc_{i,t-1}'\vdelta^{(h)}
+
u_{i,t}^{(h)},
\end{equation}
for $h=0,\ldots,10$, where the vector of control variables is
\begin{equation*}
\vc_{i,t-1}
=
\left(
x_{i,t-1}^{FD},
\ldots,
x_{i,t-4}^{FD},
y_{i,t-1},
\ldots,
y_{i,t-4}
\right)'.
\end{equation*}
The dependent variable $y_{i,t+h}$ is either the logarithm of real GDP or the unemployment rate in country $i$ at time $t+h$.\footnote{The linear model in equation \eqref{eq::PLP_RR} includes horizon-specific time effects, $g_{t}^{(h)}$. We remove these effects by cross-sectionally demeaning the data before estimation.}

To obtain the effects of a typical financial-distress shock, \citet{Romer2017} multiply the estimates of $\theta^{(h)}$ by $7$, corresponding to a ``moderate crisis (minus).'' We follow the same convention when reporting results in the tables and figures.

\begin{figure}[htbp!]
    \centering
    \includegraphics[width=0.9\linewidth]{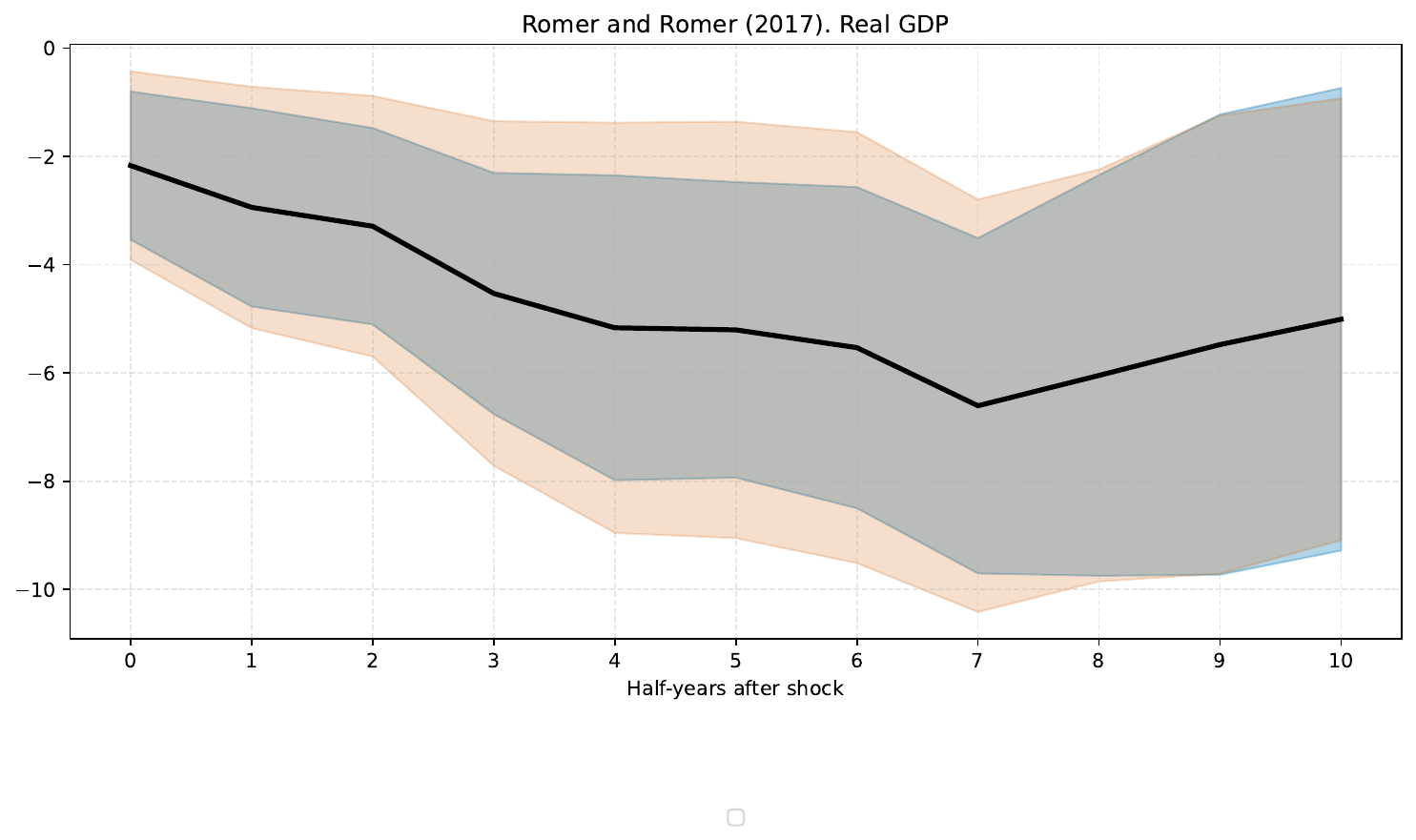}
    \caption{Impulse responses of real GDP to financial distress: \citet{Romer2017}.}
        \caption*{\footnotesize Notes. The black solid line represents the HPJ estimate. The blue shaded area represents $95\%$ confidence intervals based on the standard errors of \citet{mei2023nickellbiaspanellocal}. The orange shaded area represents confidence intervals based on the bias-correction-adjusted standard errors. The grey area indicates the overlap between the two confidence intervals.}
    \label{fig::RR_1}
\end{figure}

Using the FE estimator, \citet{Romer2017} find that output declines and unemployment increases moderately but persistently following financial crises, with statistically significant effects. \citet{mei2023nickellbiaspanellocal} find that these effects become generally larger after HPJ bias correction. Figure \ref{fig::RR_1} displays the HPJ estimates for the logarithm of real GDP, while Table \ref{tab:RR_2017_1} compares the associated standard errors.

For horizons up to and including $h=8$, the bias-correction-adjusted standard errors are substantially larger than those of \citet{mei2023nickellbiaspanellocal}, resulting in wider confidence intervals. Nevertheless, none of the resulting confidence intervals includes zero, confirming the qualitative conclusions of \citet{Romer2017} and \citet{mei2023nickellbiaspanellocal}. For this dataset, the procedures of \citet{mei2023nickellbiaspanellocal} and \citet{GalvaoKato2014} yield very similar confidence intervals, as illustrated in Table \ref{tab:RR_2017_1}.

\begin{figure}[htbp!]
    \centering
    \includegraphics[width=0.9\linewidth]{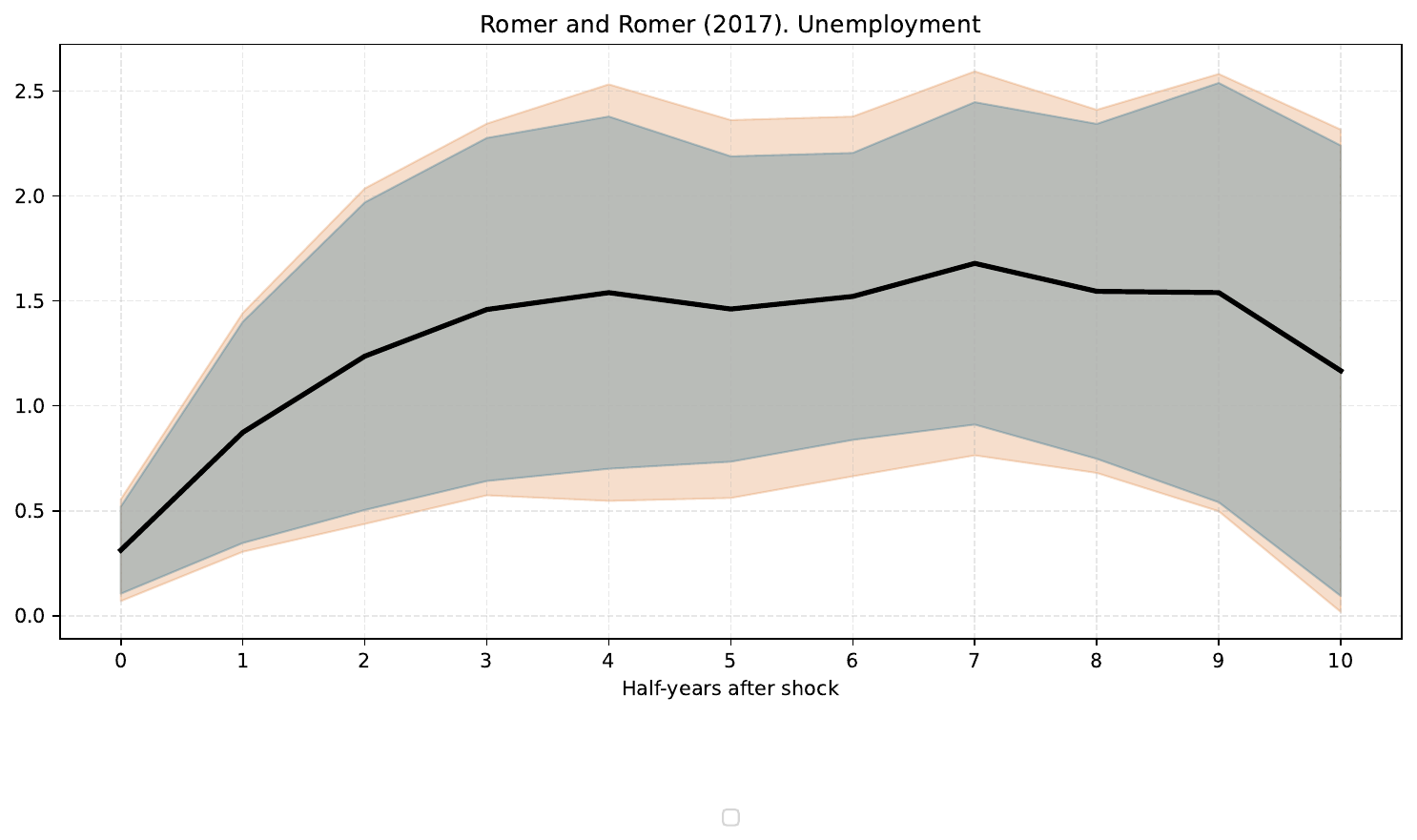}
    \caption{Impulse responses of the unemployment rate to financial distress: \citet{Romer2017}. }
            \caption*{\footnotesize Notes. The black solid line represents the HPJ estimate. The blue shaded area represents $95\%$ confidence intervals based on the standard errors of \citet{mei2023nickellbiaspanellocal}. The orange shaded area represents confidence intervals based on the bias-correction-adjusted standard errors. The grey area indicates the overlap between the two confidence intervals.}
    \label{fig::RR_2}
\end{figure}

For the unemployment rate, the same pattern continues to hold. In particular, confidence intervals based on the bias-correction-adjusted standard errors are generally wider than those based on the method of \citet{mei2023nickellbiaspanellocal}. Again, the differences are most pronounced at intermediate horizons.\footnote{We remove the first two years of observations, corresponding to the first four half-years, to obtain a balanced panel. The resulting estimates in Figure \ref{fig::RR_2} are therefore not numerically identical to those of \citet{mei2023nickellbiaspanellocal}, but they are qualitatively similar. Omitting the first four half-years somewhat reduces the HPJ point estimates and tightens the confidence intervals. Consequently, zero is excluded from the confidence intervals at all horizons in our results. By contrast, in \citet{mei2023nickellbiaspanellocal}, zero is included in the confidence intervals at $h=9,10$.}

\begin{table}[htbp!]
  \renewcommand{\arraystretch}{.9}
  \addtolength{\tabcolsep}{-2pt}
  \centering
  \footnotesize
  \caption{Ratios of standard errors: \citet{Romer2017}, real GDP.}
\begin{tabular}{c|ccccccccccc}
\hline
\hline
&\multicolumn{11}{c}{$h$}\\
\hline
&0&1&2&3&4&5&6&7&8&9&10\\
\hline
FE      & 1.0167 & 1.0470 & 1.0572 & 1.0688 & 1.0258 & 1.0377 & 0.9975 & 0.9432 & 0.8439 & 0.8030 & 0.7871 \\
GK2014  & 1.0004 & 0.9935 & 0.9931 & 1.0102 & 0.9945 & 0.9940 & 0.9860 & 0.9920 & 0.9486 & 0.8860 & 0.9016 \\
MSS2026 & 1      & 1      & 1      & 1      & 1      & 1      & 1      & 1      & 1      & 1      & 1 \\
BCA     & 1.2719 & 1.2174 & 1.3273 & 1.4278 & 1.3457 & 1.4088 & 1.3417 & 1.2298 & 1.0283 & 0.9953 & 0.9548 \\
\hline
\hline
\end{tabular}
\caption*{Notes. See Table \ref{tab:CS_2008}.}
\label{tab:RR_2017_1}
\end{table}

\begin{table}[htbp!]
  \renewcommand{\arraystretch}{.9}
  \addtolength{\tabcolsep}{-2pt}
  \centering
  \footnotesize
  \caption{Ratios of standard errors: \citet{Romer2017}, unemployment rate.}
\begin{tabular}{c|ccccccccccc}
\hline
\hline
&\multicolumn{11}{c}{$h$}\\
\hline
&0&1&2&3&4&5&6&7&8&9&10\\
\hline
FE      & 1.0283 & 1.0014 & 0.9944 & 0.9877 & 1.0013 & 0.9892 & 0.9199 & 0.8588 & 0.8449 & 0.7690 & 0.7783 \\
GK2014  & 0.9977 & 0.9964 & 0.9930 & 0.9886 & 0.9888 & 0.9801 & 0.9705 & 0.9733 & 0.9764 & 0.9664 & 0.9677 \\
MSS2026 & 1      & 1      & 1      & 1      & 1      & 1      & 1      & 1      & 1      & 1      & 1 \\
BCA     & 1.1696 & 1.0793 & 1.0908 & 1.0834 & 1.1826 & 1.2376 & 1.2535 & 1.1909 & 1.0843 & 1.0428 & 1.0708 \\
\hline
\hline
\end{tabular}
\caption*{Notes. See Table \ref{tab:CS_2008}.}
\label{tab:RR_2017_2}
\end{table}

\section{Monte Carlo study}
\label{section:montecarlo}

\subsection{The setup}

In this section, we expand on the preliminary numerical findings presented in Section \ref{ssection:numerical_illustration}. In particular, we consider the DGP
\begin{align}
y_{i,t}
&=
\mu_{i}
+
u_{i,t},
\\
u_{i,t}
&=
\alpha_{0}u_{i,t-1}
+
\theta_{0}x_{i,t}
+
\epsi_{i,t},
\label{eq::GDP1_MC}
\end{align}
and set $\mu_{i}=0$ for $i=1,\ldots,N$ without loss of generality. We generate
$\epsi_{i,t}\overset{\scriptstyle\text{iid}}{\sim}\calN(0,1)$
and
$x_{i,t}\overset{\scriptstyle\text{iid}}{\sim}\calN(0,1)$,
and set $\theta_{0}=1$, as in \citet{HerbstJohannsen2024}. We also set $u_{i,0}=0$ for all $i$.

We consider the following fixed values of $\alpha_{0}$:
\begin{equation*}
\alpha_{0}
\in
\left\{
0.70,
0.85,
0.95,
0.99,
1.00
\right\}.
\end{equation*}
Rather than imposing the drifting sequences in equation \eqref{eq::drifting}, we use fixed values of $\alpha_{0}$ to assess the empirical relevance of the theoretical framework developed in this paper. Finally, we consider the following sample sizes:
\begin{equation*}
T
\in
\left\{
30,
50,
100
\right\},
\quad
N
\in
\left\{
30,
50,
100
\right\}.
\end{equation*}
These values are chosen to reflect the typical dimensions of the datasets considered in Section \ref{section:empirical}.

For each combination of design parameters, we report the scaled bias\footnote{Bias is measured relative to the true value $\theta_{0}^{(h)}$ and scaled by $\sqrt{NT}$.} of the FE, HJ, and HPJ estimators, as well as empirical coverage rates for confidence intervals based on a nominal coverage level of $95\%$. We do not report coverage rates for the uncorrected FE estimator.

For the HJ approach, we construct two-sided percentile-bootstrap confidence intervals, following \citet{goncalves2013} and \citet{GalvaoKato2014}. We denote this procedure by $B_{HJ}$. For the HPJ estimator, we consider three standard-error constructions for conventional two-sided confidence intervals:
\begin{itemize}
\item $t_{GK}$: the HPJ-CCM variance estimator of \citet{GalvaoKato2014};

\item $t_{MSS}$: the variance estimator proposed by \citet{mei2023nickellbiaspanellocal};

\item $t_{BCA}$: the \emph{bias-correction-adjusted} procedure defined in equation \eqref{eq::baa_ccm}.
\end{itemize}
The number of Monte Carlo replications is set to $M=4000$, and the number of bootstrap draws is set to $B=200$.

\subsection{Results}

Tables \ref{tab::MC_N30_bias}--\ref{tab::MC_N100_bias} report bias results, whereas Tables \ref{tab::MC_N30_coverage}--\ref{tab::MC_N100_coverage} report coverage rates. We first discuss the bias results.

Consistent with the results in Section \ref{section:asymptotic}, the bias of the FE estimator is generally small at $h=0$ when $\alpha_{0}$ is low, but becomes nonnegligible otherwise. At nonzero horizons and for lower values of $\alpha_{0}$, the HJ correction removes most of the bias. However, a nonnegligible bias remains as $\alpha_{0}$ approaches one. By contrast, the HPJ procedure is generally more successful in removing the leading bias when $\alpha_{0}\approx 1$, in line with Corollary \ref{lemma::expansionHPJ}.

The asymptotic approximation performs less well in designs where $N$ substantially exceeds $T$. In particular, for $N=100$ and $T=30$, a substantial bias remains that is not captured by the first-order asymptotic theory developed in this paper. This bias is especially pronounced at longer horizons, confirming the findings of \citet{everaert2026LP}. Restricting attention to $T=50$, the reported absolute bias is largest when $\alpha_{0}=0.95$. This pattern is consistent with the predictions of Proposition \ref{proposition::HPJ_distribution_theta}.

Turning to the coverage rates reported in Tables \ref{tab::MC_N30_coverage}--\ref{tab::MC_N100_coverage}, the $B_{HJ}$ procedure tends to yield coverage below the nominal level. This pattern is largely explained by the remaining bias of the HJ estimator. The coverage gap increases with both $h$ and $N$, and is particularly pronounced at $h=4$. The value of $\alpha_{0}$ is especially important at nonzero horizons, whereas its effect is less pronounced when $h=0$.

For the HPJ estimator, the choice of standard errors is important. In the reported designs, the $t_{BCA}$ procedure yields coverage rates that are consistently closer to the nominal level than those obtained using $t_{GK}$ and, to a lesser extent, $t_{MSS}$. The gains from the bias-correction-adjusted procedure become more pronounced as $h$, $\alpha_{0}$, or $N$ increases. Overall, the results support the use of bias-correction-adjusted standard errors in panel LP applications with moderate values of $T$, even though the three approaches are first-order asymptotically equivalent under the relevant regularity conditions.
  \begin{sidewaystable}[htbp!]
  \renewcommand{\arraystretch}{.9}
  \addtolength{\tabcolsep}{-2pt}
  \centering
 \footnotesize
  \caption{Monte Carlo bias results for $\theta^{(h)}$, $h=0,\ldots,4$, and $N=30$.}
  \begin{tabular}{cc|ccc|ccc|ccc|ccc|ccc}
  \hline
  \hline
  &&\multicolumn{3}{c|}{$h=0$}&\multicolumn{3}{c|}{$h=1$}&\multicolumn{3}{c|}{$h=2$}&\multicolumn{3}{c|}{$h=3$}&\multicolumn{3}{c}{$h=4$}\\
  \hline
  $T$&$\alpha_{0}$&\textnormal{FE}&\textnormal{HJ}&\textnormal{HPJ}&\textnormal{FE}&\textnormal{HJ}&\textnormal{HPJ}&\textnormal{FE}&\textnormal{HJ}&\textnormal{HPJ}&\textnormal{FE}&\textnormal{HJ}&\textnormal{HPJ}&\textnormal{FE}&\textnormal{HJ}&\textnormal{HPJ}\\
   \hline
\hline
 30 & 0.70 & -0.10 & -0.10 &  0.19 & -1.35 & -0.31 &  0.51 & -2.21 & -0.44 &  0.62 & -2.86 & -0.56 &  0.55 &  -3.32 & -0.64 &  0.27 \\
  & 0.85 & -0.22 & -0.22 &  0.29 & -1.72 & -0.69 &  0.78 & -3.01 & -1.10 &  1.00 & -4.18 & -1.51 &  0.97 &  -5.19 & -1.87 &  0.58 \\
  & 0.95 & -0.51 & -0.51 &  0.19 & -2.59 & -1.57 &  0.43 & -4.60 & -2.62 &  0.34 & -6.57 & -3.69 & -0.09 &  -8.47 & -4.74 & -0.94 \\
  & 0.99 & -0.71 & -0.71 &  0.02 & -3.21 & -2.20 & -0.10 & -5.67 & -3.68 & -0.53 & -8.14 & -5.20 & -1.32 & -10.58 & -6.71 & -2.55 \\
  & 1.00 & -0.76 & -0.76 & -0.02 & -3.34 & -2.33 & -0.22 & -5.89 & -3.90 & -0.74 & -8.49 & -5.52 & -1.62 & -11.05 & -7.15 & -2.93 \\
  \hline
 50 & 0.70 & -0.05 & -0.05 &  0.09 & -0.93 & -0.14 &  0.26 & -1.56 & -0.21 &  0.32 & -2.02 & -0.27 &  0.30 &  -2.35 & -0.31 &  0.23 \\
 & 0.85 & -0.11 & -0.11 &  0.17 & -1.11 & -0.32 &  0.51 & -1.98 & -0.51 &  0.71 & -2.75 & -0.70 &  0.82 &  -3.42 & -0.87 &  0.82 \\
 & 0.95 & -0.30 & -0.30 &  0.21 & -1.68 & -0.89 &  0.59 & -3.03 & -1.50 &  0.82 & -4.35 & -2.11 &  0.90 &  -5.64 & -2.72 &  0.81 \\
  & 0.99 & -0.54 & -0.54 &  0.04 & -2.39 & -1.61 &  0.08 & -4.25 & -2.71 & -0.05 & -6.12 & -3.82 & -0.34 &  -7.99 & -4.96 & -0.82 \\
 & 1.00 & -0.60 & -0.60 & -0.02 & -2.57 & -1.79 & -0.09 & -4.57 & -3.02 & -0.33 & -6.57 & -4.26 & -0.73 &  -8.59 & -5.54 & -1.32 \\
 \hline
100 & 0.70 & -0.01 & -0.01 &  0.04 & -0.59 & -0.04 &  0.10 & -1.01 & -0.07 &  0.12 & -1.31 & -0.09 &  0.12 &  -1.57 & -0.15 &  0.05 \\
& 0.85 & -0.03 & -0.03 &  0.07 & -0.65 & -0.10 &  0.21 & -1.20 & -0.17 &  0.30 & -1.67 & -0.23 &  0.36 &  -2.12 & -0.34 &  0.34 \\
 & 0.95 & -0.11 & -0.11 &  0.16 & -0.89 & -0.34 &  0.46 & -1.66 & -0.58 &  0.71 & -2.40 & -0.81 &  0.92 &  -3.16 & -1.10 &  1.03 \\
 & 0.99 & -0.32 & -0.32 &  0.07 & -1.53 & -0.98 &  0.20 & -2.74 & -1.65 &  0.26 & -3.96 & -2.33 &  0.27 &  -5.23 & -3.07 &  0.17 \\
& 1.00 & -0.41 & -0.41 & -0.00 & -1.80 & -1.25 & -0.03 & -3.20 & -2.11 & -0.12 & -4.62 & -2.98 & -0.27 &  -6.09 & -3.91 & -0.53 \\

\hline
\hline
    \end{tabular}%
    \caption*{Notes.  Here ``Bias'' refers to the Monte Carlo expectation of $\sqrt{NT}(\widehat{\theta}_{HPJ}^{(h)}-\theta_{0}^{(h)})$ for $\theta_{0}^{(h)}=\theta_{0}\alpha_{0}^{h}$. ``FE'' stands for the fixed-effect estimator, ``HJ'' is the bias-correction procedure of \citet{HerbstJohannsen2024}, while ``HPJ'' is the half-panel jackknife of \citet{SPJ}. All results are obtained using $M=4000$ Monte Carlo replications.}
    \label{tab::MC_N30_bias}
  \end{sidewaystable}
  \begin{sidewaystable}[htbp!]
  \renewcommand{\arraystretch}{.9}
  \addtolength{\tabcolsep}{-2pt}
  \centering
 \footnotesize
  \caption{Monte Carlo bias results for $\theta^{(h)}$, $h=0,\ldots,4$, and $N=50$.}
  \begin{tabular}{cc|ccc|ccc|ccc|ccc|ccc}
  \hline
  \hline
  &&\multicolumn{3}{c|}{$h=0$}&\multicolumn{3}{c|}{$h=1$}&\multicolumn{3}{c|}{$h=2$}&\multicolumn{3}{c|}{$h=3$}&\multicolumn{3}{c}{$h=4$}\\
  \hline
  $T$&$\alpha_{0}$&\textnormal{FE}&\textnormal{HJ}&\textnormal{HPJ}&\textnormal{FE}&\textnormal{HJ}&\textnormal{HPJ}&\textnormal{FE}&\textnormal{HJ}&\textnormal{HPJ}&\textnormal{FE}&\textnormal{HJ}&\textnormal{HPJ}&\textnormal{FE}&\textnormal{HJ}&\textnormal{HPJ}\\
   \hline
\hline
 30 & 0.70 & -0.15 & -0.15 &  0.22 & -1.70 & -0.37 &  0.68 & -2.84 & -0.55 &  0.81 &  -3.65 & -0.68 &  0.74 &  -4.28 & -0.83 &  0.35 \\
  & 0.85 & -0.30 & -0.30 &  0.35 & -2.18 & -0.86 &  1.03 & -3.86 & -1.39 &  1.31 &  -5.34 & -1.89 &  1.29 &  -6.69 & -2.39 &  0.77 \\
  & 0.95 & -0.67 & -0.67 &  0.22 & -3.31 & -1.99 &  0.57 & -5.89 & -3.34 &  0.47 &  -8.41 & -4.68 & -0.05 & -10.89 & -6.06 & -1.17 \\
  & 0.99 & -0.94 & -0.94 &  0.01 & -4.10 & -2.79 & -0.09 & -7.26 & -4.69 & -0.64 & -10.41 & -6.61 & -1.63 & -13.57 & -8.58 & -3.21 \\
  & 1.00 & -0.99 & -0.99 & -0.04 & -4.26 & -2.96 & -0.25 & -7.55 & -4.97 & -0.90 & -10.85 & -7.02 & -2.00 & -14.18 & -9.13 & -3.68 \\
 \hline
 50 & 0.70 & -0.05 & -0.05 &  0.12 & -1.19 & -0.17 &  0.31 & -2.00 & -0.25 &  0.39 &  -2.60 & -0.34 &  0.37 &  -3.01 & -0.38 &  0.30 \\
  & 0.85 & -0.12 & -0.12 &  0.23 & -1.42 & -0.40 &  0.63 & -2.54 & -0.64 &  0.90 &  -3.54 & -0.89 &  1.04 &  -4.39 & -1.09 &  1.06 \\
  & 0.95 & -0.36 & -0.36 &  0.27 & -2.15 & -1.14 &  0.74 & -3.89 & -1.91 &  1.02 &  -5.61 & -2.71 &  1.13 &  -7.25 & -3.48 &  1.04 \\
  & 0.99 & -0.67 & -0.67 &  0.07 & -3.07 & -2.06 &  0.09 & -5.46 & -3.46 & -0.07 &  -7.88 & -4.91 & -0.45 & -10.26 & -6.35 & -1.02 \\
  & 1.00 & -0.75 & -0.75 & -0.00 & -3.29 & -2.29 & -0.11 & -5.85 & -3.85 & -0.42 &  -8.45 & -5.46 & -0.94 & -11.02 & -7.07 & -1.65 \\
 \hline
100 & 0.70 & -0.02 & -0.02 &  0.04 & -0.74 & -0.03 &  0.15 & -1.26 & -0.04 &  0.20 &  -1.64 & -0.07 &  0.20 &  -1.94 & -0.12 &  0.15 \\
 & 0.85 & -0.04 & -0.04 &  0.09 & -0.82 & -0.11 &  0.29 & -1.49 & -0.17 &  0.44 &  -2.09 & -0.24 &  0.53 &  -2.63 & -0.33 &  0.55 \\
 & 0.95 & -0.14 & -0.14 &  0.20 & -1.13 & -0.41 &  0.62 & -2.08 & -0.68 &  0.98 &  -3.01 & -0.97 &  1.28 &  -3.95 & -1.28 &  1.48 \\
& 0.99 & -0.42 & -0.42 &  0.10 & -1.94 & -1.23 &  0.29 & -3.48 & -2.07 &  0.41 &  -5.03 & -2.92 &  0.45 &  -6.61 & -3.82 &  0.39 \\
& 1.00 & -0.54 & -0.54 & -0.01 & -2.30 & -1.59 & -0.02 & -4.08 & -2.67 & -0.11 &  -5.88 & -3.77 & -0.27 &  -7.72 & -4.91 & -0.54 \\

\hline
\hline
    \end{tabular}%
    \caption*{Notes.  See Table \ref{tab::MC_N30_bias}.}
    \label{tab::MC_N50_bias}
  \end{sidewaystable}
    \begin{sidewaystable}[htbp!]
  \renewcommand{\arraystretch}{.9}
  \addtolength{\tabcolsep}{-2pt}
  \centering
 \footnotesize
  \caption{Monte Carlo bias results for $\theta^{(h)}$, $h=0,\ldots,4$, and $N=100$.}
  \begin{tabular}{cc|ccc|ccc|ccc|ccc|ccc}
  \hline
  \hline
  &&\multicolumn{3}{c|}{$h=0$}&\multicolumn{3}{c|}{$h=1$}&\multicolumn{3}{c|}{$h=2$}&\multicolumn{3}{c|}{$h=3$}&\multicolumn{3}{c}{$h=4$}\\
  \hline
  $T$&$\alpha_{0}$&\textnormal{FE}&\textnormal{HJ}&\textnormal{HPJ}&\textnormal{FE}&\textnormal{HJ}&\textnormal{HPJ}&\textnormal{FE}&\textnormal{HJ}&\textnormal{HPJ}&\textnormal{FE}&\textnormal{HJ}&\textnormal{HPJ}&\textnormal{FE}&\textnormal{HJ}&\textnormal{HPJ}\\
   \hline
\hline
 30 & 0.70 & -0.16 & -0.16 &  0.36 & -2.38 & -0.50 &  0.95 &  -4.05 & -0.81 &  1.05 &  -5.24 & -1.04 &  0.92 &  -6.10 &  -1.22 &  0.39 \\
  & 0.85 & -0.37 & -0.37 &  0.54 & -3.05 & -1.17 &  1.45 &  -5.48 & -1.98 &  1.76 &  -7.61 & -2.72 &  1.70 &  -9.50 &  -3.43 &  0.97 \\
 & 0.95 & -0.89 & -0.89 &  0.37 & -4.61 & -2.76 &  0.83 &  -8.32 & -4.70 &  0.60 & -11.91 & -6.63 & -0.16 & -15.40 &  -8.58 & -1.72 \\
  & 0.99 & -1.26 & -1.26 &  0.07 & -5.72 & -3.87 & -0.09 & -10.23 & -6.59 & -0.92 & -14.70 & -9.31 & -2.32 & -19.15 & -12.08 & -4.51 \\
  & 1.00 & -1.34 & -1.34 &  0.00 & -5.94 & -4.10 & -0.30 & -10.63 & -6.98 & -1.27 & -15.30 & -9.88 & -2.82 & -19.99 & -12.84 & -5.16 \\
 \hline
 50 & 0.70 & -0.08 & -0.08 &  0.16 & -1.67 & -0.23 &  0.45 &  -2.78 & -0.31 &  0.61 &  -3.63 & -0.43 &  0.59 &  -4.23 &  -0.52 &  0.45 \\
& 0.85 & -0.18 & -0.18 &  0.32 & -1.99 & -0.55 &  0.91 &  -3.53 & -0.85 &  1.32 &  -4.93 & -1.18 &  1.54 &  -6.16 &  -1.49 &  1.54 \\
 & 0.95 & -0.52 & -0.52 &  0.38 & -3.02 & -1.59 &  1.05 &  -5.43 & -2.63 &  1.50 &  -7.83 & -3.73 &  1.67 & -10.17 &  -4.83 &  1.53 \\
 & 0.99 & -0.95 & -0.95 &  0.09 & -4.30 & -2.88 &  0.14 &  -7.62 & -4.80 & -0.04 & -11.00 & -6.80 & -0.52 & -14.37 &  -8.83 & -1.33 \\
 & 1.00 & -1.05 & -1.05 & -0.00 & -4.62 & -3.20 & -0.15 &  -8.17 & -5.34 & -0.51 & -11.79 & -7.57 & -1.19 & -15.43 &  -9.84 & -2.19 \\
 \hline
100 & 0.70 & -0.02 & -0.02 &  0.06 & -1.08 & -0.07 &  0.17 &  -1.86 & -0.13 &  0.20 &  -2.36 & -0.14 &  0.23 &  -2.78 &  -0.20 &  0.17 \\
 & 0.85 & -0.06 & -0.06 &  0.13 & -1.19 & -0.18 &  0.37 &  -2.19 & -0.31 &  0.52 &  -3.00 & -0.39 &  0.67 &  -3.77 &  -0.51 &  0.71 \\
 & 0.95 & -0.20 & -0.20 &  0.29 & -1.61 & -0.60 &  0.84 &  -3.01 & -1.04 &  1.28 &  -4.31 & -1.42 &  1.71 &  -5.62 &  -1.85 &  2.00 \\
 & 0.99 & -0.58 & -0.58 &  0.14 & -2.76 & -1.76 &  0.39 &  -4.97 & -2.97 &  0.51 &  -7.13 & -4.16 &  0.59 &  -9.35 &  -5.40 &  0.52 \\
& 1.00 & -0.74 & -0.74 &  0.00 & -3.25 & -2.25 & -0.03 &  -5.80 & -3.80 & -0.19 &  -8.31 & -5.32 & -0.39 & -10.89 &  -6.91 & -0.74 \\

\hline
\hline
    \end{tabular}%
    \caption*{Notes.  See Table \ref{tab::MC_N30_bias}.}
    \label{tab::MC_N100_bias}
  \end{sidewaystable}

 \begin{sidewaystable}[htbp!]
  \renewcommand{\arraystretch}{.9}
  \addtolength{\tabcolsep}{-2pt}
  \centering
 \footnotesize
  \caption{Monte Carlo coverage results for $\theta^{(h)}$, $h=0,\ldots,4$, and $N=30$.}
  \begin{tabular}{cc|cccc|cccc|cccc|cccc|cccc}
  \hline
  \hline
  &&\multicolumn{4}{c|}{$h=0$}&\multicolumn{4}{c|}{$h=1$}&\multicolumn{4}{c|}{$h=2$}&\multicolumn{4}{c|}{$h=3$}&\multicolumn{4}{c}{$h=4$}\\
  \hline
  $T$&$\alpha_{0}$&$B_{HJ}$&$t_{GK}$&$t_{MSS}$&$t_{BCA}$&$B_{HJ}$&$t_{GK}$&$t_{MSS}$&$t_{BCA}$&$B_{HJ}$&$t_{GK}$&$t_{MSS}$&$t_{BCA}$&$B_{HJ}$&$t_{GK}$&$t_{MSS}$&$t_{BCA}$&$B_{HJ}$&$t_{GK}$&$t_{MSS}$&$t_{BCA}$\\
   \hline
\hline
 30& 0.70 & 0.83 & 0.92 & 0.93 & 0.93 & 0.91 & 0.90 & 0.92 & 0.93 & 0.93 & 0.89 & 0.92 & 0.94 & 0.93 & 0.88 & 0.92 & 0.94 & 0.92 & 0.87 & 0.92 & 0.94 \\
  & 0.85 & 0.85 & 0.92 & 0.93 & 0.93 & 0.92 & 0.88 & 0.90 & 0.92 & 0.93 & 0.87 & 0.90 & 0.94 & 0.93 & 0.86 & 0.90 & 0.94 & 0.93 & 0.84 & 0.90 & 0.94 \\
  & 0.95 & 0.89 & 0.92 & 0.93 & 0.93 & 0.93 & 0.90 & 0.92 & 0.94 & 0.92 & 0.88 & 0.91 & 0.94 & 0.90 & 0.85 & 0.90 & 0.94 & 0.87 & 0.81 & 0.88 & 0.91 \\
 & 0.99 & 0.93 & 0.92 & 0.94 & 0.93 & 0.90 & 0.91 & 0.92 & 0.94 & 0.86 & 0.87 & 0.90 & 0.93 & 0.80 & 0.82 & 0.87 & 0.90 & 0.73 & 0.74 & 0.82 & 0.86 \\
  & 1.00 & 0.93 & 0.93 & 0.93 & 0.94 & 0.89 & 0.91 & 0.92 & 0.93 & 0.82 & 0.87 & 0.90 & 0.92 & 0.73 & 0.81 & 0.86 & 0.89 & 0.65 & 0.72 & 0.80 & 0.84 \\
 \hline
 50 & 0.70 & 0.88 & 0.94 & 0.94 & 0.94 & 0.91 & 0.92 & 0.93 & 0.93 & 0.92 & 0.91 & 0.92 & 0.93 & 0.93 & 0.90 & 0.92 & 0.94 & 0.93 & 0.89 & 0.92 & 0.93 \\
  & 0.85 & 0.88 & 0.93 & 0.94 & 0.94 & 0.91 & 0.91 & 0.91 & 0.92 & 0.92 & 0.89 & 0.91 & 0.92 & 0.93 & 0.87 & 0.90 & 0.93 & 0.93 & 0.86 & 0.90 & 0.93 \\
  & 0.95 & 0.90 & 0.93 & 0.93 & 0.93 & 0.92 & 0.90 & 0.91 & 0.93 & 0.92 & 0.88 & 0.90 & 0.92 & 0.92 & 0.87 & 0.90 & 0.93 & 0.91 & 0.86 & 0.89 & 0.93 \\
  & 0.99 & 0.93 & 0.93 & 0.94 & 0.94 & 0.92 & 0.91 & 0.92 & 0.93 & 0.88 & 0.90 & 0.91 & 0.93 & 0.86 & 0.87 & 0.90 & 0.93 & 0.82 & 0.85 & 0.88 & 0.92 \\
  & 1.00 & 0.93 & 0.94 & 0.94 & 0.94 & 0.90 & 0.91 & 0.92 & 0.94 & 0.85 & 0.90 & 0.91 & 0.93 & 0.81 & 0.87 & 0.89 & 0.92 & 0.76 & 0.84 & 0.88 & 0.91 \\
 \hline
100 & 0.70 & 0.89 & 0.93 & 0.93 & 0.93 & 0.91 & 0.93 & 0.93 & 0.93 & 0.92 & 0.94 & 0.94 & 0.94 & 0.92 & 0.93 & 0.94 & 0.94 & 0.93 & 0.92 & 0.93 & 0.94 \\
 & 0.85 & 0.89 & 0.93 & 0.93 & 0.93 & 0.91 & 0.92 & 0.92 & 0.93 & 0.92 & 0.93 & 0.94 & 0.94 & 0.92 & 0.93 & 0.94 & 0.94 & 0.92 & 0.92 & 0.93 & 0.94 \\
 & 0.95 & 0.90 & 0.93 & 0.93 & 0.93 & 0.91 & 0.91 & 0.91 & 0.92 & 0.92 & 0.91 & 0.92 & 0.94 & 0.92 & 0.90 & 0.91 & 0.94 & 0.92 & 0.89 & 0.91 & 0.93 \\
 & 0.99 & 0.92 & 0.93 & 0.93 & 0.93 & 0.92 & 0.92 & 0.92 & 0.93 & 0.92 & 0.92 & 0.92 & 0.94 & 0.92 & 0.91 & 0.92 & 0.94 & 0.90 & 0.89 & 0.91 & 0.93 \\
 & 1.00 & 0.92 & 0.93 & 0.93 & 0.93 & 0.91 & 0.92 & 0.92 & 0.93 & 0.89 & 0.92 & 0.92 & 0.93 & 0.87 & 0.91 & 0.92 & 0.93 & 0.85 & 0.89 & 0.90 & 0.93 \\

\hline
\hline
    \end{tabular}%
    \caption*{Notes. ``$B_{HJ}$'' is the empirical coverage rate of percentile bootstrap-based confidence intervals for the bias-corrected estimator of \citet{HerbstJohannsen2024};  ``$t_{GK}$'' is the empirical coverage rate of confidence intervals based on the HPJ-CCM estimator of \citet{GalvaoKato2014}; ``$t_{MSS}$'' is the empirical coverage rate of confidence intervals based on the approach of \citet{mei2023nickellbiaspanellocal};
    ``$t_{BCA}$'' is the empirical coverage rate of confidence intervals based on bias-correction-adjusted SEs from Section \ref{ssection:inference}. All results are obtained using $M=4000$ Monte Carlo replications. The nominal coverage rate for the true value of $\theta_{0}^{(h)}$ is set at $95\%$.}
    \label{tab::MC_N30_coverage}
  \end{sidewaystable}
 \begin{sidewaystable}[htbp!]
  \renewcommand{\arraystretch}{.9}
  \addtolength{\tabcolsep}{-2pt}
  \centering
 \footnotesize
  \caption{Monte Carlo coverage results for $\theta^{(h)}$, $h=0,\ldots,4$, and $N=50$.}
  \begin{tabular}{cc|cccc|cccc|cccc|cccc|cccc}
  \hline
  \hline
  &&\multicolumn{4}{c|}{$h=0$}&\multicolumn{4}{c|}{$h=1$}&\multicolumn{4}{c|}{$h=2$}&\multicolumn{4}{c|}{$h=3$}&\multicolumn{4}{c}{$h=4$}\\
  \hline
  $T$&$\alpha_{0}$&$B_{HJ}$&$t_{GK}$&$t_{MSS}$&$t_{BCA}$&$B_{HJ}$&$t_{GK}$&$t_{MSS}$&$t_{BCA}$&$B_{HJ}$&$t_{GK}$&$t_{MSS}$&$t_{BCA}$&$B_{HJ}$&$t_{GK}$&$t_{MSS}$&$t_{BCA}$&$B_{HJ}$&$t_{GK}$&$t_{MSS}$&$t_{BCA}$\\
   \hline
\hline
 30 & 0.70 & 0.80 & 0.93 & 0.94 & 0.93 & 0.91 & 0.90 & 0.92 & 0.92 & 0.92 & 0.88 & 0.92 & 0.93 & 0.94 & 0.87 & 0.92 & 0.94 & 0.94 & 0.87 & 0.93 & 0.94 \\
 & 0.85 & 0.82 & 0.92 & 0.93 & 0.93 & 0.92 & 0.87 & 0.89 & 0.91 & 0.93 & 0.84 & 0.88 & 0.92 & 0.94 & 0.83 & 0.89 & 0.93 & 0.94 & 0.84 & 0.91 & 0.94 \\
 & 0.95 & 0.89 & 0.92 & 0.93 & 0.93 & 0.93 & 0.89 & 0.91 & 0.93 & 0.91 & 0.87 & 0.91 & 0.94 & 0.89 & 0.85 & 0.90 & 0.94 & 0.85 & 0.81 & 0.89 & 0.92 \\
  & 0.99 & 0.93 & 0.93 & 0.94 & 0.94 & 0.89 & 0.90 & 0.92 & 0.94 & 0.81 & 0.86 & 0.90 & 0.92 & 0.73 & 0.81 & 0.87 & 0.90 & 0.62 & 0.72 & 0.81 & 0.84 \\
 & 1.00 & 0.93 & 0.93 & 0.94 & 0.94 & 0.86 & 0.90 & 0.92 & 0.94 & 0.74 & 0.86 & 0.90 & 0.92 & 0.64 & 0.80 & 0.85 & 0.88 & 0.51 & 0.69 & 0.78 & 0.82 \\
 \hline
 50 & 0.70 & 0.84 & 0.94 & 0.94 & 0.94 & 0.91 & 0.93 & 0.94 & 0.94 & 0.92 & 0.91 & 0.93 & 0.94 & 0.93 & 0.91 & 0.93 & 0.94 & 0.93 & 0.90 & 0.94 & 0.95 \\
& 0.85 & 0.84 & 0.93 & 0.94 & 0.94 & 0.91 & 0.91 & 0.92 & 0.93 & 0.92 & 0.89 & 0.91 & 0.93 & 0.94 & 0.88 & 0.91 & 0.93 & 0.93 & 0.87 & 0.91 & 0.94 \\
& 0.95 & 0.88 & 0.93 & 0.93 & 0.94 & 0.93 & 0.90 & 0.91 & 0.93 & 0.93 & 0.87 & 0.90 & 0.93 & 0.93 & 0.87 & 0.90 & 0.94 & 0.92 & 0.86 & 0.91 & 0.94 \\
& 0.99 & 0.93 & 0.93 & 0.94 & 0.94 & 0.92 & 0.92 & 0.93 & 0.94 & 0.88 & 0.89 & 0.92 & 0.93 & 0.84 & 0.88 & 0.91 & 0.94 & 0.80 & 0.85 & 0.89 & 0.92 \\
 & 1.00 & 0.94 & 0.94 & 0.94 & 0.94 & 0.90 & 0.92 & 0.93 & 0.94 & 0.82 & 0.90 & 0.92 & 0.93 & 0.76 & 0.87 & 0.90 & 0.93 & 0.68 & 0.83 & 0.87 & 0.90 \\
 \hline
100 & 0.70 & 0.88 & 0.95 & 0.95 & 0.95 & 0.93 & 0.93 & 0.94 & 0.94 & 0.92 & 0.92 & 0.93 & 0.93 & 0.94 & 0.93 & 0.94 & 0.95 & 0.93 & 0.92 & 0.93 & 0.94 \\
 & 0.85 & 0.88 & 0.95 & 0.95 & 0.95 & 0.92 & 0.93 & 0.94 & 0.94 & 0.92 & 0.92 & 0.93 & 0.93 & 0.93 & 0.92 & 0.93 & 0.94 & 0.93 & 0.90 & 0.92 & 0.93 \\
 & 0.95 & 0.89 & 0.94 & 0.95 & 0.95 & 0.93 & 0.92 & 0.92 & 0.93 & 0.93 & 0.90 & 0.91 & 0.92 & 0.94 & 0.88 & 0.90 & 0.92 & 0.93 & 0.88 & 0.90 & 0.92 \\
 & 0.99 & 0.93 & 0.94 & 0.95 & 0.95 & 0.93 & 0.92 & 0.92 & 0.94 & 0.92 & 0.91 & 0.92 & 0.93 & 0.91 & 0.90 & 0.92 & 0.93 & 0.90 & 0.89 & 0.91 & 0.94 \\
 & 1.00 & 0.94 & 0.94 & 0.94 & 0.94 & 0.91 & 0.93 & 0.93 & 0.94 & 0.88 & 0.91 & 0.92 & 0.93 & 0.85 & 0.90 & 0.92 & 0.93 & 0.82 & 0.89 & 0.91 & 0.93 \\

\hline
\hline
    \end{tabular}%
    \caption*{Notes.  See Table \ref{tab::MC_N30_coverage}.}
    \label{tab::MC_N50_coverage}
  \end{sidewaystable}
 \begin{sidewaystable}[htbp!]
  \renewcommand{\arraystretch}{.9}
  \addtolength{\tabcolsep}{-2pt}
  \centering
 \footnotesize
  \caption{Monte Carlo coverage results for $\theta^{(h)}$, $h=0,\ldots,4$, and $N=100$.}
  \begin{tabular}{cc|cccc|cccc|cccc|cccc|cccc}
  \hline
  \hline
  &&\multicolumn{4}{c|}{$h=0$}&\multicolumn{4}{c|}{$h=1$}&\multicolumn{4}{c|}{$h=2$}&\multicolumn{4}{c|}{$h=3$}&\multicolumn{4}{c}{$h=4$}\\
  \hline
  $T$&$\alpha_{0}$&$B_{HJ}$&$t_{GK}$&$t_{MSS}$&$t_{BCA}$&$B_{HJ}$&$t_{GK}$&$t_{MSS}$&$t_{BCA}$&$B_{HJ}$&$t_{GK}$&$t_{MSS}$&$t_{BCA}$&$B_{HJ}$&$t_{GK}$&$t_{MSS}$&$t_{BCA}$&$B_{HJ}$&$t_{GK}$&$t_{MSS}$&$t_{BCA}$\\
   \hline
\hline
 30.00 & 0.70 & 0.65 & 0.91 & 0.93 & 0.92 & 0.89 & 0.88 & 0.91 & 0.92 & 0.92 & 0.86 & 0.91 & 0.93 & 0.93 & 0.86 & 0.92 & 0.94 & 0.93 & 0.86 & 0.92 & 0.94 \\
  & 0.85 & 0.70 & 0.89 & 0.91 & 0.90 & 0.90 & 0.83 & 0.87 & 0.89 & 0.93 & 0.81 & 0.87 & 0.90 & 0.94 & 0.82 & 0.89 & 0.92 & 0.93 & 0.83 & 0.91 & 0.94 \\
& 0.95 & 0.84 & 0.91 & 0.93 & 0.92 & 0.94 & 0.88 & 0.91 & 0.93 & 0.90 & 0.86 & 0.91 & 0.94 & 0.84 & 0.84 & 0.91 & 0.94 & 0.77 & 0.79 & 0.87 & 0.90 \\
 & 0.99 & 0.92 & 0.92 & 0.94 & 0.93 & 0.87 & 0.90 & 0.92 & 0.94 & 0.70 & 0.86 & 0.90 & 0.92 & 0.54 & 0.76 & 0.85 & 0.88 & 0.38 & 0.62 & 0.73 & 0.78 \\
& 1.00 & 0.93 & 0.92 & 0.94 & 0.93 & 0.80 & 0.90 & 0.92 & 0.93 & 0.58 & 0.85 & 0.89 & 0.91 & 0.40 & 0.73 & 0.81 & 0.86 & 0.24 & 0.58 & 0.68 & 0.74 \\
 \hline
 50 & 0.70 & 0.76 & 0.94 & 0.94 & 0.94 & 0.90 & 0.92 & 0.94 & 0.94 & 0.92 & 0.91 & 0.93 & 0.94 & 0.93 & 0.90 & 0.93 & 0.94 & 0.93 & 0.90 & 0.94 & 0.94 \\
 & 0.85 & 0.76 & 0.93 & 0.94 & 0.93 & 0.91 & 0.90 & 0.91 & 0.92 & 0.92 & 0.87 & 0.90 & 0.91 & 0.93 & 0.86 & 0.90 & 0.92 & 0.94 & 0.85 & 0.90 & 0.92 \\
& 0.95 & 0.83 & 0.92 & 0.93 & 0.93 & 0.93 & 0.88 & 0.90 & 0.92 & 0.94 & 0.85 & 0.89 & 0.92 & 0.93 & 0.85 & 0.89 & 0.93 & 0.92 & 0.85 & 0.90 & 0.94 \\
& 0.99 & 0.92 & 0.93 & 0.94 & 0.94 & 0.91 & 0.91 & 0.93 & 0.94 & 0.85 & 0.90 & 0.93 & 0.95 & 0.77 & 0.88 & 0.92 & 0.94 & 0.67 & 0.85 & 0.90 & 0.92 \\
 & 1.00 & 0.94 & 0.94 & 0.95 & 0.94 & 0.86 & 0.91 & 0.93 & 0.94 & 0.74 & 0.90 & 0.93 & 0.94 & 0.62 & 0.86 & 0.90 & 0.93 & 0.48 & 0.81 & 0.86 & 0.90 \\
  \hline
100 & 0.70 & 0.84 & 0.94 & 0.94 & 0.94 & 0.92 & 0.94 & 0.94 & 0.94 & 0.93 & 0.93 & 0.94 & 0.94 & 0.93 & 0.94 & 0.95 & 0.95 & 0.94 & 0.93 & 0.94 & 0.94 \\
& 0.85 & 0.84 & 0.94 & 0.94 & 0.94 & 0.91 & 0.93 & 0.94 & 0.94 & 0.93 & 0.92 & 0.94 & 0.94 & 0.94 & 0.93 & 0.94 & 0.94 & 0.94 & 0.91 & 0.93 & 0.94 \\
 & 0.95 & 0.85 & 0.93 & 0.94 & 0.94 & 0.92 & 0.90 & 0.91 & 0.92 & 0.94 & 0.88 & 0.90 & 0.91 & 0.94 & 0.87 & 0.89 & 0.91 & 0.94 & 0.85 & 0.88 & 0.91 \\
 & 0.99 & 0.92 & 0.94 & 0.94 & 0.94 & 0.93 & 0.92 & 0.93 & 0.94 & 0.92 & 0.91 & 0.92 & 0.94 & 0.90 & 0.90 & 0.92 & 0.94 & 0.88 & 0.89 & 0.92 & 0.94 \\
 & 1.00 & 0.94 & 0.94 & 0.94 & 0.94 & 0.91 & 0.93 & 0.94 & 0.94 & 0.84 & 0.92 & 0.93 & 0.94 & 0.78 & 0.91 & 0.92 & 0.94 & 0.71 & 0.90 & 0.92 & 0.93 \\

\hline
\hline
    \end{tabular}%
    \caption*{Notes.  See Table \ref{tab::MC_N30_coverage}.}
    \label{tab::MC_N100_coverage}
  \end{sidewaystable}

\section{Conclusions}
\label{section:conclusions}
In this paper, we analyze the bias properties of the fixed-effect estimator in panel local projections with many cross-sectional units and a long time-series dimension. We contribute to the growing literature on panel LPs by focusing on settings in which the outcome and control variables exhibit moderate to high persistence, as is common in empirical applications.

Our results show that, even in the stylized setting with observed shocks, the analytical bias-correction procedure of \citet{HerbstJohannsen2024} is not valid when the control variable is sufficiently persistent. We also show that half-panel jackknife bias correction of the FE estimator is not uniformly valid over the full parameter space considered. Nevertheless, outside the local-to-unity boundary, the HPJ procedure is applicable over a substantially broader range of persistence regimes than previously recognized in the panel LP literature.

For the coefficient on the shock of interest, the HPJ estimator has the same first-order asymptotic variance as the original FE estimator. However, finite-sample inference can differ substantially, especially when $T$ is small and persistence is high. In particular, the HPJ correction can affect the variability of the estimated coefficients on persistent controls, which remains relevant for inference on the coefficient of interest in finite samples. To address this issue, we propose \emph{bias-correction-adjusted} standard errors that explicitly account for the linear combination of full- and half-panel estimators underlying the HPJ procedure.

Our Monte Carlo results broadly support the theoretical predictions based on the asymptotic expansions developed in this paper. Although the HPJ procedure is not uniformly valid, coverage rates tend to move toward their nominal levels at shorter projection horizons and as $T$ increases. The quality of the asymptotic approximation can deteriorate when $N$ grows too quickly relative to $T$, particularly in highly persistent settings.

Finally, our theoretical analysis is specific to a stylized framework in which structural shocks are observed by the researcher.
Nevertheless, one might conjecture that the mechanisms identified in the paper also operate in more general panel LP settings. In particular, the HPJ procedure, combined with appropriately constructed standard errors, could provide a useful default approach in applications with multiple persistent control variables or population-residualized regressors, as in \citet{https://doi.org/10.3982/ECTA18756}. A formal analysis of these more general settings is left for future research.

\clearpage
\bibliography{SPJ_corrected.bib}

\newpage
\appendix
\section{Discussion of standard errors}
\label{appendix:inference}

In this appendix, we discuss the alternative standard-error estimators proposed by \citet{GalvaoKato2014} and \citet{mei2023nickellbiaspanellocal} for confidence intervals based on the HPJ estimator. We also explain why the bias-correction-adjusted procedure proposed in Section \ref{ssection:inference} is more generally applicable.

Under the conditions of \citet{GalvaoKato2014} and \citet{SPJ},

$
\lim_{T\to\infty}T Avar_{\sqrt{N}\vbeta_{HPJ}^{(h)}}
=
\lim_{T\to\infty}T Avar_{\sqrt{N}\vbeta_{FE}^{(h)}}.
$
Thus, the most direct variance estimator for the HPJ estimator sets
$
\widehat{Avar}_{\sqrt{N}\vbeta_{HPJ}^{(h)}}
=
\widehat{Avar}_{\sqrt{N}\vbeta_{FE}^{(h)}}.
$
However, as noted by \citet{SPJ}, this approach can be unreliable for small values of $T$, thereby motivating finite-sample adjustments.

As a remedy, \citet{GalvaoKato2014} propose the HPJ-CCM procedure, which replaces $\widehat{\vbeta}_{FE}^{(h)}$ by $\widehat{\vbeta}_{HPJ}^{(h)}$ when constructing residuals. That is,
\begin{equation}
\label{eq::hpj_ccm}
\widehat{\vs}_{i,T}^{(h)}
\left(
\widehat{\vbeta}_{HPJ}^{(h)}
\right)
=
\frac{1}{T-h}
\sum_{t=1}^{T-h}
\widetilde{\vx}_{i,t}
\left(
\widetilde{y}_{i,t+h}
-
\widetilde{\vx}_{i,t}'
\widehat{\vbeta}_{HPJ}^{(h)}
\right).
\end{equation}
The resulting variance estimator is
\begin{align*}
\widehat{Avar}_{\sqrt{N}\vbeta_{HPJ}^{(h)}}^{GK}
&=
\left(
\frac{1}{N}
\sum_{i=1}^{N}
\mB_{i,T}^{(h)}
\right)^{-1}
\left(
\frac{1}{N}
\sum_{i=1}^{N}
\widehat{\vs}_{i,T}^{(h)}
\left(
\widehat{\vbeta}_{HPJ}^{(h)}
\right)
\widehat{\vs}_{i,T}^{(h)}
\left(
\widehat{\vbeta}_{HPJ}^{(h)}
\right)'
\right)
\left(
\frac{1}{N}
\sum_{i=1}^{N}
\mB_{i,T}^{(h)}
\right)^{-1}.
\label{eq::avar_hpj_gk}
\end{align*}

Using the HPJ estimator to construct residuals may initially appear natural because the HPJ estimator generally has a smaller first-order bias than the FE estimator. However, using the HPJ estimator, or another bias-corrected estimator, in this construction can render the resulting variance estimator not valid uniformly over $\gamma$ for some elements of $\widehat{\vbeta}^{(h)}$.

To illustrate this point, consider the setting of Proposition \ref{proposition::HPJ_distribution_delta}. Following arguments similar to those used in the proof of that proposition, it can be shown that
\begin{align}
T^2
\widehat{Avar}_{\sqrt{N}\vdelta_{FE}^{(0)}}
&\pto
\frac{51}{5},
\quad
N,T\to\infty,
\\
T^2
\widehat{Avar}_{\sqrt{N}\vdelta_{HPJ}^{(0)}}^{GK}
&\pto
\frac{60}{5},
\quad
N,T\to\infty.
\end{align}
Thus, the HPJ-CCM estimator is consistent neither for the variance of the FE estimator nor for that of the HPJ estimator. By contrast, the standard CCM estimator, based on FE residuals, remains consistent for the variance of the FE estimator.

The failure of the HPJ-CCM approach in this setting can be understood by noting that, for any $\ddot{\vbeta}$, the estimated influence function $\widehat{\vs}_{i,T}^{(h)}(\ddot{\vbeta})$ can be written as
\begin{equation}
\label{eq::influence_est}
\begin{split}
\widehat{\vs}_{i,T}^{(h)}
\left(
\ddot{\vbeta}
\right)
={}&
\va_{i,T}^{(h)}
-
\E\left[
\va_{i,T}^{(h)}
\right]
-
\left(
\mB_{i,T}^{(h)}
-
\E\left[
\mB_{i,T}^{(h)}
\right]
\right)
\left(
\ddot{\vbeta}
-
\vbeta_{0}
\right)
\\
&+
\E\left[
\va_{i,T}^{(h)}
\right]
-
\E\left[
\mB_{i,T}^{(h)}
\right]
\left(
\ddot{\vbeta}
-
\vbeta_{0}
\right).
\end{split}
\end{equation}
Equivalently,
\begin{equation*}
\begin{split}
\widehat{\vs}_{i,T}^{(h)}
\left(
\ddot{\vbeta}
\right)
-
\vs_{i,T}^{(h)}
={}&
-
\left(
\mB_{i,T}^{(h)}
-
\E\left[
\mB_{i,T}^{(h)}
\right]
\right)
\left(
\ddot{\vbeta}
-
\vbeta_{0}
-
\E\left[
\mB_{i,T}^{(h)}
\right]^{-1}
\E\left[
\va_{i,T}^{(h)}
\right]
\right)
\\
&+
\E\left[
\va_{i,T}^{(h)}
\right]
-
\E\left[
\mB_{i,T}^{(h)}
\right]
\left(
\ddot{\vbeta}
-
\vbeta_{0}
\right).
\end{split}
\end{equation*}

When $\gamma=0$, only the first term,
$
\va_{i,T}^{(h)}
-
\E\left[
\va_{i,T}^{(h)}
\right],
$
contributes to the asymptotic variance of the FE estimator, because
$
\E\left[
\va_{i,T}^{(h)}
\right]
=
\largeO\left(
T^{-1}
\right)
$
and
$
\E\left[
\mB_{i,T}^{(h)}
\right]
=
\largeO(1).
$
Consequently, any estimator satisfying
$
\sqrt{T}
\left(
\ddot{\vbeta}
-
\vbeta_{0}
\right)
=
\smallO_{P}(1)
$
can be used to consistently estimate the asymptotic variance of both the FE and HPJ estimators.

This conclusion does not carry over to the setting of Proposition \ref{proposition::HPJ_distribution_delta}. When $\gamma=\infty$, all components in equation \eqref{eq::influence_est} are asymptotically nonnegligible. It is therefore essential that $\ddot{\vbeta}$ satisfy
\begin{equation*}
T
\left(
\ddot{\vbeta}
-
\vbeta_{0}
-
\E\left[
\mB_{i,T}^{(h)}
\right]^{-1}
\E\left[
\va_{i,T}^{(h)}
\right]
\right)
=
\smallO_{P}(1).
\end{equation*}
The biased FE estimator satisfies this requirement, whereas a bias-corrected estimator fails to do so by construction. Consequently, the HPJ-CCM approach not only fails to account for the potential variance contribution associated with $\mB_{i,T}^{(h)}$, but also retains the nonzero mean component $\E[\va_{i,T}^{(h)}]$.\footnote{If one instead used a centered version of the CCM estimator to eliminate this mean component, the HPJ-CCM procedure would substantially underestimate the sample variance: the corresponding scaled limit would equal $3$.}

Among the procedures proposed in the literature, the method of \citet{mei2023nickellbiaspanellocal} is closest to the bias-correction-adjusted standard errors proposed in this paper.\footnote{This method is implemented in the \textit{pLP} R package provided by the authors.} Their estimator also takes the form given in equation \eqref{eq::baa_ccm}, but defines $\widehat{\vq}_{i,T}^{(h)}$ as
\begin{align*}
\widehat{\vq}_{i,T}^{(h)}
&:=
\left(
\frac{1}{N}
\sum_{i=1}^{N}
\mB_{i,T}^{(h)}
\right)^{-1}
\left[
2
\widehat{\vs}_{i,T}^{(h)}
\left(
\widehat{\vbeta}_{HPJ}^{(h)}
\right)
\right.
\\
&\quad\quad\quad\left.
-
\frac{1}{2}
\left(
\widehat{\vs}_{i,T_{1}}^{(h)}
\left(
\widehat{\vbeta}_{HPJ}^{(h)}
\right)
+
\widehat{\vs}_{i,T_{2}}^{(h)}
\left(
\widehat{\vbeta}_{HPJ}^{(h)}
\right)
\right)
\right].
\end{align*}
Like the HPJ-CCM procedure, this estimator is not consistent for the variance of the HPJ estimator when $T$ is fixed. It is asymptotically justified for the full vector $\widehat{\vbeta}_{HPJ}^{(h)}$ in the stationary setting, but it is not generally valid across the persistence regimes considered here, as the exact unit-root example demonstrates. In particular, it may asymptotically either underestimate or overestimate the variance of $\widehat{\delta}_{HPJ}^{(h)}$.

\section{Proofs}
\label{appendix::proofs}
\subsection*{Notation}
Given the DGP in equation \eqref{eq::GDP1}, the model for $y_{i,t+h}$ is of the form
\begin{equation*}
y_{i,t+h}=\left(1-\alpha_{T}^{h+1}\right)\mu_{i}+\theta_{0}\alpha_{T}^{h}x_{i,t}+\alpha_{T}^{h+1}y_{i,t-1}+\left(\alpha_{T}^{h}\epsi_{i,t}+\sum_{q=0}^{h-1}\alpha_{T}^{q}(\epsi_{i,t+h-q}+\theta_{0}x_{i,t+h-q})\right).
\end{equation*}
Alternatively, using the notation $\nu_{i,t}:= \epsi_{i,t}+\theta_{0}x_{i,t}$, we can expand this as
\begin{equation*}
y_{i,t+h}=\left(1-\alpha_{T}^{h+1}\right)\mu_{i}+\theta_{0}\alpha_{T}^{h}x_{i,t}+\alpha_{T}^{h+1}y_{i,t-1}+\left(\alpha_{T}^{h}\epsi_{i,t}+\sum_{q=0}^{h-1}\alpha_{T}^{q}\nu_{i,t+h-q}\right).
\end{equation*}
For some derivations, an alternative expression (which treats $x_{i,t}$ as fully unobserved), 
\begin{equation*}
y_{i,t+h}=\left(1-\alpha_{T}^{h+1}\right)\mu_{i}+\alpha_{T}^{h+1}y_{i,t-1}+\sum_{q=0}^{h}\alpha_{T}^{q}\nu_{i,t+h-q},
\end{equation*}
is also useful. For $h=0$, this gives
\begin{align*}
y_{i,t}&=\left(1-\alpha_{T}\right)\mu_{i}+\alpha_{T}y_{i,t-1}+\nu_{i,t}\\
&=
    \mu_{i}+\sum_{q=0}^{t-1}\alpha_{T}^{q}\nu_{i,t-q}.
\end{align*}
We also follow the notational convention
\begin{equation*}
    \sum_{q=a}^{b}x_{q}=0,
\end{equation*}
for any $b<a$. If we define $\mB:= \E[\mB_{i,T}]$ and $\va^{(h)}:= \E[\va_{i,T}^{(h)}]$, then
\begin{equation*}
    b_{T,\theta}^{(h)}=\frac{1}{b_{11}b_{22}-b_{12}^{2}}(b_{22}a_{1}^{(h)}-b_{12}a_{2}^{(h)})=\left(\frac{a_{1}^{(h)}}{b_{11}}-\frac{b_{12}a_{2}^{(h)}}{b_{11}b_{22}}\right)\frac{1}{1-\frac{b_{12}^{2}}{b_{11}b_{22}}}.
\end{equation*}
Here, all the elements denoted by $b$ and $a$ are the corresponding elements of $\mB$ and $\va^{(h)}$, respectively. Similarly,
\begin{equation*}
    b_{T,\delta}^{(h)}=\frac{1}{b_{11}b_{22}-b_{12}^{2}}(b_{11}a_{2}^{(h)}-b_{12}a_{1}^{(h)})=\left(\frac{a_{2}^{(h)}}{b_{22}}-\frac{b_{12}a_{1}^{(h)}}{b_{11}b_{22}}\right)\frac{1}{1-\frac{b_{12}^{2}}{b_{11}b_{22}}}.
\end{equation*}
We will use this decomposition extensively in the bias expansions in the next section.

\section*{Appendix B.1.  Auxiliary lemmas for the bias analysis}
\begin{lemma}
\label{aux::lemma1}
    Let $\overline{u_{i}^{(h)}}:= \frac{1}{T}\sum_{t=1}^{T}u_{i,t}^{(h)}$. Then, under the DGP in equation \eqref{eq::GDP1},
    \begin{equation*}
        \overline{u_{i}^{(h)}}=\alpha_{T}^{h}\frac{1}{T}\sum_{t=1}^{T}\epsi_{i,t}+\frac{1}{T}\sum_{t=h+1}^{T+1}\sum_{q=0}^{h-1}\alpha_{T}^{q}\nu_{i,t}+\frac{1}{T}\sum_{t=2}^{h}\sum_{q=h-t+1}^{h-1}\alpha_{T}^{q}\nu_{i,t}+\frac{1}{T}\sum_{t=T+2}^{T+h}\sum_{q=0}^{T+h-t}\alpha_{T}^{q}\nu_{i,t}.
    \end{equation*}
\end{lemma}
\begin{lemma}
\label{aux::lemma2}
    Let $\overline{y_{i,-1}}:= \frac{1}{T}\sum_{t=1}^{T}y_{i,t-1}$. Then 
    \begin{equation*}
        \overline{y_{i,-1}}= \frac{1}{T}\sum_{t=1}^{T}y_{i,t-1}=\mu_{i}+\frac{1}{T}\sum_{t=1}^{T-1}\left(\sum_{q=0}^{T-1-t}\alpha_{T}^{q}\right)\nu_{i,t}.
    \end{equation*}
\end{lemma}
\begin{lemma}
\label{aux::lemma3}
    Under Assumption \ref{ass::rv}, we have
    \begin{equation*}        \E\left[\overline{u_{i}^{(h)}}\overline{x}_{i}\right]=\theta_{0}\sigma_{x}^{2}\left(\frac{T-h}{T^{2}}\sum_{q=0}^{h-1}\alpha_{T}^{q}+\frac{1}{T^{2}}\sum_{t=2}^{h}\sum_{q=h-t+1}^{h-1}\alpha_{T}^{q}\right).
    \end{equation*}
    In addition, it holds
        \begin{equation*}        
        \E\left[\overline{u_{i}^{(h)}}\overline{x}_{i}\right]=\frac{1}{T}\left(\theta_{0}\sigma_{x}^{2}\sum_{q=0}^{h-1}\alpha_{T}^{q}\right)+\largeO(T^{-2}),
    \end{equation*}
    for all $h=0,\ldots,H$ and all $\gamma\in[0,\infty]$.
\end{lemma}
\begin{lemma}
\label{aux::lemma4}
    Under Assumption \ref{ass::rv}, we have
    \begin{equation*}        \E\left[\overline{u_{i}^{(h)}}\overline{u_{i,-1}}\right]=\frac{\alpha_{T}^{h}\sigma_{\epsi}^{2}}{T^{2}}\sum_{t=1}^{T-1}\left(\sum_{q=0}^{T-1-t}\alpha_{T}^{q}\right)+\frac{\sigma_{\nu}^{2}}{T^{2}}\left(\sum_{q=0}^{h-1}\alpha_{T}^{q}\right)\sum_{t=h+1}^{T-1}\left(\sum_{q=0}^{T-1-t}\alpha_{T}^{q}\right)+\frac{\sigma_{\nu}^{2}}{T^{2}}\sum_{t=2}^{h}\sum_{q=h-t+1}^{h-1}\sum_{s=0}^{T-1-t}\alpha_{T}^{s+q}.
    \end{equation*}
\end{lemma}
\begin{lemma}
\label{aux::lemma5}
    Under Assumption \ref{ass::rv}, we have
    \begin{equation*}        
    \E\left[\overline{x}_{i}\overline{u_{i,-1}}\right]=\frac{\theta_{0}\sigma_{x}^{2}}{T^{2}}\sum_{t=0}^{T-2}\left(\sum_{q=0}^{t}\alpha_{T}^{q}\right).
     \end{equation*}      
\end{lemma}
\begin{lemma}
\label{aux::lemma6}
    Under Assumption \ref{ass::rv}, we have
    \begin{equation*}        
    \E\left[(\overline{y_{i,-1}}-\mu_{i})^{2}\right]=\frac{\sigma_{\nu}^{2}}{T^{2}}\sum_{t=0}^{T-2}\left(\sum_{q=0}^{t}\alpha_{T}^{q}\right)^{2}.
     \end{equation*}      
\end{lemma}
\begin{lemma}
\label{aux::lemma7}
    Under Assumption \ref{ass::rv}, we have
    \begin{equation*}        
    \frac{1}{T}\sum_{t=1}^{T}\E\left[\left(y_{i,t-1}-\mu_{i}\right)^{2}\right]=\frac{\sigma_{\nu}^{2}}{T}\sum_{t=0}^{T-2}\sum_{q=0}^{t}\alpha_{T}^{2q}.
     \end{equation*}      
\end{lemma}
\begin{lemma}
\label{aux::lemma8}
    Under Assumptions \ref{ass::drifting}--\ref{ass::rv}, we have
    \begin{equation*}
     \frac{1}{1-\frac{b_{12}^{2}}{b_{11}b_{22}}}=1+\largeO(T^{-1}),
    \end{equation*}
    for all $\gamma\in[0,\infty]$.
\end{lemma}

\begin{proof}[\bf Proof of Lemma \ref{aux::lemma1}]
By construction,
\begin{equation*}
u_{i,t}^{(h)}=\alpha_T u_{i,t}^{(h-1)}+\nu_{i,t+h},
\quad \nu_{i,t+h}:=\epsi_{i,t+h}+\theta_0 x_{i,t+h},
\end{equation*}
for $h\geq 1$ and $u_{i,t}^{(0)}=\epsi_{i,t}$ for $h=0$.
Hence, we obtain
\begin{equation*}
u_{i,t}^{(h)}=\alpha_T^h \epsi_{i,t}+\sum_{q=0}^{h-1}\alpha_T^q \nu_{i,t+h-q}.
\end{equation*}
Therefore,
\begin{align*}
\overline{u_i^{(h)}}
&=\frac{1}{T}\sum_{t=1}^T u_{i,t}^{(h)}\\
&=\alpha_T^h \frac{1}{T}\sum_{t=1}^T \epsi_{i,t}
+\frac{1}{T}\sum_{t=1}^T\sum_{q=0}^{h-1}\alpha_T^q \nu_{i,t+h-q}\\
&=I+II.
\end{align*}
Now set \(s=t+h-q\). For each fixed \(s\), the admissible values of \(q\) are
\begin{equation*}
\max\{0,h-s+1\}\le q \le \min\{h-1,T+h-s\},
\end{equation*}
with \(s\in\{2,\dots,T+h\}\). Hence
\begin{equation*}
II=
\frac{1}{T}\sum_{s=2}^{T+h}\sum_{q=\max\{0,h-s+1\}}^{\min\{h-1,T+h-s\}}
\alpha_T^q \nu_{i,s}.
\end{equation*}
Splitting the range of \(s\) into \(s=2,\dots,h\), \(s=h+1,\dots,T+1\), and \(s=T+2,\dots,T+h\) yields
\begin{equation*}
II
=
\frac{1}{T}\sum_{t=2}^{h}\sum_{q=h-t+1}^{h-1}\alpha_T^q \nu_{i,t}
+\frac{1}{T}\sum_{t=h+1}^{T+1}\sum_{q=0}^{h-1}\alpha_T^q \nu_{i,t}
+\frac{1}{T}\sum_{t=T+2}^{T+h}\sum_{q=0}^{T+h-t}\alpha_T^q \nu_{i,t}.
\end{equation*}
Substituting this expression back gives the desired result
\begin{equation*}
\overline{u_{i}^{(h)}}=\alpha_{T}^{h}\frac{1}{T}\sum_{t=1}^{T}\epsi_{i,t}+\frac{1}{T}\sum_{t=h+1}^{T+1}\sum_{q=0}^{h-1}\alpha_{T}^{q}\nu_{i,t}+\frac{1}{T}\sum_{t=2}^{h}\sum_{q=h-t+1}^{h-1}\alpha_{T}^{q}\nu_{i,t}+\frac{1}{T}\sum_{t=T+2}^{T+h}\sum_{q=0}^{T+h-t}\alpha_{T}^{q}\nu_{i,t}.
\end{equation*}
\end{proof}

\begin{proof}[\bf Proof of Lemma \ref{aux::lemma2}]
Since \(y_{i,t}=\mu_i+u_{i,t}\) and \(u_{i,0}=0\), we have
\begin{equation*}
\overline{y_{i,-1}}
:= \frac{1}{T}\sum_{t=1}^T y_{i,t-1}
= \mu_i + \frac{1}{T}\sum_{t=1}^T u_{i,t-1}
= \mu_i + \frac{1}{T}\sum_{t=0}^{T-1} u_{i,t}.
\end{equation*}
Because \(u_{i,0}=0\), this reduces to
\begin{equation*}
\overline{y_{i,-1}}=\mu_i+\frac{1}{T}\sum_{t=1}^{T-1}u_{i,t}.
\end{equation*}
By recursion,
\begin{equation*}
u_{i,t}=\sum_{q=0}^{t-1}\alpha_T^q \nu_{i,t-q},
\end{equation*}
so we obtain
\begin{equation*}
\frac{1}{T}\sum_{t=1}^{T-1}u_{i,t}
=\frac{1}{T}\sum_{t=1}^{T-1}\sum_{q=0}^{t-1}\alpha_T^q \nu_{i,t-q}.
\end{equation*}
Now re-index with \(s=t-q\). Equivalently, for each \(s\in\{1,\dots,T-1\}\), the admissible values of \(q\) are \(0,\dots,T-1-s\). Hence
\begin{equation*}
\frac{1}{T}\sum_{t=1}^{T-1}u_{i,t}
=\sum_{s=1}^{T-1}\left(\frac{1}{T}\sum_{q=0}^{T-1-s}\alpha_T^q\right)\nu_{i,s}.
\end{equation*}
Renaming \(s\) as \(t\) yields the desired result
\begin{equation*}
\overline{y_{i,-1}}
=\mu_{i}+\frac{1}{T}\sum_{t=1}^{T-1}\left(\sum_{q=0}^{T-1-t}\alpha_{T}^{q}\right)\nu_{i,t}.
\end{equation*}
\end{proof}

\begin{proof}[\bf Proof of Lemma \ref{aux::lemma3}]
From Lemma \ref{aux::lemma1},
\begin{equation*}
\overline{u_i^{(h)}}
=\alpha_T^h\frac{1}{T}\sum_{t=1}^T \epsi_{i,t}
+\frac{1}{T}\sum_{t=h+1}^{T+1}\sum_{q=0}^{h-1}\alpha_T^q \nu_{i,t}
+\frac{1}{T}\sum_{t=2}^{h}\sum_{q=h-t+1}^{h-1}\alpha_T^q \nu_{i,t}
+\frac{1}{T}\sum_{t=T+2}^{T+h}\sum_{q=0}^{T+h-t}\alpha_T^q \nu_{i,t}.
\end{equation*}
Multiplying by \(\overline{x}_i=T^{-1}\sum_{s=1}^T x_{i,s}\) and taking expectations, 
\begin{equation*}
\E\left[\overline{u_i^{(h)}}\overline{x}_i\right]=I+II+III+IV.
\end{equation*}
Here the first term $I=0$, because \(x_{i,s}\) is independent of \(\epsi_{i,t}\) and has mean zero. Similarly, $IV=0$, since it only contains \(\nu_{i,t}\) with \(t>T\), which are independent of \(\overline{x}_i\). Hence
\begin{align*}
\E\left[\overline{u_i^{(h)}}\overline{x}_i\right]
&=
\frac{1}{T^2}\E\left[
\left(
\sum_{t=h+1}^{T+1}\sum_{q=0}^{h-1}\alpha_T^q \nu_{i,t}
+\sum_{t=2}^{h}\sum_{q=h-t+1}^{h-1}\alpha_T^q \nu_{i,t}
\right)
\sum_{s=1}^T x_{i,s}
\right].
\end{align*}
By definition \(\nu_{i,t}=\epsi_{i,t}+\theta_0 x_{i,t}\). Since both $\nu_{i,t}$ and $x_{i,t}$ are independent over $t$, 
\begin{equation*}
\E[\nu_{i,t}x_{i,s}]
=
\theta_0\,\E[x_{i,t}x_{i,s}]
=
\theta_0\sigma_x^2\,\mathit{I}(t=s).
\end{equation*}
Therefore, only matching indices \(t=s\in\{1,\dots,T\}\) contribute.  Thus
\begin{equation*}
\E\left[\overline{u_i^{(h)}}\overline{x}_i\right]
=
\theta_{0}\sigma_{x}^{2}\left(\frac{T-h}{T^{2}}\sum_{q=0}^{h-1}\alpha_{T}^{q}
+\frac{1}{T^{2}}\sum_{t=2}^{h}\sum_{q=h-t+1}^{h-1}\alpha_{T}^{q}\right),
\end{equation*}
as desired. The final statement of the lemma follows directly from the fact that all summations involve only $h$ and are therefore finite.
\end{proof}

\begin{proof}[\bf Proof of Lemma \ref{aux::lemma4}]
From Lemmas \ref{aux::lemma1}--\ref{aux::lemma2},
\begin{equation*}
\overline{u_i^{(h)}}
=
\alpha_T^h \frac{1}{T}\sum_{t=1}^T \epsi_{i,t}
+\frac{1}{T}\sum_{t=h+1}^{T+1}\sum_{q=0}^{h-1}\alpha_T^q \nu_{i,t}
+\frac{1}{T}\sum_{t=2}^{h}\sum_{q=h-t+1}^{h-1}\alpha_T^q \nu_{i,t}
+\frac{1}{T}\sum_{t=T+2}^{T+h}\sum_{q=0}^{T+h-t}\alpha_T^q \nu_{i,t}
\end{equation*}
and
\begin{equation*}
\overline{u_{i,-1}}
=\frac{1}{T}\sum_{s=1}^{T-1}\left(\sum_{r=0}^{T-1-s}\alpha_T^r\right)\nu_{i,s}.
\end{equation*}
Thus,
\begin{equation*}
\E\left[\overline{u_{i}^{(h)}}\overline{u_{i,-1}}\right]=I+II+III+IV.
\end{equation*}
As in the proof of Lemma \ref{aux::lemma3}, it follows immediately that $IV=0$, whereas the remaining three components are nonzero.

Using the definition of \(\nu_{i,s}\), \(\E[\epsi_{i,s}x_{i,r}]=0\), and \(\E[\epsi_{i,s}\epsi_{i,r}]=\sigma_\epsi^2 \mathit{I}(s=r)\), we get
\begin{equation*}
I=\alpha_{T}^{h}\E\left[\left(\frac{1}{T}\sum_{t=1}^T \epsi_{i,t}\right) \frac{1}{T}\sum_{s=1}^{T-1}\left(\sum_{r=0}^{T-1-s}\alpha_T^r\right)\nu_{i,s}\right]
=
\alpha_{T}^{h}\frac{\sigma_\epsi^2}{T^2}\sum_{s=1}^{T-1}\left(\sum_{r=0}^{T-1-s}\alpha_T^r\right).
\end{equation*}
For $II$, only the terms with \(t=s\in\{h+1,\dots,T-1\}\) are nonzero, and
\begin{equation*}
\E[\nu_{i,t}\nu_{i,s}]=\sigma_\nu^2 \mathit{I}(t=s).
\end{equation*}
Hence
\begin{equation*}
II=\E\left[
\left(\frac{1}{T}\sum_{t=h+1}^{T+1}\sum_{q=0}^{h-1}\alpha_T^q \nu_{i,t}\right)
\frac{1}{T}\sum_{s=1}^{T-1}\left(\sum_{r=0}^{T-1-s}\alpha_T^r\right)\nu_{i,s}
\right]
=
\frac{\sigma_{\nu}^2}{T^2}
\left(\sum_{q=0}^{h-1}\alpha_T^q\right)
\sum_{s=h+1}^{T-1}\left(\sum_{r=0}^{T-1-s}\alpha_T^r\right).
\end{equation*}
Finally, for the remaining term, we obtain
\begin{equation*}
III=\E\left[\left(
\frac{1}{T}\sum_{t=2}^{h}\sum_{q=h-t+1}^{h-1}\alpha_T^q \nu_{i,t}\right)
\left(\frac{1}{T}\sum_{s=1}^{T-1}\left(\sum_{r=0}^{T-1-s}\alpha_T^r\right)\nu_{i,s}
\right)\right]
=
\frac{\sigma_{\nu}^2}{T^2}\sum_{t=2}^{h}\sum_{q=h-t+1}^{h-1}\sum_{r=0}^{T-1-t}\alpha_T^{q+r}.
\end{equation*}
Combining the three nonzero contributions yields the desired result
\begin{equation*}
\E\left[\overline{u_{i}^{(h)}}\overline{u_{i,-1}}\right]
=
\frac{\alpha_{T}^{h}\sigma_{\epsi}^{2}}{T^{2}}\sum_{t=1}^{T-1}\left(\sum_{q=0}^{T-1-t}\alpha_{T}^{q}\right)
+\frac{\sigma_{\nu}^{2}}{T^{2}}\left(\sum_{q=0}^{h-1}\alpha_{T}^{q}\right)\sum_{t=h+1}^{T-1}\left(\sum_{q=0}^{T-1-t}\alpha_{T}^{q}\right)
+\frac{\sigma_{\nu}^{2}}{T^{2}}\sum_{t=2}^{h}\sum_{q=h-t+1}^{h-1}\sum_{s=0}^{T-1-t}\alpha_{T}^{s+q}.
\end{equation*}
\end{proof}

\begin{proof}[\bf Proof of Lemma \ref{aux::lemma5}]
Using Lemma \ref{aux::lemma2},
\begin{equation*}
\overline{u_{i,-1}}=\frac{1}{T}\sum_{s=1}^{T-1}\left(\sum_{q=0}^{T-1-s}\alpha_T^q\right)\nu_{i,s},
\quad
\overline{x}_i=\frac{1}{T}\sum_{r=1}^T x_{i,r},
\end{equation*}
we have
\begin{equation*}
\E[\overline{x}_i\,\overline{u_{i,-1}}]
=
\frac{1}{T^2}\sum_{s=1}^{T-1}\sum_{q=0}^{T-1-s}\alpha_T^q
\sum_{r=1}^T \E[x_{i,r}\nu_{i,s}].
\end{equation*}
Following steps analogous to those
in the proofs of Lemmas \ref{aux::lemma3}--\ref{aux::lemma4}, we obtain
\begin{equation*}
\E[\overline{x}_i\,\overline{u_{i,-1}}]
=
\frac{\theta_0\sigma_x^2}{T^2}
\sum_{s=1}^{T-1}\sum_{q=0}^{T-1-s}\alpha_T^q.
\end{equation*}
Renaming the summation index \(s\) as \(t\), and using
\begin{equation*}
\sum_{s=1}^{T-1}\sum_{q=0}^{T-1-s}\alpha_T^q
=
\sum_{t=0}^{T-2}\sum_{q=0}^{t}\alpha_T^q,
\end{equation*}
we obtain the desired result
\begin{equation*}
\E\left[\overline{x}_{i}\overline{u_{i,-1}}\right]
=
\frac{\theta_{0}\sigma_{x}^{2}}{T^{2}}\sum_{t=0}^{T-2}\left(\sum_{q=0}^{t}\alpha_{T}^{q}\right).
\end{equation*}
\end{proof}

\begin{proof}[\bf Proof of Lemma \ref{aux::lemma6}]
From Lemma \ref{aux::lemma2},
\[
\overline{y_{i,-1}}-\mu_i
=
\frac{1}{T}\sum_{t=1}^{T-1}\left(\sum_{q=0}^{T-1-t}\alpha_T^q\right)\nu_{i,t}.
\]
Therefore,
\[
\E\left[(\overline{y_{i,-1}}-\mu_i)^2\right]
=
\frac{1}{T^2}\E\left[
\left(\sum_{t=1}^{T-1}\sum_{q=0}^{T-1-t}\alpha_T^q \nu_{i,t}\right)^2
\right].
\]
Since \(\nu_{i,t}=\epsi_{i,t}+\theta_0 x_{i,t}\) is iid over \(t\) with variance \(\sigma_{\nu}^2\), cross-terms vanish and only diagonal terms remain:
\[
\E\left[(\overline{y_{i,-1}}-\mu_i)^2\right]
=
\frac{\sigma_{\nu}^2}{T^2}\sum_{t=1}^{T-1}\left(\sum_{q=0}^{T-1-t}\alpha_T^q\right)^2.
\]
After reindexing,
\[
\E\left[(\overline{y_{i,-1}}-\mu_i)^2\right]
=
\frac{\sigma_{\nu}^2}{T^{2}}\sum_{t=0}^{T-2}\left(\sum_{q=0}^{t}\alpha_{T}^{q}\right)^{2}.
\]
\end{proof}

\begin{proof}[\bf Proof of Lemma \ref{aux::lemma7}]
From Lemma \ref{aux::lemma2},
\[
\overline{y_{i,-1}}-\mu_i
=
\frac{1}{T}\sum_{s=1}^{T-1}\left(\sum_{q=0}^{T-1-s}\alpha_T^q\right)\nu_{i,s},
\]
we get
\begin{align*}    
\frac{1}{T}\sum_{t=1}^{T}\E\left[\left(y_{i,t-1}-\mu_i\right)^2\right]
&=
\frac{1}{T}\sum_{t=1}^{T}\E\left[u_{i,t-1}^2\right]\\
&=
\frac{1}{T}\sum_{t=0}^{T-2}\E\left[\left(\sum_{q=0}^{t}\alpha_T^q \nu_{i,t+1-q}\right)^2\right]\\
&=\frac{\sigma_{\nu}^2}{T}\sum_{t=0}^{T-2}\sum_{q=0}^{t}\alpha_T^{2q}.
\end{align*}
Here, the final line follows from the fact that \(\nu_{i,t}\) is iid over \(t\) with variance \(\sigma_{\nu}^2\).
Therefore,
\[
\frac{1}{T}\sum_{t=1}^{T}\E\left[\left(y_{i,t-1}-\mu_i\right)^{2}\right]
=
\frac{\sigma_{\nu}^{2}}{T}\sum_{t=0}^{T-2}\sum_{q=0}^{t}\alpha_{T}^{2q}.
\]
\end{proof}

\begin{proof}[\bf Proof of Lemma \ref{aux::lemma8}]
Note that 
\begin{equation*}
y_{i,t-1}-\overline{y}_{i,-1}=(y_{i,t-1}-\mu_{i})-(\overline{y}_{i,-1}-\mu_{i})=u_{i,t-1}-\overline{u}_{i,-1}.
\end{equation*}
Thus, 
\begin{align*}
    b_{11}&=\frac{1}{T}\E\left[\sum_{t=1}^{T}(x_{i,t}-\overline{x}_{i})^{2}\right],\\
b_{22}&=\frac{1}{T}\E\left[\sum_{t=1}^{T}(y_{i,t-1}-\overline{y_{i,-1}})^{2}\right],\\
b_{12}&=-\E\left[\overline{x}_{i}\overline{u_{i,-1}}\right].
\end{align*}
Using the fact that $x_{i,t}$ is iid over $t$, we obtain
\begin{equation*}
    b_{11}=\frac{1}{T}\E\left[\sum_{t=1}^{T}(x_{i,t}-\overline{x}_{i})^{2}\right]=\frac{T-1}{T}\sigma_{x}^{2}=\sigma_{x}^{2}+\largeO(T^{-1}),
\end{equation*}
for all $\gamma\in[0,\infty]$. For $b_{12}$ and $b_{22}$, Lemmas \ref{aux::lemma5}--\ref{aux::lemma7} give
\begin{equation*}
 b_{12}= -\frac{\theta_{0}\sigma_{x}^{2}}{T^{2}}\sum_{t=0}^{T-2}\left(\sum_{q=0}^{t}\alpha_{T}^{q}\right)
\end{equation*}
and 
\begin{equation*}
b_{22}=\frac{\sigma_{\nu}^{2}}{T}\sum_{t=0}^{T-2}\sum_{q=0}^{t}\alpha_{T}^{2q}-\frac{\sigma_{\nu}^{2}}{T^{2}}\sum_{t=0}^{T-2}\left(\sum_{q=0}^{t}\alpha_{T}^{q}\right)^{2}.    
\end{equation*}
 We first derive results for three regimes of $\gamma$ separately, which are used in later proofs.
Then we summarize all results. First, note that
\begin{align*}
1-\alpha_{T}&=\frac{c}{T^{\gamma}}+\largeO(T^{-2\gamma}),\\
1-\alpha_{T}^{\psi}&=\frac{c\psi}{T^{\gamma}}+\largeO(T^{-2\gamma}),\\
(1-\alpha_{T})^{2}&=\frac{c^{2}}{T^{2\gamma}}+\largeO(T^{-3\gamma}), 
\end{align*}
for all $\gamma\in [0,\infty]$, for each fixed positive integer $\psi$.

\textit{Case III: $\gamma\in[0,1)$.}
Using the finite geometric-sum formula,
we obtain
\begin{align*}
\frac{1}{ \sigma_{x}^{2}}b_{12}&=-\frac{\theta_{0}}{T(1-\alpha_{T})}\left(\frac{T-1}{T}-\frac{\alpha_{T}}{T(1-\alpha_{T})}(1-\alpha_{T}^{T-1})\right)\\
&=-\frac{\theta_{0}}{T(1-\alpha_{T})}+\largeO(T^{-2+\gamma})+\largeO(T^{-2+2\gamma})\\
&=-\frac{\theta_{0}}{T(1-\alpha_{T})}+\largeO(T^{-2+2\gamma}).
\end{align*}
Analogously,
\begin{align*}
\frac{1}{\sigma_{\nu}^{2}}b_{22}&= \frac{1}{1-\alpha_{T}^{2}}\left(\frac{T-1}{T}-\frac{\alpha_{T}^{2}}{T(1-\alpha_{T}^{2})}\left(1-\alpha_{T}^{2(T-1)}\right)\right)\\
&-\frac{1}{T(1-\alpha_{T})^{2}}\left(\frac{T-1}{T}-2\frac{\alpha_{T}}{T(1-\alpha_{T})}\left(1-\alpha_{T}^{(T-1)}\right)+\frac{\alpha_{T}^{2}}{T(1-\alpha_{T}^{2})}\left(1-\alpha_{T}^{2(T-1)}\right)\right)\\
&=\frac{1}{1-\alpha_{T}^{2}}+\largeO(T^{-2+\gamma})+\largeO(T^{-1+2\gamma})+\largeO(T^{-2+2\gamma})+\largeO(T^{-2+3\gamma})\\
&=\frac{1}{1-\alpha_{T}^{2}}+\largeO(T^{-1+2\gamma})+\largeO(T^{-2+3\gamma})\\
&=\frac{1}{1-\alpha_{T}^{2}}+\largeO(T^{-1+2\gamma}).
\end{align*}
In particular,
\begin{equation}
\frac{b_{12}^{2}}{b_{11}b_{22}}=\frac{1}{T^{2}}\frac {(1+\alpha_{T})}{(1-\alpha_{T})}\frac{\theta_{0}^{2}\sigma_{x}^{2}}{\sigma_{\nu}^{2}}+\largeO(T^{-3+2\gamma})=\largeO(T^{-2+\gamma})
\end{equation}
for all $\gamma\in[0,1)$.

{\textit{Case II: $\gamma=1$.}}
For $\gamma=1$, we evaluate the corresponding expressions slightly differently. In particular,
\begin{align*}
\frac{1}{T}\frac{1}{\sigma_{\nu}^{2}}b_{22}&=\frac{1}{T^{2}}\sum_{t=0}^{T-2}\sum_{l=0}^{t}\alpha_{T}^{2l}-\frac{1}{T}\sum_{t=0}^{T-2}\left(\frac{1}{T}\sum_{l=0}^{t}\alpha_{T}^{l}\right)^{2}\\
&=\int_{0}^{1}\left(\int_{0}^{x}\exp(-2cy)dy\right)dx-\int_{0}^{1}\left(\int_{0}^{x}\exp(-cy)dy\right)^{2}dx+\largeO(T^{-1}).\\
&=\frac{1}{4c^{2}}\left(\exp(-2c)-1+2c\right)-\frac{1}{c^{2}}\int_{0}^{1}\left(\exp(-2cx)-2\exp(-cx)+1\right)dx+\largeO(T^{-1})\\
&=\frac{1}{4c^{2}}\left(\exp(-2c)-1+2c\right)-\frac{1}{c^{2}}\left(-\frac{1}{2c}(\exp(-2c)-1)+\frac{2}{c}(\exp(-c)-1)+1\right)+\largeO(T^{-1})\\
&=\frac{1}{4c^{2}}\left(\exp(-2c)-1+2c\right)-\frac{1}{2c^{3}}\left(-(\exp(-2c)-1)+4(\exp(-c)-1)+2c\right)+\largeO(T^{-1})\\
&=\frac{1}{4c^{2}}\left(\exp(-2c)-1+2c\right)-\frac{1}{2c^{3}}\left(4\exp(-c)-\exp(-2c)-3+2c\right)+\largeO(T^{-1}).
\end{align*}
Here the leading term is nonnegative for all $c>0$. Similarly,
\begin{align*}
\frac{1}{\sigma_{x}^{2}}b_{12}&=
-\theta_{0}\frac{1}{T^{2}}\sum_{t=0}^{T-2}\sum_{l=0}^{t}\alpha_{T}^{l}\\
&=-\theta_{0}\frac{1}{T^{2}}\sum_{t=0}^{T-2}(T-t-1)\alpha_{T}^{t}\\
&=-\theta_{0}\frac{1}{T}\sum_{t=0}^{T-2}\left(1-\frac{t+1}{T}\right)\exp\left(-c\frac{t}{T}\right)\\
&=-\theta_{0}\int_{0}^{1}(1-x)\exp\left(-c x\right)dx+\largeO(T^{-1})\\
&=-\theta_{0}\frac{1}{c^{2}}\left(\exp(-c)-1+c\right)+\largeO(T^{-1}).
\end{align*}
Combining these expressions gives
\begin{equation*}
\frac{b_{12}^{2}}{b_{11}b_{22}}=\largeO(T^{-1})=\smallO(1)
\end{equation*}
for $\gamma=1$.

{\textit{Case I: $\gamma\in(1,\infty]$.}}
Note that the results derived for $\gamma=1$ hold uniformly for $0 \leq c< C$, including drifting sequences $0\leq c_{T}<C$ with $c_{T}\to 0$. The expressions for $b_{12}$ and $b_{22}$ can alternatively be expressed using the series expansions
\begin{align}
\label{eq::appendixBcaseIII}
\frac{1}{T}\frac{1}{\sigma_{\nu}^{2}}b_{22}&=\sum_{k=0}^{\infty}\frac{(-2c)^{k}}{(k+2)!}+\sum_{k=0}^{\infty}\frac{1}{(k+3)!}\left(2(-c)^{k}-4(-2c)^{k}\right)+\largeO(T^{-1})
\end{align}
and 
\begin{align*}
\frac{1}{\sigma_{x}^{2}}b_{12}&=-\theta_{0}\sum_{k=0}^{\infty}\frac{(-c)^{k}}{(k+2)!}+\largeO(T^{-1}).
\end{align*}
In particular, upon setting $c_{T}:= c T^{1-\gamma}$, we obtain
\begin{align*}
\frac{1}{T}\frac{1}{\sigma_{\nu}^{2}}b_{22}&=\frac{1}{6}+\largeO(c_{T})+\largeO(T^{-1}),\\
\frac{1}{\sigma_{x}^{2}}b_{12}&=-\frac{1}{2}\theta_{0}+\largeO(c_{T})+\largeO(T^{-1}).
\end{align*}
Combining these expressions gives
\begin{equation*}
\frac{b_{12}^{2}}{b_{11}b_{22}}=\largeO(T^{-1})=\smallO(1).
\end{equation*}

{\textit{Summary of orders.}}
For future reference, we note that the orders of $b_{12}$ and $b_{22}$ can be compactly written as
\begin{align*}
b_{12}&=\largeO(T^{-1+\min(\gamma,1)}),
\end{align*}
while
\begin{align*}
\frac{b_{22}}{T^{\min(\gamma,1)}}&=C_{c,\gamma}+o(1),
\end{align*}
for $0<C_{c,\gamma}<\infty$ (where the constant is regime-specific).
Thus,
\begin{equation}
\label{eq::ref_orders_lemma8}
T\frac{b_{12}^{2}}{b_{11}b_{22}}=\largeO(T^{-1+\min(\gamma,1)}),
\end{equation}
for all $\gamma$. From here the final conclusion follows immediately.

\end{proof}

\section*{Appendix B.2.  Bias expansions of the FE estimator}

\begin{proof}[\bf Proof of Proposition \ref{lemma::expansionFE}]
Using Lemma \ref{aux::lemma8}, it is sufficient to focus on the terms $\frac{a_{1}^{(h)}}{b_{11}}$ and $\frac{b_{12}a_{2}^{(h)}}{b_{11}b_{22}}$ while establishing the corresponding rates. We first focus on $\frac{a_{1}^{(h)}}{b_{11}}$ as it is the term whose time-series properties are not affected by $\alpha_{T}$. In particular, note that
\begin{equation*}
    a_{1}^{(h)}=-\E\left[\overline{u_{i}^{(h)}}\overline{x}_{i}\right]=-\frac{1}{T}\left(\theta_{0}\sigma_{x}^{2}\sum_{q=0}^{h-1}\alpha_{T}^{q}\right)+\largeO(T^{-2}),
\end{equation*}
where the first equality follows from the fact that $\E[x_{i,t}u_{i,t}^{(h)}]=0$ for all $i$ and $t=1,\ldots,T$.

As for $b_{11}$, in Lemma \ref{aux::lemma8} we showed that 
\begin{equation*}
    b_{11}=\sigma_{x}^{2}+\largeO(T^{-1}).
\end{equation*}
Combining these two expressions, we conclude that
\begin{equation*}
\frac{a_{1}^{(h)}}{b_{11}}=-\frac{\theta_{0}}{T}\left(\sum_{q=0}^{h-1}\alpha_{T}^{q}\right)+\largeO(T^{-2}). 
\end{equation*}
It remains to investigate the properties of the remaining term. The quantities entering this term can be written as
\begin{align*}
b_{22}&=\frac{1}{T}\E\left[\sum_{t=1}^{T}(y_{i,t-1}-\overline{y_{i,-1}})^{2}\right]=\frac{1}{T}\E\left[\sum_{t=1}^{T}(u_{i,t-1}-\overline{u_{i,-1}})^{2}\right],\\
b_{12}&=-\E\left[\overline{x}_{i}\overline{u_{i,-1}}\right],\\
a_{2}^{(h)}&=-\E\left[\overline{u_{i}^{(h)}}\overline{u_{i,-1}}\right].
\end{align*}
Here, as in the derivation of $a_{1}^{(h)}$, we used the fact that $u_{i,t-1}$ is uncorrelated with $u_{i,t}^{(h)}$ when deriving $a_{2}^{(h)}$, while for $b_{12}$ we used the fact that $\E[x_{i,t}u_{i,t-1}]=0$  and $\E[u_{i,t}^{(h)}u_{i,t-1}]=0$ also holds under Assumption \ref{ass::rv}.

{\textit{Case III: $\gamma\in[0,1)$.}}
Using the derivations in Lemma \ref{aux::lemma8} we conclude that
\begin{align*}
    b_{22}&=\largeO(T^{\gamma}),\\
    b_{12}&=\largeO(T^{-1+\gamma}).
\end{align*}
Similarly, comparing the expressions in Lemmas \ref{aux::lemma4}--\ref{aux::lemma5}, we see that $a_{2}^{(h)}$ is of similar order to $b_{12}$, i.e.,
\begin{align*}
    a_{2}^{(h)}=\largeO(T^{-1+\gamma}).
\end{align*}
Combining these results gives
\begin{align*}
b_{T,\theta}^{(h)}&=-\frac{\theta_{0}}{T}\left(\sum_{q=0}^{h-1}\alpha_{T}^{q}\right)+\largeO(T^{-2})+\largeO(T^{-2+\gamma})\\
&=-\frac{\theta_{0}}{T}\left(\sum_{q=0}^{h-1}\alpha_{T}^{q}\right)+\largeO(T^{-2+\gamma}),
\end{align*}
for $\gamma\in[0,1)$. Hence, by the definition of $b_{\theta}^{(h,1)}$
\begin{equation*}
b_{\theta}^{(h,1)}=-\theta_{0}\left(\sum_{q=0}^{h-1}\alpha_{T}^{q}\right).
\end{equation*} 

{\textit{Case II: $\gamma=1$.}}
Note that the rates we derived for $b_{12},b_{22},a_{2}^{(h)}$ also continue to hold for $\gamma=1$ (as is evident from Lemma \ref{aux::lemma8}). As a result,
\begin{equation*}
    \frac{b_{12}a_{2}^{(h)}}{b_{11}b_{22}}=\largeO(T^{-1})
\end{equation*}
for $\gamma=1$, so this cannot be neglected asymptotically.  Using the same steps as for $b_{12}$ it follows immediately that
\begin{align*}
\frac{1}{\sigma_{\nu}^{2}}a_{2}^{(h)}=-\left(\frac{\sigma_{\epsi}^{2}}{\sigma_{\nu}^{2}}+h\right)\frac{1}{c^{2}}\left(\exp(-c)-1+c\right)+\largeO(T^{-1}),
\end{align*}
where we used the fact that $\lim_{T\to\infty}\alpha_{T}^{h}=1$. Recalling the definition of $\xi_{\theta}(\cdot)$
and combining this with 
\begin{equation*}
\frac{a_{1}^{(h)}}{b_{11}}=-\frac{\theta_{0}}{T}h+\largeO(T^{-2}),
\end{equation*}
we conclude that
\begin{align*}
 b_{T,\theta}^{(h)}&=-\frac{\theta_{0}}{T}\left(h+\xi_{\theta}(c)\left(\frac{\sigma_{\epsi}^{2}}{\sigma_{\nu}^{2}}+h\right)\right)+\largeO(T^{-2}).
\end{align*}
Thus, the leading bias term can be written as
\begin{equation*}
 b_{\theta}^{(h,1)}=-\theta_{0}\left(h+\xi_{\theta}(c)\left(\frac{\sigma_{\epsi}^{2}}{\sigma_{\nu}^{2}}+h\right)\right).
\end{equation*}

{\textit{Case I: $\gamma\in(1,\infty]$.}}
Finally, we consider the setting with $\gamma>1$. As in the proof of Lemma \ref{aux::lemma8} we note that the results derived for $\gamma=1$ hold uniformly for $0 \leq c< C$, including drifting sequences $0\leq c_{T}<C$ with $c_{T}\to 0$. In particular, upon setting $c_{T}:= c T^{1-\gamma}$, we note that
\begin{align*}
\frac{1}{T}\frac{1}{\sigma_{\nu}^{2}}b_{22}&=\frac{1}{6}+\largeO(c_{T})+\largeO(T^{-1}),\\
\frac{1}{\sigma_{\nu}^{2}}a_{2}^{(h)}&=-\frac{1}{2}\left(\frac{\sigma_{\epsi}^{2}}{\sigma_{\nu}^{2}}+h\right)+\largeO(c_{T})+\largeO(T^{-1}),\\
\frac{1}{\sigma_{x}^{2}}b_{12}&=-\frac{1}{2}\theta_{0}+\largeO(c_{T})+\largeO(T^{-1}).
\end{align*}
Here, we used the series expansion to write $\frac{1}{\sigma_{\nu}^{2}}a_{2}^{(h)}$ as
\begin{align*}
\frac{1}{\sigma_{\nu}^{2}}a_{2}^{(h)}=-\left(\frac{\sigma_{\epsi}^{2}}{\sigma_{\nu}^{2}}+h\right)\sum_{k=0}^{\infty}\frac{(-c_T)^{k}}{(k+2)!}+\largeO(T^{-1}).
\end{align*}
As a result
\begin{align*}
\frac{b_{12}a_{2}^{(h)}}{b_{11}b_{22}}&=\frac{3\theta_{0}}{T}\left(\frac{1}{2}\left(\frac{\sigma_{\epsi}^{2}}{\sigma_{\nu}^{2}}+h\right)\right)+\largeO(c_{T}T^{-1})+\largeO(T^{-2})\\
&=\frac{3\theta_{0}}{T}\left(\frac{1}{2}\left(\frac{\sigma_{\epsi}^{2}}{\sigma_{\nu}^{2}}+h\right)\right)+\largeO(T^{-(1+\min(\gamma-1,1))}).
\end{align*}
Combining this with
\begin{equation*}
\frac{a_{1}^{(h)}}{b_{11}}=-\frac{\theta_{0}}{T}h+\largeO(T^{-2}), 
\end{equation*}
we obtain the expansion
\begin{align*}
 b_{T,\theta}^{(h)}&=-\frac{\theta_{0}}{T}\left(h+\frac{3}{2}\left(\frac{\sigma_{\epsi}^{2}}{\sigma_{\nu}^{2}}+h\right)\right)+\largeO(T^{-(1+\min(\gamma-1,1))}).
\end{align*}
By the definition of $b_{\theta}^{(h,1)}$ this leads to
\begin{align*}
 b_{\theta}^{(h,1)}
 &=-\theta_{0}\left(h+\frac{3}{2}\left(\frac{\sigma_{\epsi}^{2}}{\sigma_{\nu}^{2}}+h\right)\right).
\end{align*}
This concludes the proof.
\end{proof}

\begin{proof}[\bf Proof of Proposition \ref{lemma::expansionFE_control}]
From the results in Proposition \ref{lemma::expansionFE}
and Lemma \ref{aux::lemma8} we know that 
\begin{equation*}
\frac{b_{12}a_{1}^{(h)}}{b_{11}b_{22}}=\largeO(T^{-2}),
\end{equation*}
for all $\gamma\in[0,\infty]$. Hence,
\begin{equation*}
    b_{T,\delta}^{(h)}=\frac{a_{2}^{(h)}}{b_{22}}+\largeO(T^{-2}),
\end{equation*}
for all $\gamma\in[0,\infty]$. The corresponding expression for $b_{\delta}^{(h,1)}$ for $\gamma=1$ and $\gamma\in(1,\infty]$ can be derived immediately using the quantities provided in Proposition \ref{lemma::expansionFE}. 

For $\gamma\in[0,1)$, Proposition \ref{lemma::expansionFE} is not informative, as only the rate of $a_{2}^{(h)}$ was provided in that proposition. From Lemma \ref{aux::lemma4} we know that
\begin{align*}
a_{2}^{(h)}&=-
\E\left[\overline{u_{i}^{(h)}}\overline{u_{i,-1}}\right]\\
&
=-\frac{\alpha_{T}^{h}\sigma_{\epsi}^{2}}{T^{2}}\sum_{t=1}^{T-1}\left(\sum_{q=0}^{T-1-t}\alpha_{T}^{q}\right)
-\frac{\sigma_{\nu}^{2}}{T^{2}}\left(\sum_{q=0}^{h-1}\alpha_{T}^{q}\right)\sum_{t=h+1}^{T-1}\left(\sum_{q=0}^{T-1-t}\alpha_{T}^{q}\right)
-\frac{\sigma_{\nu}^{2}}{T^{2}}\sum_{t=2}^{h}\sum_{q=h-t+1}^{h-1}\sum_{s=0}^{T-1-t}\alpha_{T}^{s+q}\\
&=I+II+III.
\end{align*}
The first term is
\begin{align*}
I&=-\frac{\alpha_{T}^{h}\sigma_{\epsilon}^{2}}{T^{2}}
\frac{T-1-T\alpha_T+\alpha_T^{T}}{(1-\alpha_T)^{2}} \\
&=-\frac{\alpha_{T}^{h}\sigma_{\epsilon}^{2}}{T}\frac{1}{1-\alpha_{T}}+\largeO(T^{-2+2\gamma}).
\end{align*}
The second term is
\begin{align*}
II&=-\left(\sum_{q=0}^{h-1}\alpha_{T}^{q}\right)\frac{\sigma_{\nu}^{2}}{T^{2}}
\frac{
\left(T-h-1-(T-h)\alpha_T+\alpha_T^{T-h}\right)}
{(1-\alpha_T)^{2}} \\
&=-\left(\sum_{q=0}^{h-1}\alpha_{T}^{q}\right)\frac{\sigma_{\nu}^{2}}{T^{2}}\frac{T-h}{1-\alpha_{T}}+\largeO(T^{-2+2\gamma})\\
&=-\left(\sum_{q=0}^{h-1}\alpha_{T}^{q}\right)\frac{\sigma_{\nu}^{2}}{T}\frac{1}{1-\alpha_{T}}+\largeO(T^{-2+\gamma})+\largeO(T^{-2+2\gamma})\\
&=-\left(\sum_{q=0}^{h-1}\alpha_{T}^{q}\right)\frac{\sigma_{\nu}^{2}}{T}\frac{1}{1-\alpha_{T}}+\largeO(T^{-2+2\gamma}).
\end{align*}
The third term is zero for $h=0,1$, while for $h>1$,
\begin{align*}
III&=-\frac{\sigma_{\nu}^{2}}{T^{2}(1-\alpha_T)^{2}}
\left[
\frac{\alpha_T(1-\alpha_T^{h-1})}{1-\alpha_T}
-(h-1)\alpha_T^{h}
-\frac{\alpha_T^{T-h+1}(1-\alpha_T^{2h-2})}{1-\alpha_T^{2}}
+\frac{\alpha_T^{T}(1-\alpha_T^{h-1})}{1-\alpha_T}
\right]\\
&=\largeO(T^{-2+2\gamma}).
\end{align*}
Combining all the terms gives
\begin{equation*}
a_{2}^{(h)}=-\frac{1}{T}\frac{1}{1-\alpha_{T}}\left(\alpha_{T}^{h}\sigma_{\epsi}^{2}+\sigma_{\nu}^{2}\left(\sum_{q=0}^{h-1}\alpha_{T}^{q}\right)\right)+\largeO(T^{-2+2\gamma}).    
\end{equation*}
From the proof of Lemma \ref{aux::lemma8},
\begin{align*}
\frac{1}{\sigma_{\nu}^{2}}b_{22}
&=\frac{1}{1-\alpha_{T}^{2}}+\largeO(T^{-1+2\gamma}).
\end{align*}
Thus,
\begin{align*}
    b_{T,\delta}^{(h)}=-\frac{1}{T}(1+\alpha_{T})\left(\alpha_{T}^{h}\frac{\sigma_{\epsi}^{2}}{\sigma_{\nu}^{2}}+\left(\sum_{q=0}^{h-1}\alpha_{T}^{q}\right)\right)+\largeO(T^{-2+\gamma}).
\end{align*}
This completes the proof.
\end{proof}

\section*{Appendix B.3.  Bias expansions of the HJ and HPJ estimators}

\begin{proof}[\bf Proof of Corollary \ref{lemma::expansionHJ}]
Consider the \citet{HerbstJohannsen2024} bias-corrected estimator
\begin{equation*}
\widehat{\theta}_{HJ}^{(h)}:=\widehat{\theta}^{(h)}+\frac{1}{T}\sum_{q=0}^{h-1}\widehat{\theta}^{(q)}.
\end{equation*}
As $\widehat{\theta}_{HJ}^{(h)}$ is a linear combination of FE estimators, we can make use of the results from Proposition \ref{lemma::expansionFE}.
In particular,
\begin{equation*}
b_{\theta_{HJ}}^{(h,1)}=b_{\theta}^{(h,1)}+\theta_{0}\sum_{q=0}^{h-1}\alpha_{T}^{q}.
\end{equation*}
Upon noting that $\lim_{T\to\infty}\sum_{q=0}^{h-1}\alpha_{T}^{q}=h$ for $\gamma>0$, the final result follows immediately for all values of $\gamma$ considered.
\end{proof}

\begin{proof}[\bf Proof of Corollary \ref{lemma::expansionHPJ}]
For a panel of length $m=T/2$ the same DGP parameter as in the full panel satisfies
$$
\alpha_T=e^{-cT^{-\gamma}}=e^{-(c2^{-\gamma})m^{-\gamma}}.
$$
Thus, to $c$ in the full panel corresponds the parameter $c2^{-\gamma}$ in the half-panels. It equals $c/2$ at $\gamma=1$, and remains at $c$ at $\gamma=0$.
The results for the first half-panel now follow directly from Proposition \ref{lemma::expansionFE} using panel length $T/2$ and persistence parameter $c2^{-\gamma}$, as in both cases one uses $u_{i,0}=0$ as the initial condition. As a result, we can immediately see that the following result holds for $\widehat{\theta}_{1,FE}^{(h)}$:
\begin{enumerate}[a)]
\setlength\itemsep{0.1em}
  \item For $\gamma\in (1,\infty]$: $b_{\theta_{1}}^{(h,1)}=-\theta_{0}\left(h+\frac{3}{2}\left(\frac{\sigma_{\epsi}^{2}}{\sigma_{\nu}^{2}}+h\right)\right),\quad R_{T}=T^{-(1+\min(\gamma-1,1))}$.
  \item For $\gamma=1$: $b_{\theta_{1}}^{(h,1)}=-\theta_{0}\left(h+\xi_{\theta}(c/2)\left(\frac{\sigma_{\epsi}^{2}}{\sigma_{\nu}^{2}}+h\right)\right),\quad R_{T}=T^{-2}$.
  \item For $\gamma\in [0,1)$: $b_{\theta_{1}}^{(h,1)}=-\theta_{0}\sum_{q=0}^{h-1}\alpha_{T}^{q},\quad R_{T}=T^{-2+\gamma}$.
\end{enumerate}

For the second half-panel, the situation is different as the implied initial condition, $u_{i,0}^{*}$, used in the second half-sample estimator is nonzero. As a result, the expansions derived for the full-sample estimator do not directly apply. Note that this initial condition is given by
\begin{equation*}
    u_{i,0}^{*}=\sum_{s=0}^{T_{1}-1}\alpha_{T}^{s}\nu_{i,T_{1}-s},
\end{equation*}
where $T_{1}=T/2$. As $u_{i,0}^{*}$ is uncorrelated with all $\nu_{i,t}$ for $t=T_{1}+1,\ldots,T$ it is easy to see that the results in Lemmas \ref{aux::lemma3}--\ref{aux::lemma5} continue to hold upon replacing $T$ with $T_{1}$. At the same time, this is not true for the corresponding results in Lemmas \ref{aux::lemma6}--\ref{aux::lemma7}. 

Let $\frac{1}{\sigma_{\nu}^{2}}b_{22}^{*}$ be the normalized denominator term of the second half-panel, and $\frac{1}{\sigma_{\nu}^{2}}b_{22}$ be the corresponding normalized denominator of the first half-panel. Then it is not difficult to see that
\begin{align*}
\frac{1}{\sigma_{\nu}^{2}}b_{22}^{*}&=\frac{1}{\sigma_{\nu}^{2}}b_{22}+\frac{\var{(u_{i,0}^{*})}}{\sigma_{\nu}^{2}}\left(\frac{1}{T_{1}}\sum_{t=0}^{T_{1}-1}\alpha_{T}^{2t}-\left(\frac{1}{T_{1}}\sum_{t=0}^{T_{1}-1}\alpha_{T}^{t}\right)^{2}\right)\\
&=\frac{1}{\sigma_{\nu}^{2}}b_{22}+\Delta B.
\end{align*}

{\textit{Case III: $\gamma\in[0,1)$.}}
Using steps analogous to those used in the proof of Lemma \ref{aux::lemma8}
\begin{align*}
\frac{1}{\sigma_{\nu}^{2}}b_{22}^{*}&=\largeO(T^{\gamma})+\largeO(T^{2\gamma-1})+\largeO(T^{3\gamma-2}).
\end{align*}
Thus, given the range of values considered of $\gamma$
\begin{align*}
\frac{1}{T^{\gamma}\sigma_{\nu}^{2}}b_{22}^{*}&=\frac{1}{T^{\gamma}\sigma_{\nu}^{2}}b_{22}+\smallO(1).
\end{align*}

{\textit{Case II: $\gamma=1$.}}
Let $\widetilde{c}:=c/2$. Consider the scaled version
\begin{align*}
\frac{1}{T_{1}\sigma_{\nu}^{2}}b_{22}^{*}&=\frac{1}{T_{1}\sigma_{\nu}^{2}}b_{22}+\frac{\var{(u_{i,0}^{*})}}{T_{1}\sigma_{\nu}^{2}}\left(\frac{1}{T_{1}}\sum_{t=0}^{T_{1}-1}\alpha_{T}^{2t}-\left(\frac{1}{T_{1}}\sum_{t=0}^{T_{1}-1}\alpha_{T}^{t}\right)^{2}\right).
\end{align*}
The expression for $\frac{1}{T_{1}\sigma_{\nu}^{2}}b_{22}$ is known from the derivations in Lemma \ref{aux::lemma8}. It remains to show that the second part is $\largeO(1)$. Let
\begin{align*}
\frac{1}{T_{1}}\Delta B &:=\left(-\frac{1}{2\widetilde{c}}(\exp(-2\widetilde{c})-1)\right)\left(\left(-\frac{1}{2\widetilde{c}}(\exp(-2\widetilde{c})-1)\right)-\left(-\frac{1}{\widetilde{c}}(\exp(-\widetilde{c})-1)\right)^{2}\right)+\largeO(T_{1}^{-1})\\
&=\psi(\widetilde{c})+\largeO(T_{1}^{-1}).
\end{align*}
Clearly, this term is nonnegative and $\largeO(1)$ for $0\leq c< C$. It is of the same order as the other term, so that
\begin{align*}
\frac{1}{T_{1}\sigma_{\nu}^{2}}b_{22}^{*}&=\frac{1}{T_{1}\sigma_{\nu}^{2}}b_{22}+\largeO(1).
\end{align*}
Hence, we can collect all terms in $\xi_{\theta}^{+}(\widetilde{c})$, which can be expressed as
\begin{align*}
\xi_{\theta}^{+}(\widetilde{c}):=
\xi_{\theta}(\widetilde{c})\left(\frac{\widetilde{c}\left(\exp(-2\widetilde{c})-1+2\widetilde{c}\right)-2\left(4\exp(-\widetilde{c})-\exp(-2\widetilde{c})-3+2\widetilde{c}\right)}{\widetilde{c}\left(\exp(-2\widetilde{c})-1+2\widetilde{c}\right)-2\left(4\exp(-\widetilde{c})-\exp(-2\widetilde{c})-3+2\widetilde{c}\right)+4\widetilde{c}^{3}\psi(\widetilde{c})}\right).
\end{align*}

{\textit{Case I: $\gamma\in(1,\infty]$.}}
Let $\widetilde{c}_{T}:={c}_{T}/2= \frac{c}{2} T^{1-\gamma}$ and consider $\psi(\widetilde{c}_T)$, which determines the leading terms in the series expressions for $\Delta B$:
\begin{align*}
\psi(\widetilde{c_T})&=\left(-\frac{1}{2\widetilde{c}_T}(\exp(-2\widetilde{c}_T)-1)\right)\left(\left(-\frac{1}{2\widetilde{c}_T}(\exp(-2\widetilde{c}_T)-1)\right)-\left(-\frac{1}{\widetilde{c}_T}(\exp(-\widetilde{c}_T)-1)\right)^{2}\right)\\
&=\left(\sum_{j=0}^{\infty}\frac{(-2\widetilde{c}_T)^{j}}{(j+1)!}\right)\left(\left(\sum_{j=0}^{\infty}\frac{(-2\widetilde{c}_T)^{j}}{(j+1)!}\right)-\left(\sum_{j=0}^{\infty}(2^{j+2}-2)\frac{(-\widetilde{c}_T)^{j}}{(j+2)!}\right)\right).
\end{align*}
Using the series expansions
\begin{align*}
\sum_{j=0}^{\infty}\frac{(-2\widetilde{c_{T}})^{j}}{(j+1)!}&=1+\largeO\left(c_{T}\right),\\
\sum_{j=0}^{\infty}(2^{j+2}-2)\frac{(-\widetilde{c_{T}})^{j}}{(j+2)!}&=1+\largeO\left(c_{T}\right),
\end{align*}
we obtain

\begin{equation*}
\psi(\widetilde{c}_T) = \largeO\left(T^{1-\gamma}\right).
\end{equation*}

As a result,
\begin{equation*}
\frac{1}{T^{\min(\gamma,1)}}\Delta B = \largeO\left(T^{1-\gamma}\right)+\largeO(T^{-1})=\smallO(1).
\end{equation*}

\textit{Summary.}
Combining all results, we see that for $\widehat{\theta}_{2,FE}^{(h)}$:
\begin{enumerate}[a)]
\setlength\itemsep{0.1em}
  \item For $\gamma\in (1,\infty]$: $b_{\theta_{2}}^{(h,1)}=-\theta_{0}\left(h+\frac{3}{2}\left(\frac{\sigma_{\epsi}^{2}}{\sigma_{\nu}^{2}}+h\right)\right),\quad R_{T}=T^{-(1+\min(\gamma-1,1))}$.
  \item For $\gamma=1$: $b_{\theta_{2}}^{(h,1)}=-\theta_{0}\left(h+\xi_{\theta}^{+}(0.5c)\left(\frac{\sigma_{\epsi}^{2}}{\sigma_{\nu}^{2}}+h\right)\right),\quad R_{T}=T^{-2}$.
  \item For $\gamma\in [0,1)$: $b_{\theta_{2}}^{(h,1)}=-\theta_{0}\sum_{q=0}^{h-1}\alpha_{T}^{q},\quad R_{T}=T^{-2+\gamma}$.
\end{enumerate}
The final result of this corollary follows immediately upon combining these results with those for $\widehat{\theta}_{1,FE}^{(h)}$ and $\widehat{\theta}_{FE}^{(h)}$.
\end{proof}

\section*{Appendix B.4. Auxiliary lemmas for the distributional analysis}
\begin{lemma}
\label{aux::lemma9}
    Under Assumptions \ref{ass::drifting}--\ref{ass::rv}, we have
    \begin{equation*}        
\frac{1}{NT}\sum_{i=1}^{N}\sum_{t=1}^{T}(x_{i,t}-\overline{x}_{i})^{2}=b_{11}+\largeO_{P}((NT)^{-1/2}).
     \end{equation*}      
\end{lemma}
\begin{lemma}
\label{aux::lemma10}
    Under Assumptions \ref{ass::drifting}--\ref{ass::rv}, we have
    \begin{equation*}        
\frac{1}{NT}\sum_{i=1}^{N}\sum_{t=1}^{T}(x_{i,t}-\overline{x}_{i})(y_{i,t-1}-\overline{y}_{i,-1})=b_{12}+\largeO_{P}((NT^{1-\min(\gamma,1)})^{-1/2}).
     \end{equation*}      
\end{lemma}
\begin{lemma}
\label{aux::lemma11}
    Under Assumptions \ref{ass::drifting}--\ref{ass::rv}, we have
    \begin{equation*}        
\frac{1}{NT}\sum_{i=1}^{N}\sum_{t=1}^{T}(y_{i,t-1}-\overline{y}_{i,-1})^{2}=b_{22}+\largeO_{P}(T^{\min(\gamma,1)}N^{-1/2}).
     \end{equation*}      
\end{lemma}
\begin{lemma}
\label{aux::lemma12}
    Under Assumptions \ref{ass::drifting}--\ref{ass::rv}, we have
    \begin{equation*}        
\frac{1}{NT}\sum_{i=1}^{N}\sum_{t=1}^{T}(y_{i,t-1}-\overline{y}_{i,-1})(u_{i,t}^{(h)}-\overline{u}^{(h)}_{i})=a_{2}^{(h)}+  \largeO_{P}((NT^{1-\min(\gamma,1)})^{-1/2}). 
     \end{equation*}      
\end{lemma}
\begin{lemma}
\label{aux::lemma13}
    Under Assumptions \ref{ass::drifting}--\ref{ass::rv}, we have
    \begin{equation*}        
\frac{1}{NT}\sum_{i=1}^{N}\sum_{t=1}^{T}(x_{i,t}-\overline{x}_{i})(u_{i,t}^{(h)}-\overline{u}^{(h)}_{i})=\frac{1}{NT}\sum_{i=1}^{N}\sum_{t=1}^{T}x_{i,t}u_{i,t}^{(h)}+a_{1}^{(h)}+\largeO_{P}(N^{-1/2}T^{-1}).
     \end{equation*}      
\end{lemma}
\begin{lemma}
\label{aux::lemma14}
    Under Assumptions \ref{ass::drifting}--\ref{ass::rv}, we have
\begin{equation*}
\frac{1}{\sqrt{NT}}\sum_{i=1}^{N}\sum_{t=1}^{T}x_{i,t}u_{i,t}^{(h)}\dto \calN\left(0,\sigma_{x}^{2}\lim_{T\to\infty}\left(\alpha_{T}^{2h}\sigma_{\epsi}^{2}+\sum_{q=0}^{h-1}\alpha_{T}^{2q}\sigma_{\nu}^{2}\right)\right),
\end{equation*}
as $N,T\to\infty$.       
\end{lemma}

\begin{proof}[\bf Proof of Lemma \ref{aux::lemma9}]
Given iid sampling of $x_{i,t}$ and $\E[|x_{i,t}|^{4}]<C<\infty$, it follows immediately that
\begin{equation*}
\frac{1}{NT}\sum_{i=1}^{N}\sum_{t=1}^{T}(x_{i,t}-\overline{x}_{i})^{2}=b_{11}+\largeO_{P}((NT)^{-1/2}),
\end{equation*}
where $b_{11}=\frac{T-1}{T}\sigma_{x}^{2}$.
\end{proof}

\begin{proof}[\bf Proof of Lemma \ref{aux::lemma10}]
Expand
\begin{align*}
    \frac{1}{NT}\sum_{i=1}^{N}\sum_{t=1}^{T}(x_{i,t}-\overline{x}_{i})(y_{i,t-1}-\overline{y}_{i,-1})&=\frac{1}{NT}\sum_{i=1}^{N}\sum_{t=1}^{T}(x_{i,t}-\overline{x}_{i})(u_{i,t-1}-\overline{u}_{i,-1})\\
    &=b_{12}+\frac{1}{NT}\sum_{i=1}^{N}\sum_{t=1}^{T}x_{i,t}u_{i,t-1}-\frac{1}{N}\sum_{i=1}^{N}(\overline{x}_{i}\overline{u}_{i,-1}+b_{12})\\
    &=I+II+III.
\end{align*}
Note that by iid sampling over $i$ and iid sampling of $x_{i,t}$ it follows that
\begin{align*}
\var(II)&=\E[(II)^{2}]\\
&=\E\left[\left(\frac{1}{NT}\sum_{i=1}^{N}\sum_{t=1}^{T}x_{i,t}u_{i,t-1}\right)^{2}\right]\\
&=\frac{1}{N^{2}}\sum_{i=1}^{N}\E\left[\left(\frac{1}{T}\sum_{t=1}^{T}x_{i,t}u_{i,t-1}\right)^{2}\right]\\
&=\frac{1}{(NT)^{2}}\sum_{i=1}^{N}\E\left[\sum_{t=1}^{T}x_{i,t}^{2}u_{i,t-1}^{2}\right]\\
&=\frac{\sigma_{x}^{2}}{NT^{2}}\E\left[\sum_{t=1}^{T}u_{i,t-1}^{2}\right]\\
&=\frac{\sigma_{x}^{2}}{NT}\largeO(T^{\min(\gamma,1)}).
\end{align*}
Here the final line uses the order derived in the proof of Lemma \ref{aux::lemma8}. As a result
\begin{equation*}
    II=\largeO_{P}((NT^{1-\min(\gamma,1)})^{-1/2}).
\end{equation*}
For the remaining term, define $d_{t}:=\sum_{q=0}^{T-1-t}\alpha_{T}^{q}$. Then
\begin{align*}
 \var(III)&= \frac{1}{N^{2}}\sum_{i=1}^{N}\E[(\overline{x}_{i}\overline{u}_{i,-1})^{2}]-\frac{1}{N}b_{12}^{2}\\
 &=\frac{1}{N}\E[(\overline{x}_{i}\overline{u}_{i,-1})^{2}]-\frac{1}{N}b_{12}^{2}\\
 &=\frac{1}{N T^{4}}\sum_{t=1}^{T}\sum_{s=1}^{T}\sum_{q=1}^{T-1}\sum_{k=1}^{T-1}d_{q}d_{k}\E[x_{i,t}x_{i,s}\nu_{i,q}\nu_{i,k}]-\frac{1}{N}b_{12}^{2}\\
  &=\frac{\sigma_{x}^{2}\sigma_{\epsi}^{2}}{N T^{3}}\left(\sum_{t=1}^{T-1}d_{t}^{2}\right)+\frac{\theta_{0}^{2}}{N T^{4}}\left(\sum_{t=1}^{T-1}d_{t}^{2}\right)\left(\E[x_{i,t}^{4}]+\sigma_{x}^{4}(T-3)\right)+\frac{1}{N}b_{12}^{2}\\
    &=\frac{\sigma_{x}^{2}\sigma_{\epsi}^{2}}{N T^{3}}\left(\sum_{t=1}^{T-1}d_{t}^{2}\right)+\frac{\theta_{0}^{2}}{N T^{4}}\left(\sum_{t=1}^{T-1}d_{t}^{2}\right)\left(\E[x_{i,t}^{4}]+\sigma_{x}^{4}(T-3)\right)+\frac{1}{N}b_{12}^{2}.
\end{align*}
From the proof of Lemma \ref{aux::lemma8} we know that
\begin{align*}
    \frac{1}{T^{3}}\sum_{t=1}^{T-1}d_{t}^{2}=\frac{1}{T^{3}}\sum_{t=0}^{T-2}\left(\sum_{q=0}^{t}\alpha_{T}^{q}\right)^{2}=\largeO(T^{-2+2\min(\gamma,1)}).
\end{align*}
Similarly, $b_{12}=\largeO(T^{-1+\min(\gamma,1)})$. As a result,
\begin{equation*}
    III=\largeO_{P}(N^{-1/2}T^{-1+\min(\gamma,1)}).
\end{equation*}
Combining all the results gives
\begin{equation*}
 \frac{1}{NT}\sum_{i=1}^{N}\sum_{t=1}^{T}(x_{i,t}-\overline{x}_{i})(y_{i,t-1}-\overline{y}_{i,-1})=b_{12}+  \largeO_{P}((NT^{1-\min(\gamma,1)})^{-1/2}). 
\end{equation*}
\end{proof}

\begin{proof}[\bf Proof of Lemma \ref{aux::lemma11}]
Using the familiar decomposition
    \begin{align*}        
\frac{1}{NT}\sum_{i=1}^{N}\sum_{t=1}^{T}(y_{i,t-1}-\overline{y}_{i,-1})^{2}&=\frac{1}{NT}\sum_{i=1}^{N}\sum_{t=1}^{T}(u_{i,t-1}-\overline{u}_{i,-1})^{2}\\
&=b_{22}+\frac{1}{N}\sum_{i=1}^{N}\left(\frac{1}{T}\sum_{t=1}^{T}(u_{i,t-1}-\overline{u}_{i,-1})^{2}-b_{22}\right)
\\
&=I+II.
     \end{align*}
By iid sampling over $i$ it follows that
\begin{align*}
\var(II)&=\E\left[\left(\frac{1}{N}\sum_{i=1}^{N}\left(\frac{1}{T}\sum_{t=1}^{T}(u_{i,t-1}-\overline{u}_{i,-1})^{2}-b_{22}\right)\right)^{2}\right]\\
&=\frac{1}{N}\E\left[\left(\frac{1}{T}\sum_{t=1}^{T}(u_{i,t-1}-\overline{u}_{i,-1})^{2}-b_{22}\right)^{2}\right]\\
&\leq\frac{1}{N}\E\left[\left(\frac{1}{T}\sum_{t=1}^{T}(u_{i,t-1}-\overline{u}_{i,-1})^{2}\right)^{2}\right]\\
&\leq\frac{1}{N}\E\left[\left(\frac{1}{T}\sum_{t=1}^{T}u_{i,t-1}^{2}\right)^{2}\right]\\
&\leq\frac{1}{N}\E\left[\left(\frac{1}{T}\sum_{t=1}^{T}u_{i,t-1}^{4}\right)\right].
\end{align*}
Since the errors \(\nu_{i,t}\) are iid over \(t\) with $\E[|\nu_{i,t}|^{4}]<C<\infty$, the AR(1) structure of the model implies that
\begin{align*}
\E[u_{i,t}^4]
&=3\sigma_\nu^4\left(\sum_{q=0}^{t-1}\alpha_T^{2q}\right)^2
+\zeta_4\sum_{q=0}^{t-1}\alpha_T^{4q}\\
&\leq C\left(\sum_{q=0}^{t-1}\alpha_T^{2q}\right)^2\\
&\leq C\left(\sum_{q=0}^{T}\alpha_T^{2q}\right)^2\\
&=\largeO(T^{2\min(\gamma,1)}),
\end{align*}
for $0<t\leq T$, where $\zeta_4:=\E[\nu_{i,t}^4]-3\sigma_\nu^4$ and $C$ is a constant satisfying $0<C<\infty$. From here 
\begin{eqnarray*}
\var(II)&=\largeO(N^{-1}T^{2\min(\gamma,1)}).
\end{eqnarray*}
Thus,
\begin{align*}
    \frac{1}{N}\sum_{i=1}^{N}\left(\frac{1}{T}\sum_{t=1}^{T}(y_{i,t-1}-\overline{y}_{i,-1})^{2}\right)=b_{22}+\largeO_{P}(T^{\min(\gamma,1)}N^{-1/2}).
\end{align*}
This completes the proof.
\end{proof}

\begin{proof}[\bf Proof of Lemma \ref{aux::lemma12}]
The proof is analogous to the proof of Lemma \ref{aux::lemma10} upon noticing that 
\begin{equation*}
u_{i,t}^{(h)}=\alpha_T^h \epsi_{i,t}+\sum_{q=0}^{h-1}\alpha_T^q \nu_{i,t+h-q}
\end{equation*}
and the fact that $h$ is finite.
\end{proof}

\begin{proof}[\bf Proof of Lemma \ref{aux::lemma13}]
Expand
\begin{align*}
    \frac{1}{NT}\sum_{i=1}^{N}\sum_{t=1}^{T}(x_{i,t}-\overline{x}_{i})(u_{i,t}^{(h)}-\overline{u}_{i}^{(h)})&=\frac{1}{NT}\sum_{i=1}^{N}\sum_{t=1}^{T}x_{i,t}u_{i,t}^{(h)}+a_{1}^{(h)}-\frac{1}{N}\sum_{i=1}^{N}(\overline{x}_{i}\overline{u}_{i}^{(h)}+a_{1}^{(h)})\\
    &=I+II+III.
\end{align*}
Upon noticing that 
\begin{equation*}
u_{i,t}^{(h)}=\alpha_T^h \epsi_{i,t}+\sum_{q=0}^{h-1}\alpha_T^q \nu_{i,t+h-q}
\end{equation*}
and the fact that $h$ is finite,
\begin{align*}
III&=-\alpha_T^h\frac{1}{N}\sum_{i=1}^{N}(\overline{x}_{i}\overline{\epsi}_{i}-\E[\overline{x}_{i}\overline{\epsi}_{i}])-\sum_{q=0}^{h-1}\alpha_T^q\left(\frac{1}{N}\sum_{i=1}^{N}(\overline{x}_{i}\overline{\nu}_{i,h-q}-\E[\overline{x}_{i}\overline{\nu}_{i,h-q}])\right)\\
&=III'+III''.
\end{align*}
Given that $h$ is finite, all terms of $III$ are of the same order. In particular,
\begin{align*}
\var(III')&=\frac{\alpha_T^{2h}}{N}
\E\left[(\overline{x}_{i}\overline{\epsi}_{i})^{2}\right]-\frac{\alpha_T^{2h}}{N}(\E[\overline{x}_{i}\overline{\epsi}_{i}])^{2}\\
&=\frac{\alpha_T^{2h}}{NT^{2}}\sigma_{x}^{2}\sigma_{\epsi}^{2}\\
&=\largeO(N^{-1}T^{-2}).
\end{align*}
Thus, $III'=\largeO_{P}(N^{-1/2}T^{-1})$.
The orders of the remaining terms can be established analogously. To see this, take, for example, $h=1$. Then
\begin{align*}
III''&=-\left(\frac{1}{N}\sum_{i=1}^{N}(\overline{x}_{i}\overline{\nu}_{i,1}-\E[\overline{x}_{i}\overline{\nu}_{i,1}])\right)\\
&=-\left(\frac{1}{N}\sum_{i=1}^{N}(\overline{x}_{i}\overline{\epsi}_{i,1}-\E[\overline{x}_{i}\overline{\epsi}_{i,1}])\right)-\theta_{0}\left(\frac{1}{N}\sum_{i=1}^{N}(\overline{x}_{i}\overline{x}_{i,1}-\E[\overline{x}_{i}\overline{x}_{i,1}])\right)\\
&=III_{1}''+III_{2}''.
\end{align*}
Given the derivations for $III'$, it follows immediately that $III_{1}''=\largeO_{P}(N^{-1/2}T^{-1})$. For the other term, we have
\begin{align*}
\var(III_{2}'')&=\frac{\theta_{0}^{2}}{N}
\E\left[(\overline{x}_{i}\overline{x}_{i,1})^{2}\right]-\frac{\theta_{0}^{2}}{N}(\E[\overline{x}_{i}\overline{x}_{i,1}])^{2}\\
&=\frac{\theta_{0}^{2}}{NT^{4}}\left((T-1)\E[x_{i,t}^{4}]+
(3(T-1)(T-2)+2T-1)
\sigma_{x}^{4}\right)-\frac{\theta_{0}^{2}}{N}\left(\frac{T-1}{T^{2}}\right)^{2}\sigma_{x}^{4}\\
&=\largeO(N^{-1}T^{-2}).
\end{align*}
Thus, $III_{2}''=\largeO_{P}(N^{-1/2}T^{-1})$. Hence, overall
\begin{equation*}
III=\largeO_{P}(N^{-1/2}T^{-1}).
\end{equation*}
\end{proof}

\begin{proof}[\bf Proof of Lemma \ref{aux::lemma14}]
Observe that
\begin{equation*}
\frac{1}{\sqrt{NT}}\sum_{i=1}^{N}\sum_{t=1}^{T}x_{i,t}u_{i,t}^{(h)}=\frac{1}{\sqrt{N}}\sum_{i=1}^{N}Z_{i,T}^{(h)},
\end{equation*}
where 
\begin{equation*}
Z_{i,T}^{(h)}:= \frac{1}{\sqrt{T}}\sum_{t=1}^{T}x_{i,t}u_{i,t}^{(h)}.
\end{equation*}
Let $\xi_{i,t}^{(h)}:=x_{i,t}u_{i,t}^{(h)}$ and define the sigma-field 
$\calG_{i,t}:=\sigma(x_{i,s},\epsi_{i,s}:s\ge t)$.
Then $\xi_{i,t}^{(h)}$ is $\calG_{i,t}$-measurable and $\E(\xi_{i,t}^{(h)}|\calG_{i,t+1})=0$, hence it is a reverse martingale difference sequence (MDS). In particular
\begin{equation*}
\E\left[\left(Z_{i,T}^{(h)}\right)^{2}\right]=\frac{1}{T}\sum_{t=1}^{T}\E\left[\left(\xi_{i,t}^{(h)}\right)^{2}\right]=   \sigma_{x}^{2}\left(\alpha_{T}^{2h}\sigma_{\epsi}^{2}+\sum_{q=0}^{h-1}\alpha_{T}^{2q}\sigma_{\nu}^{2}\right).
\end{equation*}
Moreover, by the Burkholder's inequality (that also applies to reverse MDS) for any $\delta>0$
\begin{equation*}
\E\left[\left|Z_{i,T}^{(h)}\right|^{2+\delta}\right]\leq C_{2+\delta} \frac{1}{T}\sum_{t=1}^{T}\E\left[\left|\xi_{i,t}^{(h)}\right|^{2+\delta}\right]=C_{2+\delta}\E\left[\left|\xi_{i,t}^{(h)}\right|^{2+\delta}\right],
\end{equation*}
for some finite constant $C_{2+\delta}>0$, as long as $\E\left[\left|\xi_{i,t}^{(h)}\right|^{2+\delta}\right]<C<\infty$ exists. Here the final equality follows from the fact that $(\xi_{i,t}^{(h)}:t=1,\ldots,T)$ are identically distributed over $t$. Take $\delta=2$ then
\begin{equation*}
\E\left[\left(\xi_{i,t}^{(h)}\right)^{4}\right]=\E[x_{i,t}^{4}]\E[(u_{i,t}^{(h)})^{4}].
\end{equation*}
Here the first term is finite by Assumption \ref{ass::rv}, while by the generalized Minkowski's inequality
\begin{equation*}
\left(\E[(u_{i,t}^{(h)})^{4}]\right)^{1/4}\leq \left(\sum_{q=0}^{h}\alpha_{T}^{q}\right)\left(\E[(\epsi_{i,t})^{4}]\right)^{1/4}+|\theta_{0}|\left(\sum_{q=0}^{h-1}\alpha_{T}^{q}\right)\left(\E[(x_{i,t})^{4}]\right)^{1/4}, 
\end{equation*}
that is finite by Assumptions \ref{ass::drifting}-\ref{ass::rv} for any finite $h$. Hence $\E[(u_{i,t}^{(h)})^{4}]<C<\infty$, and as a result 
\begin{equation*}
\sup_{T}\E\left[\left|Z_{i,T}^{(h)}\right|^{2+\delta}\right]<C<\infty,
\end{equation*}
for $\delta\in (0,2]$. Hence, given that $(Z_{i,T}^{(h)}:i=1,\ldots,N)$ are iid over $i$ the Lyapunov's condition is satisfied immediately. Using Theorem 2 of \citet{phillipsMoon1999}
\begin{equation*}
\frac{1}{\sqrt{NT}}\sum_{i=1}^{N}\sum_{t=1}^{T}x_{i,t}u_{i,t}^{(h)}\dto \calN\left(0,\sigma_{x}^{2}\lim_{T\to\infty}\left(\alpha_{T}^{2h}\sigma_{\epsi}^{2}+\sum_{q=0}^{h-1}\alpha_{T}^{2q}\sigma_{\nu}^{2}\right)\right),
\end{equation*}
as $N,T\to\infty$.   


\end{proof}

\section*{Appendix B.5. Asymptotic distribution of $\widehat{\theta}_{FE}^{(h)}$}

\begin{proof}[\bf Proof of Proposition \ref{proposition::FE_distribution_theta}]
By the Frisch--Waugh--Lovell theorem,
\begin{align*}
\widehat{\theta}_{FE}^{(h)}=&\theta_{0}^{(h)}+\left(\left(\frac{1}{N}\sum_{i=1}^{N}\mB_{i,T (1,1)}\right)-\left(\frac{1}{N}\sum_{i=1}^{N}\mB_{i,T (1,2)}\right)^{2}\left(\frac{1}{N}\sum_{i=1}^{N}\mB_{i,T (2,2)}\right)^{-1}\right)^{-1}\\
&\times\left(\left(\frac{1}{N}\sum_{i=1}^{N}\va_{i,T (1)}^{(h)}\right)-\left(\frac{1}{N}\sum_{i=1}^{N}\mB_{i,T (1,2)}\right)\left(\frac{1}{N}\sum_{i=1}^{N}\mB_{i,T (2,2)}\right)^{-1}\left(\frac{1}{N}\sum_{i=1}^{N}\va_{i,T (2)}^{(h)}\right)\right)\\
=&\theta_{0}^{(h)}+D_{N,T}^{-1} n_{N,T}^{(h)}.
\end{align*}
Alternatively, using the desired convergence rate $\sqrt{NT}$,
\begin{equation*}
\sqrt{NT}\left(\widehat{\theta}_{FE}^{(h)}-\theta_{0}^{(h)}\right)= \left(D_{N,T}\right)^{-1}\left(\sqrt{NT}n_{N,T}^{(h)}\right).
\end{equation*}

{\textit{Denominator.}}
We first investigate the denominator, $D_{N,T}$. Using Lemmas \ref{aux::lemma9}--\ref{aux::lemma13}, we show that, as $N,T\to\infty$,
\begin{equation*}
D_{N,T}=b_{11}+\largeO_{P}(r_{NT})
\end{equation*}
for $b_{11}=\frac{T-1}{T}\sigma_{x}^{2}>0$ and some $r_{NT}$. Expanding the denominator gives
\begin{align*}
D_{N,T}&=b_{11}+\left(\frac{1}{N}\sum_{i=1}^{N}\mB_{i,T (1,1)}-b_{11}\right)-\left(\frac{1}{N}\sum_{i=1}^{N}\mB_{i,T (1,2)}\right)^{2}\left(\frac{1}{N}\sum_{i=1}^{N}\mB_{i,T (2,2)}\right)^{-1}\\
&=I+II+III.
\end{align*}
For $II$ we make use of Lemma \ref{aux::lemma9}:
\begin{equation*}
    II=\frac{1}{N}\sum_{i=1}^{N}\mB_{i,T (1,1)}-b_{11}=\largeO_{P}((NT)^{-1/2}).
\end{equation*}
After rearranging, $III$ is of the form
\begin{align*}
III=-    \left(\sqrt{T^{-\min(\gamma,1)}}\frac{1}{N}\sum_{i=1}^{N}\mB_{i,T (1,2)}\right)^{2}\left(\frac{1}{T^{\min(\gamma,1)}}\frac{1}{N}\sum_{i=1}^{N}\mB_{i,T (2,2)}\right)^{-1}.
\end{align*}
Next, using Lemma \ref{aux::lemma10}, we have
\begin{align*}
\frac{\sqrt{T^{-\min(\gamma,1)}}}{N}\sum_{i=1}^{N}\mB_{i,T (1,2)}
&=\sqrt{T^{-\min(\gamma,1)}}b_{12}
+\largeO_{P}((NT)^{-1/2}).
\end{align*}
Lemma \ref{aux::lemma11}, on the other hand, implies that
    \begin{equation*}        
\frac{1}{N{T^{\min(\gamma,1)}}}\sum_{i=1}^{N}\mB_{i,T (2,2)}=\frac{b_{22}}{T^{\min(\gamma,1)}}+\largeO_{P}(N^{-1/2}).
     \end{equation*}      
Using the result from Lemma \ref{aux::lemma8} that states
\begin{equation*}
\frac{b_{22}}{T^{\min(\gamma,1)}}=C_{c,\gamma}+o(1)+\largeO_{P}(N^{-1/2}),
\end{equation*}
as $T\to\infty$ for $0<C_{c,\gamma}<\infty$ (where the constant is regime-specific), we conclude that
\begin{align*}
III=&-T^{-\min(\gamma,1)}b_{12}^{2}\left(\frac{1}{T^{\min(\gamma,1)}}b_{22}+\largeO_{P}(N^{-1/2})\right)^{-1}+\largeO_{P}(N^{-1/2}T^{-3/2+\min(\gamma/2;1/2)}) \\
&+\largeO_{P}(N^{-1}T^{-1})\\
=&-\frac{b_{12}^{2}}{b_{22}}-T^{-\min(\gamma,1)}b_{12}^{2}\left(\left(\frac{1}{T^{\min(\gamma,1)}}b_{22}+\largeO_{P}(N^{-1/2})\right)^{-1}-\left(\frac{1}{T^{\min(\gamma,1)}}b_{22}\right)^{-1}\right)\\
&+\largeO_{P}(N^{-1/2}T^{-3/2+\min(\gamma/2;1/2)})+\largeO_{P}(N^{-1}T^{-1})\\
=&-\frac{b_{12}^{2}}{b_{22}}+\largeO_{P}(N^{-1/2}T^{-2+\min(\gamma,1)})+\largeO_{P}(N^{-1/2}T^{-3/2+\min(\gamma/2;1/2)})+\largeO_{P}(N^{-1}T^{-1})\\
=&-\frac{b_{12}^{2}}{b_{22}}+\largeO_{P}(N^{-1/2}T^{-1})\\
=&\largeO_{P}(T^{-2+\min(\gamma,1)})+\largeO_{P}(N^{-1/2}T^{-1}).
\end{align*}
In the last line, we used equation \eqref{eq::ref_orders_lemma8} from the proof of Lemma \ref{aux::lemma8}.
Combining these expressions gives
\begin{equation*}
D_{N,T}=b_{11}+\largeO_{P}((NT)^{-1/2})+\largeO_{P}(T^{-2+\min(\gamma,1)}).
\end{equation*}
Thus, by the continuous mapping theorem (CMT),
\begin{equation}
\label{eq::denominator_distribution}
D_{N,T}^{-1}=b_{11}^{-1}+\largeO_{P}((NT)^{-1/2})+\largeO_{P}(T^{-2+\min(\gamma,1)}).
\end{equation}

{\textit{Numerator.}} For the numerator,
\begin{align*}
\sqrt{NT}n_{N,T}^{(h)}&=\left(\frac{\sqrt{NT}}{N}\sum_{i=1}^{N}\va_{i,T (1)}^{(h)}\right)-\sqrt{NT}\left(\frac{1}{N}\sum_{i=1}^{N}\mB_{i,T (1,2)}\right)\left(\frac{1}{N}\sum_{i=1}^{N}\mB_{i,T (2,2)}\right)^{-1}\left(\frac{1}{N}\sum_{i=1}^{N}\va_{i,T (2)}^{(h)}\right)\\
&=I'+II'.
\end{align*}
Using Lemma \ref{aux::lemma13},
\begin{equation*}
I'=\frac{1}{\sqrt{NT}}\sum_{i=1}^{N}\sum_{t=1}^{T}x_{i,t}u_{i,t}^{(h)}+\sqrt{NT}a_{1}^{(h)}+\largeO_{P}(T^{-1/2}).
\end{equation*}
As for the second term, after rearranging we get
\begin{align*}
II'=-\sqrt{NT}    \left(\sqrt{T^{-\min(\gamma,1)}}\frac{1}{N}\sum_{i=1}^{N}\mB_{i,T (1,2)}\right)\left(\frac{1}{T^{\min(\gamma,1)}}\frac{1}{N}\sum_{i=1}^{N}\mB_{i,T (2,2)}\right)^{-1}\left(\sqrt{T^{-\min(\gamma,1)}}\frac{1}{N}\sum_{i=1}^{N}\va_{i,T (2)}^{(h)}\right).
\end{align*}
Using Lemma \ref{aux::lemma12},
\begin{align*}
\frac{\sqrt{T^{-\min(\gamma,1)}}}{N}\sum_{i=1}^{N}\va_{i,T (2)}
&=\sqrt{T^{-\min(\gamma,1)}}a_{2}^{(h)}
+\largeO_{P}((NT)^{-1/2}).
\end{align*}
Analogous steps to those used to derive the term III of the denominator give
\begin{align*}
II'&=-\sqrt{NT}\frac{b_{12}a_{2}^{(h)}}{b_{22}}+\largeO_{P}(\sqrt{T}T^{-2+\min(\gamma,1)})+\largeO_{P}(T^{-1+\min(\gamma/2;1/2)})  +\largeO_{P}((NT)^{-1/2}),\\
&=-\sqrt{NT}\frac{b_{12}a_{2}^{(h)}}{b_{22}}+\largeO_{P}(T^{-1/2}).
\end{align*}
As a result,
\begin{align*}
\sqrt{NT}n_{N,T}^{(h)}&=\frac{1}{\sqrt{NT}}\sum_{i=1}^{N}\sum_{t=1}^{T}x_{i,t}u_{i,t}^{(h)}+\sqrt{NT}\left(a_{1}^{(h)}-\frac{b_{12}a_{2}^{(h)}}{b_{22}}\right)+\largeO_{P}(T^{-1/2}),\\
&=\frac{1}{\sqrt{NT}}\sum_{i=1}^{N}\sum_{t=1}^{T}x_{i,t}u_{i,t}^{(h)}+\sqrt{NT}\frac{b_{11}}{T}b_{\theta}^{(h,1)}+\largeO(\sqrt{NT}R_{T})+\largeO_{P}(T^{-1/2}).
\end{align*}
Here the remainder rate $R_{T}$ follows from Proposition \ref{lemma::expansionFE}. Thus, under Assumption \ref{ass::expansion},
\begin{equation*}
\sqrt{NT}\largeO(R_{T})=\smallO(1), 
\end{equation*}
for 
\begin{equation*}
\kappa
>
\left\{
\begin{array}{lr}
        3^{-1}, & \text{for } \gamma\in\{1\}\cup[2,\infty],\\
        (3-2\gamma)^{-1}, & \text{for } \gamma\in[0,1),\\
        (2\gamma-1)^{-1}, & \text{for } \gamma\in(1,2).
\end{array}
\right.
\end{equation*}
As a result, the numerator simplifies to
\begin{equation}
\label{eq::proof_numerator}
\sqrt{NT}n_{N,T}^{(h)}
=\frac{1}{\sqrt{NT}}\sum_{i=1}^{N}\sum_{t=1}^{T}x_{i,t}u_{i,t}^{(h)}+\sqrt{NT}\frac{b_{11}}{T}b_{\theta}^{(h,1)}+\smallO_{P}(1).
\end{equation}
From Lemma \ref{aux::lemma14}, $\frac{1}{\sqrt{NT}}\sum_{i=1}^{N}\sum_{t=1}^{T}x_{i,t}u_{i,t}^{(h)}=\largeO_{P}(1)$. Hence, we can make use of the result for the denominator term in equation \eqref{eq::denominator_distribution} to conclude that
\begin{align*}
\sqrt{NT}\left(\widehat{\theta}_{FE}^{(h)}-\theta_{0}^{(h)}-\frac{1}{T}b_{\theta}^{(h,1)}\right)=&\frac{1}{b_{11}}\frac{1}{\sqrt{NT}}\sum_{i=1}^{N}\sum_{t=1}^{T}x_{i,t}u_{i,t}^{(h)}+\largeO_{P}((NT)^{-1/2})+\largeO_{P}(T^{-2+\min(\gamma,1)})\\
&+\largeO_{P}(T^{-1})+\largeO_{P}(N^{1/2}T^{-5/2+\min(\gamma,1)})+\smallO_{P}(1)\\
=&\frac{1}{b_{11}}\frac{1}{\sqrt{NT}}\sum_{i=1}^{N}\sum_{t=1}^{T}x_{i,t}u_{i,t}^{(h)}+\smallO_{P}(1)\\
=&\frac{1}{\sigma_{x}^{2}}\frac{1}{\sqrt{NT}}\sum_{i=1}^{N}\sum_{t=1}^{T}x_{i,t}u_{i,t}^{(h)}+\smallO_{P}(1).
\end{align*}
The final line follows by the CMT upon noting that $b_{11}=\sigma_{x}^{2}+\largeO(T^{-1})$. The final conclusion follows using the result of Lemma \ref{aux::lemma14} and the CMT.
\end{proof}

\begin{proof}[\bf Proof of Proposition \ref{proposition::HPJ_distribution_theta}]
This proof follows analogous steps to the proof of Proposition \ref{proposition::FE_distribution_theta}. In particular, the results in Lemmas \ref{aux::lemma9}--\ref{aux::lemma13} can be modified accordingly to also apply to both half-panels. For the second half, its initial condition $u_{i,T_{1}}$ is independent of subsequent innovations and
satisfies $\E[u_{i,T_{1}}^{2}]=\largeO(T^{\min(\gamma,1)})$ and $\E[u_{i,T_{1}}^{4}]=\largeO(T^{2\min(\gamma,1)})$. These bounds preserve the orders used in
Lemmas \ref{aux::lemma9}--\ref{aux::lemma13}.

Using the linearity of the HPJ estimator, it follows immediately that
\begin{align*}
\sqrt{NT}\left(\widehat{\theta}_{HPJ}^{(h)}-\theta_{0}^{(h)}-\frac{1}{T}b_{\theta_{HPJ}}^{(h,1)}\mathit{I}(\gamma=1)\right)=&\frac{1}{b_{11}}\frac{1}{\sqrt{NT}}\left(2\sum_{i=1}^{N}\sum_{t=1}^{T}x_{i,t}u_{i,t}^{(h)}\right)\\
&-\frac{1}{b_{11}}\frac{1}{\sqrt{NT}}\left(\sum_{i=1}^{N}\sum_{t=1}^{T_{1}}x_{i,t}u_{i,t}^{(h)}+\sum_{i=1}^{N}\sum_{t=T_{1}+1}^{T}x_{i,t}u_{i,t}^{(h)}\right)+\smallO_{P}(1)\\
=&\frac{1}{b_{11}}\frac{1}{\sqrt{NT}}\left(\sum_{i=1}^{N}\sum_{t=1}^{T}x_{i,t}u_{i,t}^{(h)}\right)+\smallO_{P}(1).
\end{align*}
The main result now follows using results from the proof of Proposition \ref{proposition::FE_distribution_theta}.
\end{proof}

\section*{Appendix B.6. Asymptotic distribution of $\widehat{\delta}_{HPJ}^{(0)}$}
This section is concerned with the proof of Proposition \ref{proposition::HPJ_distribution_delta}. When $\theta_{0}=0$ is known and $\alpha_{T}=1$, the model reduces to
\begin{equation*}
y_{i,t}=y_{i,t-1}+\epsi_{i,t},
\end{equation*}
Proposition \ref{proposition::HPJ_distribution_delta} is closely related to Theorem 4 in \citet{Hahn2002}. As a result, we use a proof strategy that closely resembles theirs. For simplicity, we set $\widehat{\delta}_{FE}^{(0)}=\widehat{\alpha}_{FE}$, and similarly for the corresponding half-panel estimators.

In the unit-root case, all three estimators can be expanded as follows:
\begin{align*}
\widehat{\alpha }_{FE}   &=1+\frac{\sum_{i}A_{iT}}{\sum_{i}B_{iT}},\\
\widehat{\alpha }_{1,FE} &=1+\frac{\sum_{i}A_{iT,1}}{\sum_{i}B_{iT,1}},\\
\widehat{\alpha }_{2,FE} &=1+\frac{\sum_{i}A_{iT,2}}{\sum_{i}B_{iT,2}}.
\end{align*}
Here
\begin{align*}
A_{iT}&=\sum_{t=1}^{T}\epsi_{i,t}u_{i,t-1}-T\bar{\epsi}_{i}\bar{u}_{i,-},\\
B_{iT}&=\sum_{t=1}^{T}u_{i,t-1}^{2}-T\bar{u}_{i,-}^{2},
\end{align*}
and similarly for the half-sample analogues. 

Let
\begin{equation*}
\mA_{T}:=\left(
\begin{array}{ccccc}
0 & 0 & 0 & \cdots & 0 \\
1 & 0 & 0 & \cdots & 0 \\
1 & 1 & 0 & \cdots & 0 \\
\vdots & \vdots & \ddots & \ddots & \vdots \\
1 & 1 & \cdots & 1 & 0%
\end{array}%
\right),
\end{equation*}
and
\begin{equation*}
\mH_{T}:=\mI_{T}-\frac{1}{T}\vones _{T}\vones _{T}^{\prime
},\quad \mH_{T}\vy_{i,-}=\mH_{T}\mA_{T}\vepsi_{i} =\mD_{T}\vepsi_{i} ,\quad
\mD_{T}=\mH_{T}\mA_{T}.
\end{equation*}

\begin{lemma}
\label{lemma::aux_jackknife}
Under the specialized model of Proposition \ref{proposition::HPJ_distribution_delta}, the following holds as $T\to\infty$:
\begin{align*}
\covar(A_{iT},A_{iT,1}) &=\covar(A_{iT},A_{iT,2})=\frac{1}{96}T^{2}+\smallO(T^{2}), \\
\covar(A_{iT},B_{iT,1}) &=\covar(B_{iT},A_{iT,2})=-\frac{1}{192}T^{3}+\smallO(T^{3}), \\
\covar(A_{iT},B_{iT,2}) &=\covar(B_{iT},A_{iT,1})=\frac{1}{192}T^{3}+\smallO(T^{3}), \\
\covar(B_{iT},B_{iT,1}) &=\covar(B_{iT},B_{iT,2})=\frac{1}{360}T^{4}+\smallO(T^{4}).
\end{align*}
\end{lemma}

\begin{proof}[\bf Proof of Lemma \ref{lemma::aux_jackknife}]
For this lemma, we use
the following covariance identity for quadratic forms in a zero-mean Gaussian vector with
identity covariance matrix
\begin{align*}
\covar(\vepsi_{i} ^{\prime }\mG\vepsi_{i} ,\vepsi_{i} ^{\prime
}\mH\vepsi_{i} ) =
\tr\left(\mG(\mH+\mH^{\prime })\right)=\tr\left((\mG+\mG^{\prime })\mH\right).
\end{align*}
It remains to express all the quantities in terms of the quadratic forms as specified above. The proof uses similar steps to those in \citet{Hahn2002}. The full-sample FE estimator can be written as follows
\begin{equation}
\label{eq::proof_FE_decomposition}
\widehat{\alpha }=1+\frac{\sum_{i}\vepsi_{i} ^{\prime }\mH_{T}\vy_{i,-}}{%
\sum_{i}\vy_{i,-}^{\prime }\mH_{T}\vy_{i,-}}=1+\frac{\sum_{i}\vepsi_{i} ^{\prime
}\mD_{T}\vepsi_{i} }{\sum_{i}\vepsi_{i} ^{\prime }\mD_{T}^{\prime
}\mD_{T}\vepsi_{i} }=1+\frac{\sum_{i}A_{iT}}{\sum_{i}B_{iT}}.
\end{equation}
Here we make use of the fact that $u_{i,0}=0$ for all $i$, and recall that $T_{1}:=T/2$.

We now want to derive a decomposition similar to equation \eqref{eq::proof_FE_decomposition} for the two half-sample estimators. Consider the following partitions:
\begin{align*}
\vy_{i,-} &=\left(
\begin{array}{c}
\vy_{i,1-} \\
\vy_{i,2-}%
\end{array}%
\right) ,\quad \vepsi_{i} =\left(
\begin{array}{c}
\vepsi _{i,1} \\
\vepsi _{i,2}%
\end{array}%
\right) =\left(
\begin{array}{c}
\mS_{1}\vepsi_{i}  \\
\mS_{2}\vepsi_{i}
\end{array}%
\right) ,
\end{align*}
where
\begin{eqnarray*}
\quad \mS_{1}:=\left(
\begin{array}{cc}
\mI_{T_{1}} & \mZeros%
\end{array}%
\right) ,\quad \mS_{2}:=\left(
\begin{array}{cc}
\mZeros & \mI_{T_{1}}%
\end{array}%
\right). 
\end{eqnarray*}
Then
\begin{align*}
\mH_{T_{1}}\vy_{i,1-} &=\mH_{T_{1}}\mA_{T_{1}}\vepsi _{i,1}=\mD_{T_{1}}\vepsi
_{i,1}=\mD_{T_{1}}\mS_{1}\vepsi_{i},  \\
\mH_{T_{1}}\vy_{i,2-} &=\mH_{T_{1}}\mA_{T_{1}}\vepsi _{i,2}=\mD_{T_{1}}\vepsi
_{i,2}=\mD_{T_{1}}\mS_{2}\vepsi_{i}.
\end{align*}
Based on these decompositions, we can expand the two half-sample FE estimators:
\begin{align*}
\widehat{\alpha }_{1,FE} &=1+\frac{\sum_{i}\vepsi _{i,1}^{\prime
}\mH_{T_{1}}\vy_{i,1-}}{\sum_{i}\vy_{i,1-}^{\prime }\mH_{T_{1}}\vy_{i,1-}}=1+\frac{%
\sum_{i}\vepsi_{i} ^{\prime }\mS_{1}^{\prime }\mD_{T_{1}}\mS_{1}\vepsi _{i}}{%
\sum_{i}\vepsi_{i} ^{\prime }\mS_{1}^{\prime }\mD_{T_{1}}^{\prime
}\mD_{T_{1}}\mS_{1}\vepsi_{i} }=1+\frac{\sum_{i}A_{iT,1}}{\sum_{i}B_{iT,1}},\\
\widehat{\alpha }_{2, FE} &=1+\frac{\sum_{i}\vepsi _{i,2}^{\prime
}\mH_{T_{1}}\vy_{i,2-}}{\sum_{i}\vy_{i,2-}^{\prime }\mH_{T_{1}}\vy_{i,2-}}=1+\frac{%
\sum_{i}\vepsi_{i} ^{\prime }\mS_{2}^{\prime }\mD_{T_{1}}\mS_{2}\vepsi _{i}}{%
\sum_{i}\vepsi_{i} ^{\prime }\mS_{2}^{\prime }\mD_{T_{1}}^{\prime
}\mD_{T_{1}}\mS_{2}\vepsi_{i} }=1+\frac{\sum_{i}A_{iT,2}}{\sum_{i}B_{iT,2}}.
\end{align*}
We can now combine the results for the quadratic forms and the definitions of $A_{iT}$, $A_{iT,1}$, $A_{iT,2}$, $B_{iT}$, $B_{iT,1}$, and $B_{iT,2}$ to derive all relevant covariance terms. In particular,
\begin{align*}
\covar(A_{iT},A_{iT,1}) &=\tr (\mS_{1}(\mD_{T}+\mD_{T}^{\prime
})\mS_{1}^{\prime }\mD_{T_{1}}), \\
\covar(A_{iT},A_{iT,2}) &=\tr (\mS_{2}(\mD_{T}+\mD_{T}^{\prime
})\mS_{2}^{\prime }\mD_{T_{1}}), \\
\covar(A_{iT},B_{iT,1}) &=\tr (\mS_{1}(\mD_{T}+\mD_{T}^{\prime
})\mS_{1}^{\prime }\mD_{T_{1}}^{\prime }\mD_{T_{1}}), \\
\covar(A_{iT},B_{iT,2}) &=\tr (\mS_{2}(\mD_{T}+\mD_{T}^{\prime
})\mS_{2}^{\prime }\mD_{T_{1}}^{\prime }\mD_{T_{1}}), \\
\covar(B_{iT},A_{iT,1}) &=2\tr (\mS_{1}\mD_{T}^{\prime }\mD_{T}\mS_{1}^{\prime
}\mD_{T_{1}}), \\
\covar(B_{iT},A_{iT,2}) &=2\tr (\mS_{2}\mD_{T}^{\prime }\mD_{T}\mS_{2}^{\prime
}\mD_{T_{1}}), \\
\covar(B_{iT},B_{iT,1}) &=2\tr (\mS_{1}\mD_{T}^{\prime }\mD_{T}\mS_{1}^{\prime
}\mD_{T_{1}}^{\prime }\mD_{T_{1}}) ,\\
\covar(B_{iT},B_{iT,2}) &=2\tr (\mS_{2}\mD_{T}^{\prime }\mD_{T}\mS_{2}^{\prime
}\mD_{T_{1}}^{\prime }\mD_{T_{1}}).
\end{align*}%
As in \citet{Hahn2002}, we have
\begin{align*}
\tr (\mD_{T}^{\prime }\mD_{T}) &=\frac{1}{6}(T^{2}-1),\\ 
\tr
(\mD_{T}^{\prime }\mD_{T})^{2}&=\frac{1}{90}T^{4}+\frac{1}{36}T^{2}-\frac{7}{180},\\
\tr (\mD_{T}^{\prime }\mD_{T}^{2})&=-\frac{1}{12}(T^{2}-1), \\
\tr (\mD_{T}+\mD_{T}^{\prime })^{2} &=\frac{1}{6}T^{2}+T-\frac{7}{6},\\
\tr (\mD_{T}^{2})&=\frac{1}{2}\tr [(\mD_{T}+\mD_{T}^{\prime })^{2}]-%
\tr (\mD_{T}^{\prime }\mD_{T})=-\frac{1}{12}T^{2}+\frac{1}{2}T-\frac{5}{12}.
\end{align*}%
Moreover, define the $T \times 1$ vectors
\begin{align*}
\vu_{T}&:=(T-1,T-2,\ldots,0)',\\
\vw_{T}&:=(\vones_{T}'\vu_{T},\ldots,\vones_{1}'\vu_{1})',\\
\text{\textrm{rev}}\vu_{T}&:=(0,1,\ldots,T-1)'.
\end{align*}
Note that
\begin{eqnarray*}
\text{\textrm{rev}}\vu_{T}+\vu_{T}&=&(T-1)\vones_{T}.
\end{eqnarray*}
We can use these definitions to establish the following algebraic results:
\begin{align*}
\vones _{T}^{\prime }\vu_{T} &=\frac{1}{2}T\left( T-1\right) ,\\
\vu_{T}^{\prime }\vu_{T} &=\frac{1}{6}T\left( 2T-1\right) \left( T-1\right),\\
\vu_{T}^{\prime }\text{\textrm{rev}}\vu_{T} &=\frac{1}{6}T\left(
T-1\right) \left( T-2\right),  \\
\vones _{T}^{\prime }\vw_{T} &=\frac{1}{6}T\left( T^{2}-1\right) ,\\
\vw_{T}^{\prime }\vw_{T}&=\frac{1}{60}T\left( T^{2}-1\right) \left(
3T^{2}-2\right),  \\
\vu_{T}^{\prime }\vw_{T} &=\frac{1}{24}T\left( T^{2}-1\right) \left(
3T-2\right) ,\\
\vw_{T}^{\prime }\text{\textrm{rev}}\vu_{T}&=\frac{1}{24}%
T\left( T^{2}-1\right) \left( T-2\right).
\end{align*}%
Furthermore,
\begin{align*}
\vones _{T}^{\prime }\mA_{T} &=\vu_{T}^{\prime },\\
 \mA_{T}\vones _{T}&=\text{
\textrm{rev}}\vu_{T},\\
\mA_{T}\vu_{T}&=\vones _{T}\vones _{T}^{\prime
}\vu_{T}-\vw_{T}, \\
\vones _{T}^{\prime }\mA_{T}\vones _{T} &=\vu_{T}^{\prime }\vones _{T},\\
\vu_{T}^{\prime }\mA_{T}\vones _{T}&=\vu_{T}^{\prime }\text{\textrm{rev}}%
\vu_{T},\\
\vu_{T}^{\prime }\mA_{T}^{\prime }\vones _{T}&=\vu_{T}^{\prime
}\vu_{T},\\
\vones _{T}^{\prime }\mA_{T}^{\prime }\mA_{T}\vones
_{T}&=\vu_{T}^{\prime }\vu_{T},
\end{align*}
and
\begin{align*}
\vu_{T}^{\prime }\mA_{T}^{\prime }\mA_{T}\vu_{T} =(\vones _{T}\vones _{T}^{\prime
}\vu_{T}-\vw_{T})^{\prime }(\vones _{T}\vones _{T}^{\prime }\vu_{T}-\vw_{T})
=T\left(
\vones _{T}^{\prime }\vu_{T}\right) ^{2}-2\vones _{T}^{\prime }\vu_{T}\vones
_{T}^{\prime }\vw_{T}+\vw_{T}^{\prime }\vw_{T}.
\end{align*}
Thus,
\begin{align*}
\mD_{T} &=\mA_{T}-\frac{1}{T}\vones _{T}\vu_{T}^{\prime },\\
\vones
_{T}^{\prime }\mD_{T}&=\vzeros_{T}, \\
\mD_{T}^{\prime }\mD_{T} &=\mA_{T}^{\prime }\mD_{T}=\mD_{T}^{\prime }\mA_{T},\\
\vu_{T}^{\prime }\mD_{T}\vones _{T} &=\vu_{T}^{\prime }\mA_{T}\vones _{T}-\frac{1}{T}%
\left( \vu_{T}^{\prime }\vones _{T}\right) ^{2}=-\frac{1}{12}T\left(
T^{2}-1\right),  \\
\vu_{T}^{\prime }\mD_{T}\vu_{T} &=(\vu_{T}^{\prime }\vones _{T})^{2}-\vu_{T}^{\prime
}\vw_{T}-\frac{1}{T}\vu_{T}^{\prime }\vones _{T}\vu_{T}^{\prime }\vu_{T}=-\frac{1}{24}%
T^{2}\left( T^{2}-1\right) =\frac{T}{2}\vu_{T}^{\prime }\mD_{T}\vones _{T}.
\end{align*}%
Using these results, we obtain
\begin{align*}
\mD_{T}\mD_{T}^{\prime } &=\mA_{T}\mA_{T}^{\prime }-\frac{1}{T}\mA_{T}\vu_{T}\vones
_{T}^{\prime }-\frac{1}{T}\vones _{T}\vu_{T}^{\prime }\mA_{T}^{\prime }+\frac{1}{%
T^{2}}\vu_{T}^{\prime }\vu_{T}\vones _{T}\vones _{T}^{\prime } ,\\ 
\mD_{T}^{\prime }\mD_{T} &=\mA_{T}^{\prime }\mA_{T}-\frac{1}{T}\mA_{T}^{\prime }\vones
_{T}\vu_{T}^{\prime }-\frac{1}{T}\vu_{T}\vones _{T}^{\prime }\mA_{T}+\frac{1}{T}%
\vu_{T}\vu_{T}^{\prime }, \\
\vones _{T}^{\prime }\mD_{T}^{\prime }\mD_{T}\vones _{T} &=\vones _{T}^{\prime }\mA_{T}^{\prime }\mD_{T}\vones _{T}=\text{\textrm{rev}}\vu_{T}^{\prime }\mD_{T}\vones _{T}=-\vu_{T}^{\prime }\mD_{T}\vones _{T},\\
\vu_{T}^{\prime }\mD_{T}^{\prime }\mD_{T}\vones _{T} &=\vu_{T}^{\prime }\mD_{T}^{\prime }\mA_{T}\vones _{T}=\vu_{T}^{\prime }\mD_{T}^{\prime }\text{\textrm{rev}}\vu_{T}=-\vu_{T}^{\prime }\mD_{T}^{\prime }\vu_{T}=-\frac{T}{2}\vu_{T}^{\prime }\mD_{T}\vones _{T},\\
\vu_{T}^{\prime }\mD_{T}^{\prime }\mD_{T}\vu_{T} &=\vu_{T}^{\prime }\mA_{T}^{\prime
}\mA_{T}\vu_{T}-\frac{1}{T}(\vu_{T}^{\prime }\vu_{T})^{2}=\frac{1}{180}T\left(
T^{2}-1\right) \left( 4T^{2}-1\right) =-\frac{1}{15}\left( 4T^{2}-1\right)
\vu_{T}^{\prime }\mD_{T}\vones _{T}.
\end{align*}%
Notice that we can use these results to derive all the expressions within the $\tr(\cdot)$ operators of the relevant covariances. In particular,
\begin{align*}
\mA_{T} &=\left(
\begin{array}{cc}
\mA_{T_{1}} & \mZeros \\
\vones _{T_{1}}\vones _{T_{1}}^{\prime } & \mA_{T_{1}}%
\end{array}%
\right),\\
\vu_{T}^{\prime }&=\left(
\begin{array}{cc}
\vu_{T_{1}}^{\prime }+\frac{T}{2}\vones _{T_{1}}^{\prime } & \vu_{T_{1}}^{\prime }%
\end{array}%
\right), \\
\mD_{T}\mS_{1}^{\prime } &=\left(
\begin{array}{c}
\mA_{T_{1}} \\
\vones _{T_{1}}\vones _{T_{1}}^{\prime }%
\end{array}%
\right) -\frac{1}{T}\vones _{T}\left( \vu_{T_{1}}^{\prime }+\frac{T}{2}\vones
_{T_{1}}^{\prime }\right) =\left(
\begin{array}{c}
\mD_{T_{1}}+\frac{1}{T}\vones _{T_{1}}\vu_{T_{1}}^{\prime }-\frac{1}{2}\vones
_{T_{1}}\vones _{T_{1}}^{\prime } \\
-\left( \frac{1}{T}\vones _{T_{1}}\vu_{T_{1}}^{\prime }-\frac{1}{2}\vones
_{T_{1}}\vones _{T_{1}}^{\prime }\right)
\end{array}%
\right) , \\
\mD_{T}\mS_{2}^{\prime } &=\left(
\begin{array}{c}
\mZeros \\
\mA_{T_{1}}%
\end{array}%
\right) -\frac{1}{T}\vones _{T}\vu_{T_{1}}^{\prime }=\left(
\begin{array}{c}
-\frac{1}{T}\vones _{T_{1}}\vu_{T_{1}}^{\prime } \\
\mD_{T_{1}}+\frac{1}{T}\vones _{T_{1}}\vu_{T_{1}}^{\prime }%
\end{array}%
\right),  \\
\mS_{1}\mD_{T}\mS_{1}^{\prime } &=\mD_{T_{1}}+\frac{1}{T}\vones _{T_{1}}\vu_{T_{1}}^{\prime
}-\frac{1}{2}\vones _{T_{1}}\vones _{T_{1}}^{\prime } ,\\
\mS_{2}\mD_{T}\mS_{2}^{\prime } &=\mD_{T_{1}}+\frac{1}{T}\vones _{T_{1}}\vu_{T_{1}}^{\prime }
=\mS_{1}\mD_{T}\mS_{1}^{\prime }+\frac{1}{2}\vones _{T_{1}}\vones _{T_{1}}^{\prime }.
\end{align*}%
Furthermore,
\begin{align*}
\mS_{1}\mD_{T}^{\prime }\mD_{T}\mS_{1}^{\prime } &=\mD_{T_{1}}^{\prime }\mD_{T_{1}}+2\left(
\frac{1}{2}\vones_{T_{1}}\vones _{T_{1}}^{\prime }-\frac{1}{T}\vones
_{T_{1}}\vu_{T_{1}}^{\prime }\right) ^{\prime }\left( \frac{1}{2}\vones _{T_{1}}\vones
_{T_{1}}^{\prime }-\frac{1}{T}\vones _{T_{1}}\vu_{T_{1}}^{\prime }\right)  \\
&=\mD_{T_{1}}^{\prime }\mD_{T_{1}}+\left(\frac{1}{T}\vu_{T_{1}}\vu_{T_{1}}^{\prime }-\frac{1}{2}%
\vu_{T_{1}}\vones _{T_{1}}^{\prime }-\frac{1}{2}\vones _{T_{1}}\vu_{T_{1}}^{\prime }+\frac{%
T}{4}\vones _{T_{1}}\vones _{T_{1}}^{\prime }\right).
\end{align*}%
Thus, 
\begin{align*}
\mS_{2}\mD_{T}^{\prime }\mD_{T}\mS_{2}^{\prime } 
&=\mD_{T_{1}}^{\prime }\mD_{T_{1}}+\frac{1}{T}\vu_{T_{1}}\vu_{T_{1}}^{\prime }\\
&=\mS_{1}\mD_{T}^{\prime }\mD_{T}\mS_{1}^{\prime }+\left(\frac{1}{2}\vu_{T_{1}}\vones _{T_{1}}^{\prime }+\frac{1}{2}\vones _{T_{1}}\vu_{T_{1}}^{\prime }-\frac{T}{4}\vones _{T_{1}}\vones _{T_{1}}^{\prime }\right).
\end{align*}%
Substituting these expressions
into the corresponding covariance terms gives
\begin{align*}
\covar(A_{iT},A_{iT,1}) =&\tr \left(\left( \mD_{T_{1}}+\mD_{T_{1}}^{\prime }+\frac{1%
}{T}\left( \vones _{T_{1}}\vu_{T_{1}}^{\prime }+\vu_{T_{1}}\vones _{T_{1}}^{\prime
}\right) -\vones _{T_{1}}\vones _{T_{1}}^{\prime }\right) \mD_{T_{1}}\right) \\
=&\tr ( \mD_{T_{1}}^{2})+\tr (\mD_{T_{1}}^{\prime }\mD_{T_{1}})+\frac{1}{T}%
\vu_{T_{1}}^{\prime }\mD_{T_{1}}\vones _{T_{1}} \\
=&\left( -\frac{1}{12}\left( T_{1}\right) ^{2}+\frac{1}{2}\left( T_{1}\right) -%
\frac{5}{12}\right)\\
&+\left( \frac{1}{6}(\left( T_{1}\right) ^{2}-1)\right) +%
\frac{1}{T}\left( -\frac{1}{12}\left( T_{1}\right) \left( \left( T_{1}\right)
^{2}-1\right) \right)  \\
=&\frac{1}{96}T^{2}+\frac{1}{4}T-\frac{13}{24}, \\
\covar(A_{iT},A_{iT,2})=&\tr \left(\left( \mD_{T_{1}}+\mD_{T_{1}}^{\prime }+\frac{1}{T}%
\left( \vones _{T_{1}}\vu_{T_{1}}^{\prime }+\vu_{T_{1}}\vones _{T_{1}}^{\prime }\right)
\right) \mD_{T_{1}}\right)\\
=&\covar(A_{iT},A_{iT,1}), \\
\covar(A_{iT},B_{iT,1}) =&\tr\left( \left( \mD_{T_{1}}+\mD_{T_{1}}^{\prime }+\frac{1%
}{T}\left( \vones _{T_{1}}\vu_{T_{1}}^{\prime }+\vu_{T_{1}}\vones _{T_{1}}^{\prime
}\right) -\vones _{T_{1}}\vones _{T_{1}}^{\prime }\right) \mD_{T_{1}}^{\prime }\mD_{T_{1}}\right) \\
=&2\tr (\mD_{T_{1}}^{\prime }\mD_{T_{1}}^{2})+\frac{2}{T}\vu_{T_{1}}^{\prime
}\mD_{T_{1}}^{\prime }\mD_{T_{1}}\vones _{T_{1}}-\vones _{T_{1}}^{\prime }\mD_{T_{1}}^{\prime
}\mD_{T_{1}}\vones _{T_{1}} \\
=&2\tr (\mD_{T_{1}}^{\prime }\mD_{T_{1}}^{2})+\frac{1}{2}\vu_{T_{1}}^{\prime
}\mD_{T_{1}}\vones _{T_{1}} \\
=&2\left( -\frac{1}{12}\left( T_{1}\right) ^{2}+\frac{1}{12}\right) +\frac{1}{%
2}\left( -\frac{1}{12}\left( T_{1}\right) \left( \left( T_{1}\right)
^{2}-1\right) \right)  \\
=&-\frac{1}{192}T^{3}-\frac{1}{24}T^{2}+\frac{1}{48}T+\frac{1}{6}, \\
\covar(A_{iT},B_{iT,2}) =&\tr \left(\left( \mD_{T_{1}}+\mD_{T_{1}}^{\prime }+\frac{1%
}{T}\left( \vones _{T_{1}}\vu_{T_{1}}^{\prime }+\vu_{T_{1}}\vones _{T_{1}}^{\prime
}\right) \right) \mD_{T_{1}}^{\prime }\mD_{T_{1}}\right) \\
=&2\tr (\mD_{T_{1}}^{\prime }\mD_{T_{1}}^{2})+\frac{2}{T}\vu_{T_{1}}^{\prime
}\mD_{T_{1}}^{\prime }\mD_{T_{1}}\vones _{T_{1}} \\
=&2\tr (\mD_{T_{1}}^{\prime }\mD_{T_{1}}^{2})-\frac{1}{2}\vu_{T_{1}}^{\prime
}\mD_{T_{1}}\vones _{T_{1}} \\
=&2\left( -\frac{1}{12}\left( T_{1}\right) ^{2}+\frac{1}{12}\right) -\frac{1}{%
2}\left( -\frac{1}{12}\left( T_{1}\right) \left( \left( T_{1}\right)
^{2}-1\right) \right)  \\
=&\frac{1}{192}T^{3}-\frac{1}{24}T^{2}-\frac{1}{48}T+\frac{1}{6}, \\
\covar(B_{iT},A_{iT,1}) =&2\tr \left(\left( \mD_{T_{1}}^{\prime }\mD_{T_{1}}+\frac{1%
}{T}\vu_{T_{1}}\vu_{T_{1}}^{\prime }-\frac{1}{2}\vu_{T_{1}}\vones _{T_{1}}^{\prime }-\frac{1%
}{2}\vones _{T_{1}}\vu_{T_{1}}^{\prime }+\frac{T}{4}\vones _{T_{1}}\vones
_{T_{1}}^{\prime }\right) \mD_{T_{1}}\right) \\
=&2\left( \tr (\mD_{T_{1}}^{\prime }\mD_{T_{1}}^{2})+\frac{1}{T}%
\vu_{T_{1}}^{\prime }\mD_{T_{1}}\vu_{T_{1}}-\frac{1}{2}\vu_{T_{1}}^{\prime }\mD_{T_{1}}\vones
_{T_{1}}\right)  \\
=&2\left( \tr (\mD_{T_{1}}^{\prime }\mD_{T_{1}}^{2})-\frac{1}{4}%
\vu_{T_{1}}^{\prime }\mD_{T_{1}}\vones _{T_{1}}\right) \\
=&\covar(A_{iT},B_{iT,2}), \\
\end{align*}%
\begin{align*}
\covar(B_{iT},A_{iT,2}) =&2\tr \left(\left( \mD_{T_{1}}^{\prime }\mD_{T_{1}}+\frac{1%
}{T}\vu_{T_{1}}\vu_{T_{1}}^{\prime }\right) \mD_{T_{1}}\right) \\
=&2\tr (\mD_{T_{1}}^{\prime }\mD_{T_{1}}^{2})+\frac{1}{2}\vu_{T_{1}}^{\prime
}\mD_{T_{1}}\vones _{T_{1}}\\
=&\covar(A_{iT},B_{iT,1}), \\
\covar(B_{iT},B_{iT,1}) =&2\tr \left(\left( \mD_{T_{1}}^{\prime }\mD_{T_{1}}+\frac{1%
}{T}\vu_{T_{1}}\vu_{T_{1}}^{\prime }-\frac{1}{2}\vu_{T_{1}}\vones _{T_{1}}^{\prime }-\frac{1%
}{2}\vones _{T_{1}}\vu_{T_{1}}^{\prime }+\frac{T}{4}\vones _{T_{1}}\vones
_{T_{1}}^{\prime }\right) \mD_{T_{1}}^{\prime }\mD_{T_{1}}\right) \\
=&2\left( \tr \left((\mD_{T_{1}}^{\prime }\mD_{T_{1}})^{2}\right)+\frac{1}{T}%
\vu_{T_{1}}^{\prime }\mD_{T_{1}}^{\prime }\mD_{T_{1}}\vu_{T_{1}}\right)  \\
=&2\left( \left( \frac{1}{90}\left( T/2\right) ^{4}+\frac{1}{36}\left(
T/2\right) ^{2}-\frac{7}{180}\right) +\frac{1}{T}\left( \frac{1}{180}\left(
T/2\right) \left( \left( T/2\right) ^{2}-1\right) \left( 4\left( T/2\right)
^{2}-1\right) \right) \right)  \\
=&\frac{1}{360}T^{4}+\frac{1}{144}T^{2}-\frac{13}{180},
\covar(B_{iT},B_{iT,2})=&2\tr \left(\left(\left( \mD_{T_{1}}^{\prime }\mD_{T_{1}}\right)+\frac{1}{T}%
\vu_{T_{1}}\vu_{T_{1}}^{\prime }\right) \mD_{T_{1}}^{\prime }\mD_{T_{1}}\right)\\
=&\covar%
(B_{iT},B_{iT,1}).
\end{align*}%

\end{proof}

\begin{proof}[\bf Proof of Proposition \ref{proposition::HPJ_distribution_delta}]
Because the HPJ estimator is a linear combination
of the three FE estimators, its distribution is determined by their joint distribution. In
particular,
\begin{align*}
\widehat{\alpha }_{HPJ} =&2\widehat{\alpha }_{FE}-\frac{1}{2}(\widehat{\alpha }_{1,FE }+%
\widehat{\alpha }_{2,FE})=1+\vj^{\prime }\vx,\quad \vj=\left(
\begin{array}{c}
2 \\
-1/2 \\
-1/2%
\end{array}%
\right),
\end{align*}
where
\begin{eqnarray*}
\quad \vx:=\left(
\begin{array}{c}
\sum_{i}A_{iT}/\sum_{i}B_{iT} \\
\sum_{i}A_{iT,1}/\sum_{i}B_{iT,1} \\
\sum_{i}A_{iT,2}/\sum_{i}B_{iT,2}%
\end{array}%
\right).
\end{eqnarray*}%
Taking sequential asymptotic limits, first as $N\to\infty$ and then as $T\to\infty$, gives
\begin{equation*}
T\sqrt{N}\left( \widehat{\alpha }_{HPJ}-1-b_{T,\alpha_{HPJ}}\right) \dto%
\calN(0,\vj'\mW\vj),
\end{equation*}
where
\begin{equation*}
 b_{T,\alpha_{HPJ}}:=   (2 b_{T,\alpha_{FE}}-0.5 b_{T,\alpha_{1,FE}}-0.5 b_{T,\alpha_{2,FE}}),
\end{equation*}
and
\begin{equation*}
\quad \mW:=\lim_{N,T\to \infty }NT^{2}\var%
\left( \vx\right) =\left(
\begin{array}{ccc}
w_{11} & w_{12} & w_{13} \\
w_{12} & w_{22} & 0 \\
w_{13} & 0 & w_{33}%
\end{array}%
\right),
\end{equation*}%
where $w_{22}=w_{33}=4w_{11}$.
Thus,
\begin{equation*}
\vj'\mW\vj=2(3w_{11}-w_{12}-w_{13}).
\end{equation*}
Here the use of sequential asymptotic arguments is inconsequential for the final results, as the joint limit theory can be established similarly to Lemma 12 in \citet{Hahn2002}.

From \citet{Hahn2002} we know that $w_{11}=51/5$, while the remaining components $w_{12}$ and $w_{13}$ remain to be determined. In particular,
\begin{align*}
w_{12} =&\lim_{T\to
\infty }T^{2}\frac{\covar(A_{iT}+\frac{3}{T+1}B_{iT},A_{iT,1}+\frac{3}{T/2+1}B_{iT,1})%
}{\E[B_{iT}]\E[B_{iT,1}]} \\
=&\lim_{T\to \infty }T^{2}\frac{\covar(A_{iT},A_{iT,1})+\frac{3}{T/2+1%
}\covar(A_{iT},B_{iT,1})}{\E[B_{iT}]\E[B_{iT,1}]}\\
&+\lim_{T\to \infty }T^{2}\frac{\frac{3}{T+1}\covar(B_{iT},A_{iT,1})+\frac{3}{T+1}\frac{%
3}{T/2+1}\covar(B_{iT},B_{iT,1})}{\E[B_{iT}]\E[B_{iT,1}]}
\end{align*}
and
\begin{align*}
w_{13} =&\lim_{T\to \infty }T^{2}\frac{\covar(A_{iT},A_{iT,2})+\frac{3}{%
T/2+1}\covar(A_{iT},B_{iT,2})}{\E[B_{iT}]\E[B_{iT,2}]}\\
&+\lim_{T\to \infty }T^{2}\frac{\frac{3}{T+1}\covar(B_{iT},A_{iT,2})+\frac{3}{T+1}%
\frac{3}{T/2+1}\covar(B_{iT},B_{iT,2})}{\E[B_{iT}]\E[B_{iT,2}]}.
\end{align*}
Since $\E[B_{iT,1}]=\E[B_{iT,2}]$, we have
\begin{align*}
w_{12}+w_{13}=&\lim_{T\to \infty }T^{2}\left(\frac{\covar(A_{iT},A_{iT,1}+A_{iT,2})+\frac{3}{T+1}\frac{3}{T/2+1}\covar(B_{iT},B_{iT,1}+B_{iT,2})}{\E[B_{iT}]\E[B_{iT,1}]}\right)\\
&+\lim_{T\to \infty }T^{2}\left(\frac{\frac{3}{T/2+1}\covar(A_{iT},B_{iT,1}+B_{iT,2})+\frac{3}{T+1}\covar(B_{iT},A_{iT,1}+A_{iT,2})}{\E[B_{iT}]\E[B_{iT,1}]}\right)\\
=& I+II.
\end{align*}
Here,
\begin{align*}
I&=\lim_{T\to \infty }T^{2}\left(\frac{\covar(A_{iT},A_{iT,1}+A_{iT,2})+18\left(\frac{1}{T}\right)^{2}\left(\covar(B_{iT},B_{iT,1}+B_{iT,2})\right)}{\E[B_{iT}]\E[B_{iT,1}]}\right)
\end{align*}
and
\begin{align*}
II&=\lim_{T\to \infty }T^{2}\left(\frac{\frac{3}{T/2+1}\covar(A_{iT},B_{iT,1}+B_{iT,2})+\frac{3}{T+1}\covar(B_{iT},A_{iT,1}+A_{iT,2})}{\E[B_{iT}]\E[B_{iT,1}]}\right).
\end{align*}
From Lemma \ref{lemma::aux_jackknife} we know that
\begin{align*}
\covar(A_{iT},A_{iT,1}) =&\covar(A_{iT},A_{iT,2})=\frac{1}{96}T^{2}+\smallO(T^{2}), \\
\covar(A_{iT},B_{iT,1}) =&\covar(B_{iT},A_{iT,2})=-\frac{1}{192}T^{3}+\smallO(T^{3}), \\
\covar(A_{iT},B_{iT,2}) =&\covar(B_{iT},A_{iT,1})=\frac{1}{192}T^{3}+\smallO(T^{3}), \\
\covar(B_{iT},B_{iT,1}) =&\covar(B_{iT},B_{iT,2})=\frac{1}{360}T^{4}+\smallO(T^{4}).
\end{align*}
Thus,
\begin{align*}
\covar(A_{iT},B_{iT,1})+\covar(A_{iT},B_{iT,2})=&\smallO(T^{3}),\\
\covar(B_{iT},A_{iT,1})+\covar(B_{iT},A_{iT,2})=&\smallO(T^{3}).
\end{align*}
This, together with
\begin{align*}
\E[B_{iT}] =& \frac{T^{2}}{6}+\smallO(T^{2}), \\
\E[B_{iT,2}] =& \frac{T^{2}}{24}+\smallO(T^{2}),
\end{align*}%
implies that the second term $II$ is of lower order:
\begin{align*}
II=&\lim_{T\to\infty}T^{2}\left(\frac{\frac{\smallO(T^{3})}{T}+\frac{\smallO(T^{3})}{T}}{\largeO(T^{4})}\right)\\
=&0.
\end{align*}
For the leading term $I$, we obtain
\begin{align*}
I&=\lim_{T\to\infty}T^{2}144\left(\frac{2\covar(A_{iT},A_{iT,1})+\frac{36}{T^{2}}\covar(B_{iT},B_{iT,1})}{T^{4}}\right)\\
&=144\left(\frac{2}{96}+\frac{36}{360}\right)\\
&=3+\frac{72}{5}\\
&=\frac{87}{5}.
\end{align*}
This gives
\begin{align*}
\vj'\mW\vj&=2(3w_{11}-w_{12}-w_{13})\\
&=2\left(\frac{153}{5}-\frac{87}{5}\right)\\
&=\frac{132}{5}.
\end{align*}
Hence,
\begin{equation*}
\sqrt{NT^{2}}\left( \widehat{\alpha }_{HPJ}-1- b_{T,\alpha_{HPJ}}\right) \dto%
\calN\left( 0,\frac{132}{5}\right).
\end{equation*}
Finally, we show that the bias is negligible asymptotically for $\kappa>1/2$. Using the result in \citet{Hahn2002} gives
\begin{align*}
 b_{T,\alpha_{HPJ}}&=   (2 b_{T,\alpha_{FE}}-0.5 b_{T,\alpha_{1,FE}}-0.5 b_{T,\alpha_{2,FE}})\\
 &=-2\frac{3}{T+1}+\frac{3}{1+T/2}\\
 &=-\frac{6}{(T+1)(T+2)}\\
 &=\largeO(T^{-2}).
\end{align*}
Consequently,
\begin{align*}
    \sqrt{NT^{2}}b_{T,\alpha_{HPJ}}=\largeO_{P}(\sqrt{N}T^{-1})=\smallO_{P}(1)
\end{align*}
for $\kappa>1/2$, so
\begin{equation*}
\sqrt{NT^{2}}\left( \widehat{\alpha }_{HPJ}-1\right) \dto%
\calN\left( 0,\frac{132}{5}\right).
\end{equation*}
 This completes the proof.
\end{proof}

\end{document}